\documentclass[dvips]{lib/iopbk2e}
\usepackage[dvips]{graphicx}

\newcommand{\Wo}{\mbox{${\rm W}_0$}}
\newcommand{\msun}{\mbox{${\rm M}_\odot$}}

\newcommand{\msunpc}{ \mbox{ ${\rm M}_\odot \,{\rm pc}^{-3}$ } }
\newcommand{\lsun}{\mbox{${\rm L}_\odot$}}
\newcommand{\rsun}{\mbox{${\rm R}_\odot$}}

\newcommand{\kms}{\mbox{${\rm km~s}^{-1}$}}

\newcommand{\nbody}{\mbox{{{\em N}-body}}}
\newcommand{\Nbody}{\mbox{{{\em N}-body}}}

\newcommand{\tcross}{\mbox{${t_{\rm hm}}$}}
\newcommand{\tcrss}{\mbox{${t_{\rm hm}}$}}
\newcommand{\trlx}{\mbox{${t_{\rm rt}}$}}

\newcommand{\tdf}{{\mbox{${t_{\rm df}}$}}}

\newcommand{\tcc}{\mbox{${t_{\rm cc}}$}}

\newcommand{\tcoll}{\mbox{${t_{\rm coll}}$}}
\newcommand{\tdisr}{\mbox{${t_{\rm disr}}$}}

\newcommand{\tlast}{\mbox{${t_{\rm last}}$}}
\newcommand{\tms}{\mbox{${t_{\rm ms}}$}}

\newcommand{\Ncoll}{\mbox{$N_{\rm coll}$}}
\newcommand{\nbf}{\mbox{$\Gamma_{\rm bf}$}}

\newcommand{\mbh}{\mbox{${m_{\rm bh}}$}}
\newcommand{\Mbh}{\mbox{${m_{\rm bh}}$}}
\newcommand{\mbar}{\mbox{${\langle m \rangle}$}}

\newcommand{\mm}{\mbox{$\langle m \rangle$}}
\newcommand{\mmean}{\mbox{${\langle m \rangle}$}}
\newcommand{\Mgal}{\mbox{${M_{\rm Gal}}$}}

\newcommand{\mdf}{\mbox{${m_{\rm df}}$}}
\newcommand{\Mrun}{\mbox{${M_{\rm run}}$}}
\newcommand{\mseed}{\mbox{${m_{\rm seed}}$}}
\newcommand{\mh}{\mbox{${m_{\rm h}}$}}
\newcommand{\Mh}{\mbox{${M_{\rm h}}$}}

\newcommand{\mdot}{\mbox{$\dot{m}$}}
\newcommand{\Mdot}{\mbox{$\dot{M}$}}
\newcommand{\mcrit}{\mbox{$\dot{m}_{\rm crit}$}}

\newcommand{\rvir}{\mbox{${r_{\rm vir}}$}}

\newcommand{\Rgc}{\mbox{${R_{\rm gc}}$}}

\newcommand{\Rdot}{\mbox{$\dot{R}$}}

\newcommand{\vorb}{\mbox{$v_{\rm c}$}}
\newcommand{\vgc}{\mbox{$v_{\rm disp}$}}

\newcommand{\vrot}{\mbox{$v_{\rm rot}$}}

\newcommand{\erf}{\mbox{${\rm erf}$}}

\newcommand{\vdisp}{\mbox{$\langle v \rangle$}}

\newcommand{\lnL}{\mbox{${\ln \Lambda}$}}

\newcommand{\ncoll}{\mbox{${\cal N}_{\rm coll}$}}

\newcommand{\fcoll}{\mbox{${\rm f}_{\rm coll}$}}
\newcommand{\dmcoll}{\mbox{${\delta m_{\rm coll}}$}}

\newcommand{\tdfg}{\mbox{${T_{\rm f}}$}}
\newcommand{\pc}{\mbox{${\rm pc}$}}

\def\apgt{\ {\raise-.5ex\hbox{$\buildrel>\over\sim$}}\ }
\def\aplt{\ {\raise-.5ex\hbox{$\buildrel<\over\sim$}}\ }

\input ./journals.def
\input ./psfig.sty

\makeindex

\title{COMO School of Physics}
\author{Some authors// all over the world}
\makeindex
\begin{document}
\ifx\href\undefined\else\hypersetup{linktocpage=true}\fi
\maketitle
\pagenumbering{roman}
\setcounter{page}{5}
\tableofcontents
\pagenumbering{arabic}
\setcounter{page}{1}
%

\Chapter[Star Cluster Ecology]
	{The ecology of black holes in star clusters}
	{S. F. Portegies Zwart\\
         University of Amsterdam 
}

\section{Introduction}

In this lecture we investigate the formation and evolution of black
holes in star clusters. 

The star clusters we investigate are generally rich, containing more
than $10^4$ stars, and with a density exceeding $10^4$ stars/pc$^3$.
Among these are young populous clusters, globular cluster and the
nuclei of galaxies.

Under usual circumstances black holes \index{black hole} are formed
from stars with a zero-age main-sequence (ZAMS) mass of at least
25--30\,\msun\, (Maeder 1992; Portegies Zwart, Verbunt \& Ergma 1997;
Ergma \& van den Heuvel 1998).\nocite{1992A&A...264..105M}
\nocite{1997A&A...321..207P}\nocite{1998A&A...331L..29E} These stars
live less than about 10\,Myr, after which a supernova results in a
black hole in the range of 5--20\,\msun\, (Fryer \& Kalogera,
2001).\nocite{2001ApJ...554..548F} The rest of the mass is lost from
the star in the windy phase proceeding the supernova or in the
explosion itself (Heger 2003).\nocite{2003ApJ...591..288H} For a Scalo
(1986)\nocite{scalo86} or Kroupa \& Weidner
(2003)\nocite{2003ApJ...598.1076K} initial mass functions (IMF),
variants of Salpeter (1955)\nocite{1955ApJ...121..161S}, one in
2300-3500 stars collapse to a black hole. So black holes are rare, but
still, the Milky-way Galaxy is populated by 30 to 40 million black
holes.  In the remainder we will refer to these relatively common type
of black holes as {\em stellar mass black holes} or simply as
BH.\index{black hole!stellar mass}

{\em Supermassive black holes} (SMBH)\index{black hole!super massive}
are, with about one per galaxy, among the rarest single objects in the
universe. They have masses of about $10^6$ to $10^{10}$\,\msun\, which
also makes them among the most massive single objects in the Universe.
The best evidence for the existence of a supermassive black hole comes
from the orbit of the star S2 which is in a 15.2\,year orbit around an
unseen object (Sch\"odel et al.\, 2002).\nocite{2002Natur.419..694S}
The mass of this black hole is $3.7 \pm 1.5 \times 10^6$\,\msun, which
puts it directly at the bottom of the mass scale for SMBH's (see also
fig.\,\ref{fig:bhmass}).

The gap in mass between stellar mass black holes and supermassive
black holes may be bridged by a third type, often referred to as {\em
intermediate mass black holes} (IMBH).\index{black hole!intermediate
mass}

In this chapter our main interest is the formation and evolution of
these black hole families in star clusters.  Also the possible
evolutionary link between stellar mass black holes, via intermediate
mass black holes to supermassive black holes will be addressed.  We
mainly focus, however, on the ecology of star clusters.\index{star
cluster!ecology} The term {\em star cluster ecology} was introduced in
1992 by Douglas Heggie\nocite{1992Natur.359..772H} to illustrate the
complicated interplay between stellar evolution and stellar dynamics,
which by their mutual interactions has similarities with biological
systems.

In the here studied ecology we mainly address the gravitational
dynamics, stellar evolution, binary evolution, external influence and
how these seemingly separate effects work together on the star
cluster, much in the same way as in an organism.

\subsection{Setting the stage}

\begin{figure}
\includegraphics[width=\linewidth]{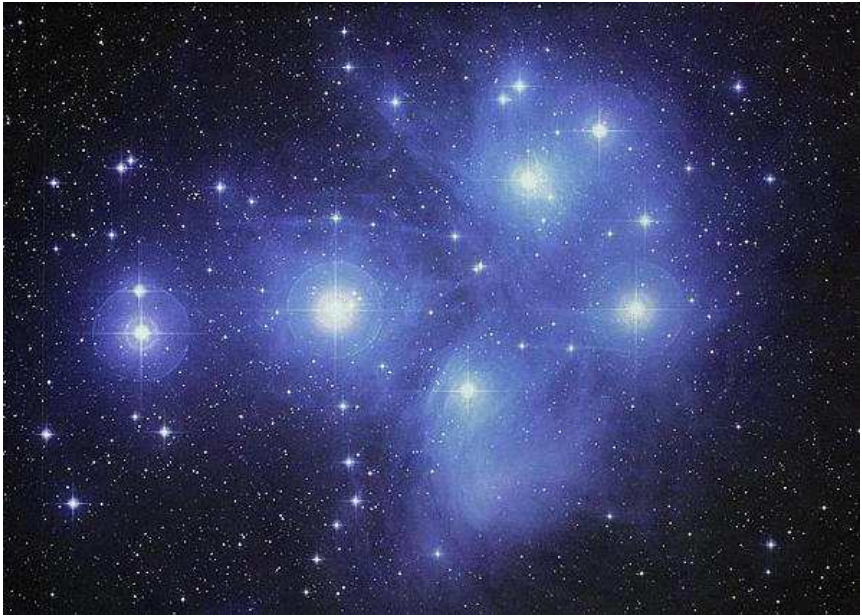}
\caption[]{The Pleiads \index{Pleiads} open star cluster (image by David Malin)
contains a few thousand stars in a volume of a several parsec cubed.
}
\label{fig:pleiades}
\end{figure}

\begin{figure}
\begin{minipage}[b]{0.45\linewidth}
\includegraphics[width=0.9\textwidth]{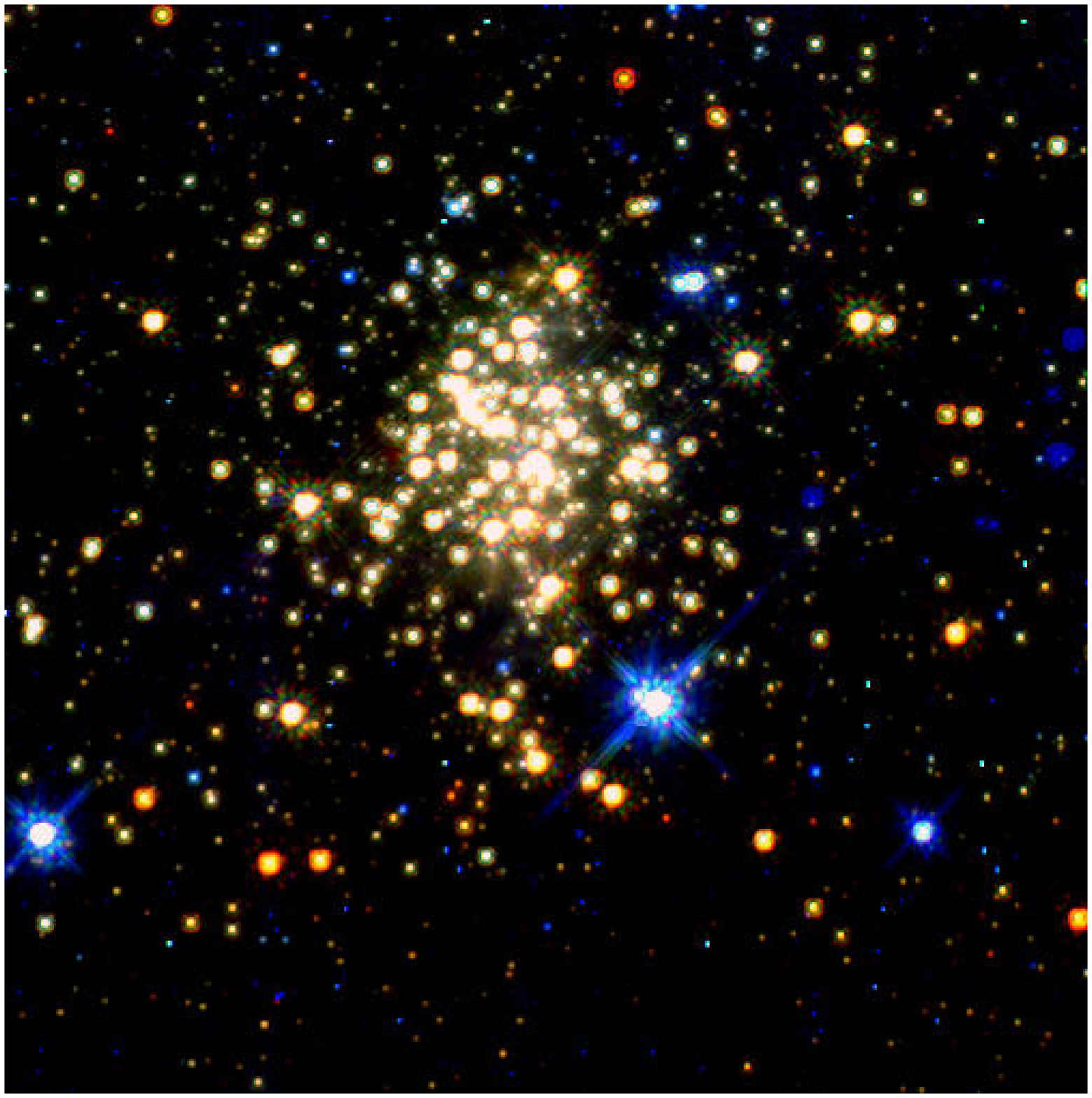}
\end{minipage}\hfill
\begin{minipage}[b]{0.45\linewidth}
\includegraphics[width=0.75\textwidth]{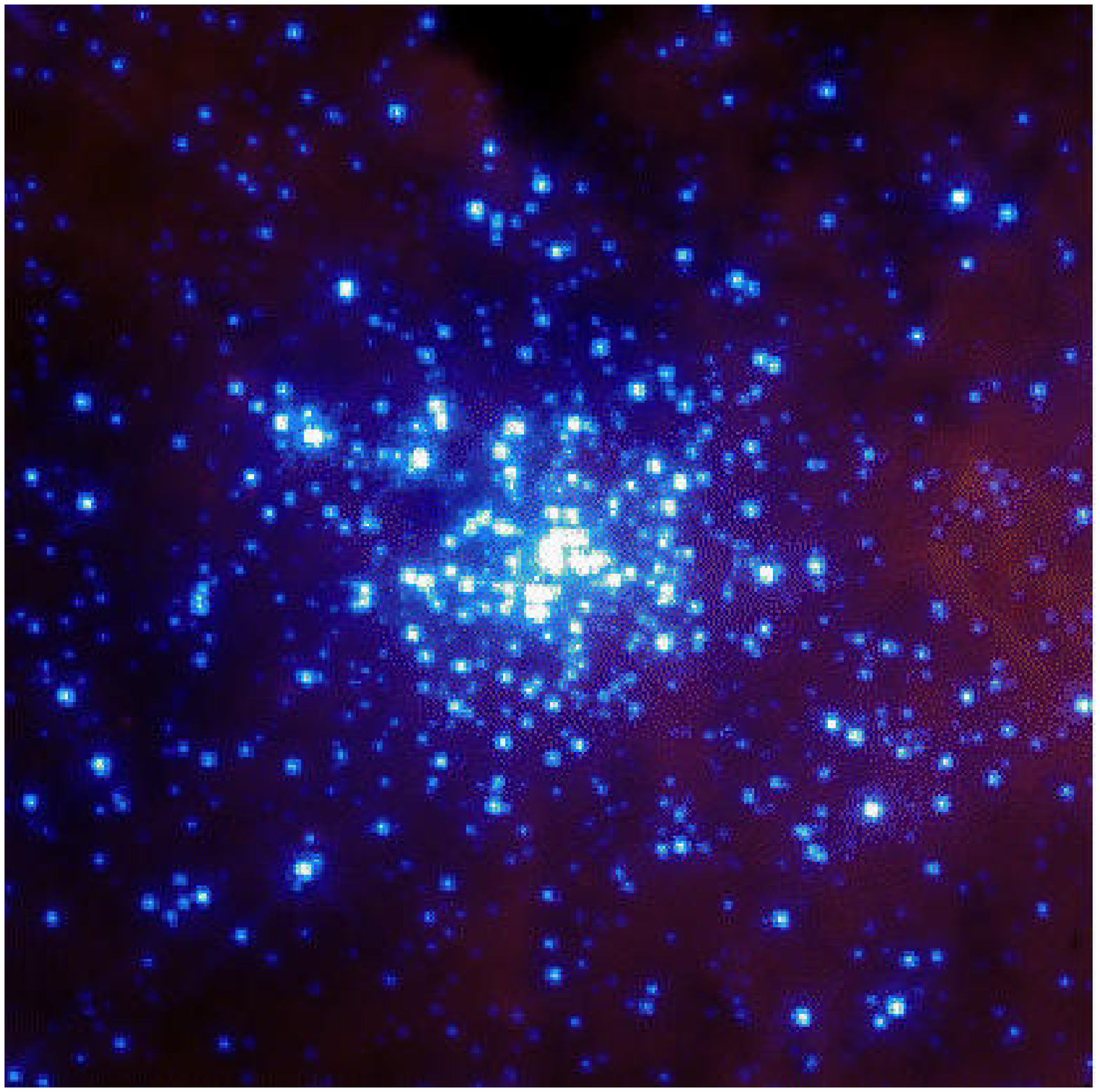}
\end{minipage}\hfill
\caption[]{
{\bf Left}: Arches star cluster (HST image by Figer et al.\,
1999).\nocite{1999ApJ...514..202F} At a projected distance of $\sim
30$\,pc from the Galactic center this cluster is strongly influenced
by external forces. The cluster on the image still looks quite
symmetric since we only observe the inner part. \\
{\bf Right}: The young dense star cluster R136 (NGC 2070, HST image by
N. Walborn)in the 30 Doradus region of the large Magellanic cloud.  }
\label{fig:YoDeC}
\end{figure}

\begin{figure}
\includegraphics[width=0.65\linewidth,angle=-90]{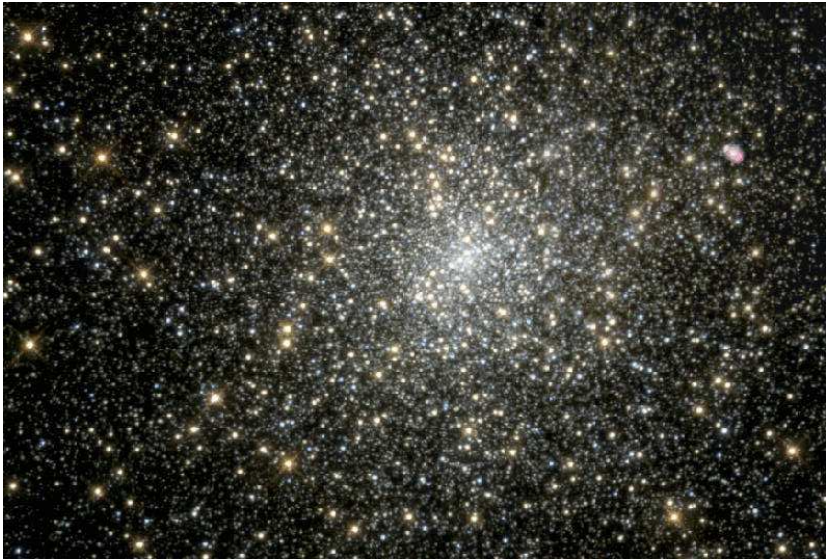}
\caption[]{The globular cluster M15 (NGC 7078, HST image by 
Guhathakurta).\index{M15} }
\label{fig:M15}
\end{figure}

\begin{figure}
\includegraphics[width=\linewidth]{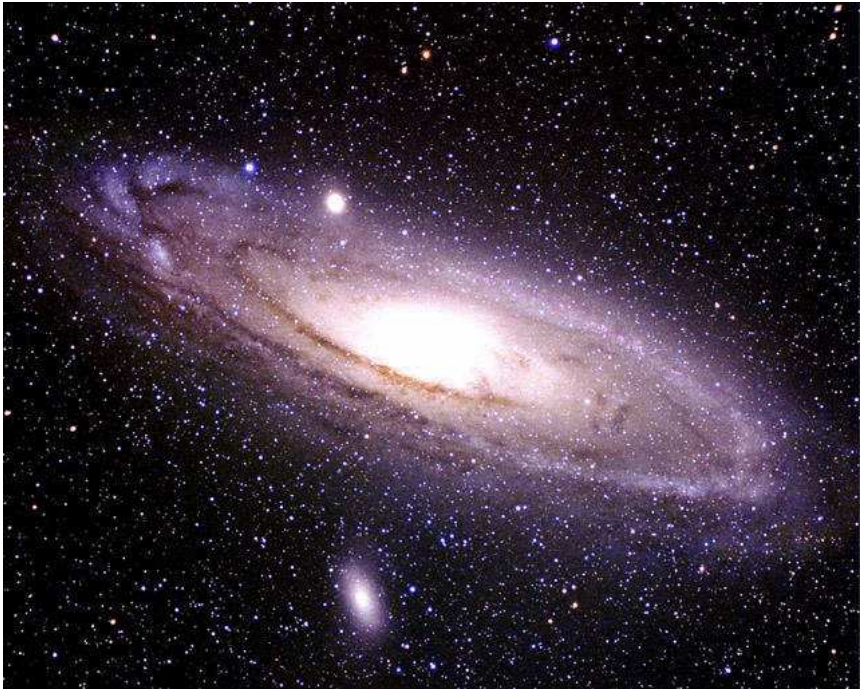}
\caption[]{The Andromeda galaxy (M31, image by Jason Ware).
}
\label{fig:M31}
\end{figure}

The main objects of our study are clusters of stars. There are a
number of distinct families of star clusters, with one common
characteristic: a star cluster\index{star cluster} is a self
gravitating system of stars, all of which have about the same age.  We
make the distinction between various types, among which are: open
cluster like Pleiads \index{Pleiads} (see fig.\,\ref{fig:pleiades}),
young dense cluster like R\,136\index{R136} in the 30 Doradus region
of the large Magellanic cloud or the Arches\index{Arches} cluster (see
fig.\,\ref{fig:YoDeC}), globular cluster like M15\index{M15} (see
fig.\,\ref{fig:M15}).  If galactic nuclei contain a population of
stars with similar ages we could include these in the definition.  The
nucleus of the Andromeda \index{Andromeda} galaxy, depicted in
Figure\,\ref{fig:M31} is an example, though the stellar population in
its nucleus has probably a large spread in ages.

Figure\,\ref{fig:pleiades} shows the $\sim 115$\,Myr old star cluster
Pleiads\index{Pleiads}\footnote{According to the Greek mythology, one
day the great hunter Orion saw the Pleiads as they walked through the
Boeotian countryside, and fancied them. He pursued them for seven
years, until Zeus answered their prayers for delivery and transformed
them into birds (doves or pigeons) placing them among the
stars. Later, when Orion was killed (many conflicting stories as to
how), he was placed in the heavens behind the Pleiads, immortalizing
the chase.}, at a distance of about 135\,pc (Pinfield et al.\,
1998;\nocite{1998MNRAS.299..955P} Raboud \& Mermilliod
1998;\nocite{1998A&A...329..101R} Bouvier et al
1998\nocite{1997A&A...323..139B}).  This cluster contains about 2000
stars, half of which are contained in a sphere with a radius of about
8\,pc.

Figure\,\ref{fig:YoDeC} shows Hubble space telescope (HST) images of
two well known young dense star clusters, Arches\index{Arches} (to the
left) and NGC\,2070\index{NGC 2070} (right) in the Large Magellanic
Cloud. These clusters are the prototypical examples of young dense
star cluster (YoDeCs),\index{star cluster!YoDeC} which are young
($\aplt 10$\,Myr), massive $\sim 10^4$\,\msun\, and dense $n \apgt
10^5$\,stars/pc$^3$. The two clusters show also that the class of
young dense star clusters hosts two families of star clusters; the
strongly tidally perturbed cluster (like Arches) and the isolated
cluster (like R\,136).

Figure\,\ref{fig:M15} give a Hubble's view of the core collapsed
globular cluster M\,15\index{M15} (Guhathakurta
1996).\nocite{1996AJ....111..267G} It is one of the old globular star
clusters in the Milky-way's halo, contains about a million stars, and
the central density is of the order of $n \apgt 10^5$ stars/pc$^3$.
No core has been measured in the density profile, and therefore this
cluster is identified as a collapsed globular \linebreak
(Harris 1996, {\tt
http://physun.physics.mcmaster.ca/Globular.html}).\nocite{1996AJ....112.1487H}

In figure\,\ref{fig:M31} we show an image of the Andromeda
Galaxy\index{Andromeda}\footnote{Andromeda was the Ethiopian princess,
whom Perseus rescued and married. She became queen of Mycenae and,
after her death, a constellation. The galaxy was later named after the
constellation.}  A total of 693 candidate globular clusters were found
in recent 2MASS MIR observations (Galleti
2004).\nocite{2004A&A...416..917G} It is still unclear why M31 has
more than trice as many globular clusters than the Milky-way Galaxy.
Note also that there are no YoDeCs observed in M31, which is also
somewhat puzzling, as the Milky-way Galaxy has at least four.

\begin{figure}
\includegraphics[width=\textwidth,angle=-90]{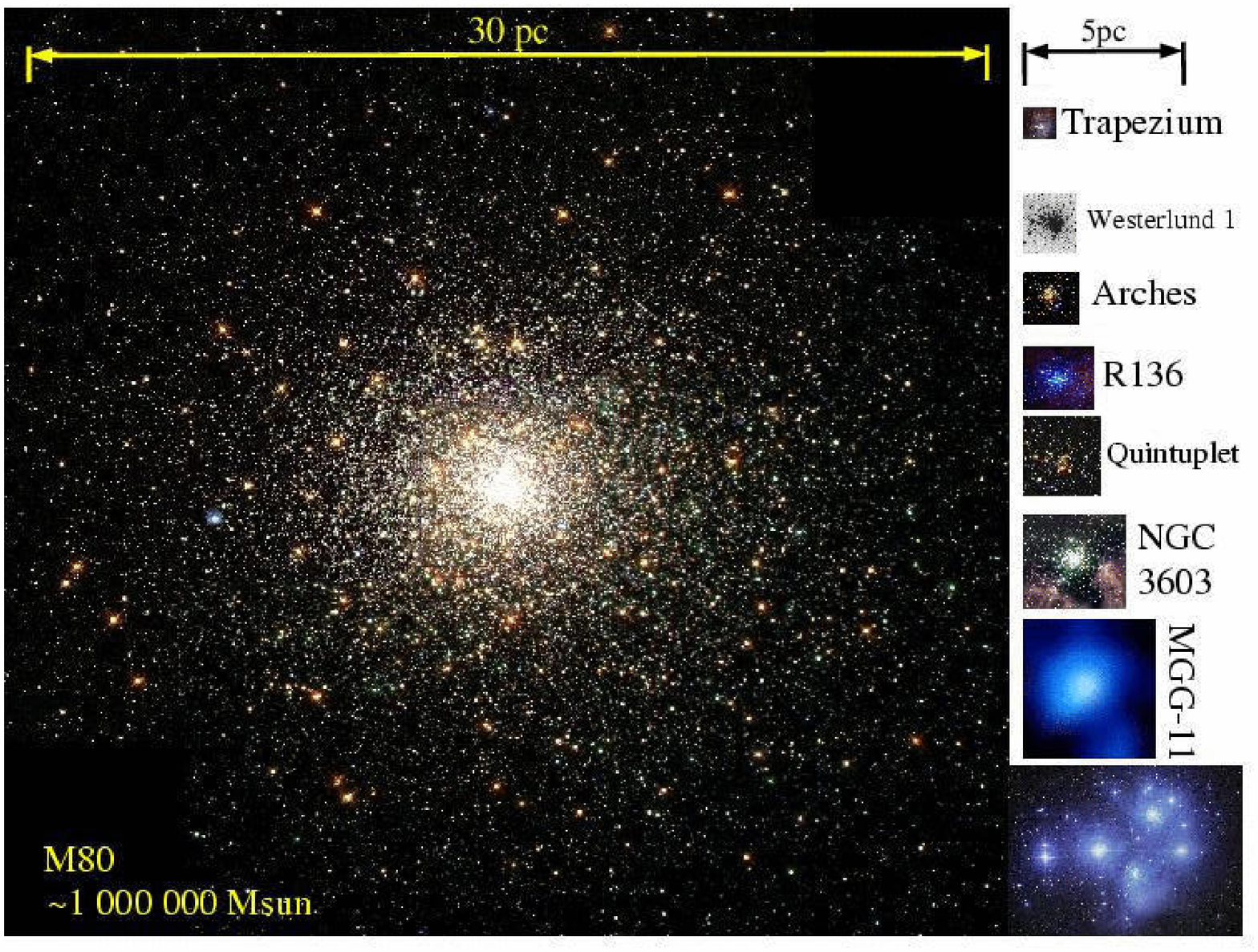}
\caption[]{Star clusters in the correct physical scale. The Galaxy
fig.\,\ref{fig:M31} was left out for practical reasons.
Images are from:
M80 (F. Ferraro);\index{M80}
Trapezium cluster \index{Trapezium}(HST image by . Bally, D. Devine, R. Sutherland, and D. Johnson);
Westerlund 1 (2MASS image by S. van Dyk);\index{Westerlund 1}
Arches (HST image by Figer, 1999);\index{Arches}
R\,136 (HST image by N. Walborn);\index{R136}
Quintuplet (HST image by Figer 1999);\index{Quintuplet}
MGG-11 (VLT image by McCrady et al 2003);\index{MGG11}
Pleiads (AAT image by D. Malin).\index{Pleiads}
}
\label{fig:composite}
\end{figure}

In figure\,\ref{fig:composite} we place a few well known star clusters
in retrospect. All cluster images are on the same scale.  It is
interesting to see is how large the globular cluster M80 is compared
to some depicted open clusters (Pleiads) and compared to the known
Galactic YoDeCs (Arches, Quintuple,\index{Quintuplet}
NGC3603\index{NGC 3603} and Westerlund 1\index{Westerlund 1}).

\begin{table}
\caption[]{Overview of selected parameters for young dense
clusters\index{star cluster} (Massey \& Hunter 1998), globular
clusters (Djorgovski \& Meylan, 1994) and galactic nuclei (Schr\"odel
et al. 2002)\nocite{2002Natur.419..694S}.  The first three columns
list the cluster type, the total mass (in solar units) and the virial
radius (in pc).  For globulars the total mass and virial radius are
given as distributions with a mean and the standard deviation around
the mean.  The orbital separation (in solar units) for a 1000\,$kT$
binary consisting of two 10\,\msun\, stars is given in the fourth
column.  The fifth and sixth columns list the expected number of
black-hole binaries that are formed by the cluster and the fraction of
these binaries which merge within 12\,Gyr (allowing $\sim 3$\,Gyr for
the formation and ejection of the binaries and assuming a 15\,Gyr old
Universe (Freedman et al.\,2001).\nocite{2001ApJ...553...47F} The
contributions to the total black hole merger rate per star clusters
per year (MR) are given in the final column (for details see
\S\,\ref{Sect:BHejection}).  The bottom row contains estimated
parameters for the zero-age population of globular clusters in the
Galaxy. \\ $^\star$ Estimate for the parameters at birth for the
population of globular clusters.  }
\medskip\begin{tabular}{lcc|crrl}  \hline
cluster  & M/\msun     &\rvir/pc  &$1000\,kT$&$N_{\rm b}$&$f_{\rm merge}$&MR\\ 
type     &[$\log$]     & [$\log$] &[\rsun]    &     & &[Myr$^{-1}$]\\ \hline
Populous & 4.5         &  -0.4    &   420     &  7.9& 7.7\%& 0.0061  \\ 
Globular & $5.5\pm0.5$ &$0.5\pm 0.3$& 315     &  150& 51 \%& 0.0064  \\ 
Nucleus  & $\sim 7$    &$\aplt 0$ &$\aplt 3.3$& 2500&100 \%& 0.21    \\ 
Globular$^\star$
         & $6.0\pm0.5$ &$0\pm 0.3$&    33     &  500& 92 \%& 0.038 \\ 
\end{tabular}
\label{Tab:clusters} 
\end{table}

There appears to be a clear relation between the number of globular
clusters, like M15 (see fig.\,\ref{fig:M15}), and the Hubble type of
the host galaxy. This relation is expressed in the specific number of
globular clusters $S_{\rm N}$.\index{star cluster!specific number of} For
open clusters and YoDeCs no such relation is known, but young star
clusters are particularly abundant in starburst and interacting
galaxies like the Antennae and M82.

Table\,\ref{Tab:galaxies} lists the space densities and specific
numbers of globular clusters $S_N$ per $M_v = -15$ magnitude (van den
Bergh 1984)\nocite{1984PASP...96..329V} for various Hubble types of
galaxies.  The values given for $S_N$ in Table \,\ref{Tab:galaxies}
are corrected for internal absorption; the absorbed component is
estimated from observations in the far infrared.
\begin{equation}
	S_N = N_{GC} 10^{0.4(M_v + 15)}.
\end{equation}
Here $N_GC$ is the total number of globular clusters in the galaxy
under consideration.  The estimated number density of globular
clusters in the local Universe is
\begin{equation}
	\phi_{GC}  = 8.4\,h^3\; {\rm Mpc^{-3}}
\end{equation}
(where $h = H_0/100~{\rm km\,s}^{-1}{\rm\,Mpc}^{-1}$), slightly
smaller than the result reported by Phinney
(1991).\nocite{1991ApJ...380L..17P}

\begin{table}
\caption[]{Galaxy morphology class,\index{Galaxy!morphology} space
densities, average absolute magnitude (Heyl et
al. 1997)\nocite{1997MNRAS.285..613H}, and the specific frequency of
globular clusters $S_N$ (from van den Bergh, 1995 and McLaughlin
1999)\nocite{1999ApJ...512L...9M}\nocite{1995AJ....110.2700V}.  The
final column gives the contribution to the total number density of
globular clusters.  The galaxy morphologies are identified as: E
(elliptical), S0-Scd (spiral galaxies), Blue E (young blue elliptical
galaxies), Sdm (blue spiral galaxies) and StarB (star burst galaxies).
}
\medskip\begin{tabular}{l|rlcr} \hline
Galaxy   & $\phi_{\rm GN}$&$M_v$          &$S_Nh^2$& GC space density\\
Type     &[$10^{-3}\,h_0$\,Mpc$^{-3}$]& &       & [$h_0^3$ Mpc$^{-3}$] \\  \hline
E--S0    & 3.49   & -20.7 & 10       & 6.65 \\
Sab      & 2.19   & -20.0 &  7       & 1.53 \\
Sbc      & 2.80   & -19.4 &  1       & 0.16 \\
Scd      & 3.01   & -19.2 &  0.2     & 0.03 \\ 
Blue E   & 1.87   & -19.6 & 14       & 1.81 \\
Sdm/StarB& 0.50   & -19.0 &  0.5     & 0.01 \\  \hline
\end{tabular}
\label{Tab:galaxies} 
\end{table}

\begin{figure}
\vspace*{2cm}
\begin{minipage}[b]{0.33\linewidth}
\includegraphics[width=0.5\linewidth]{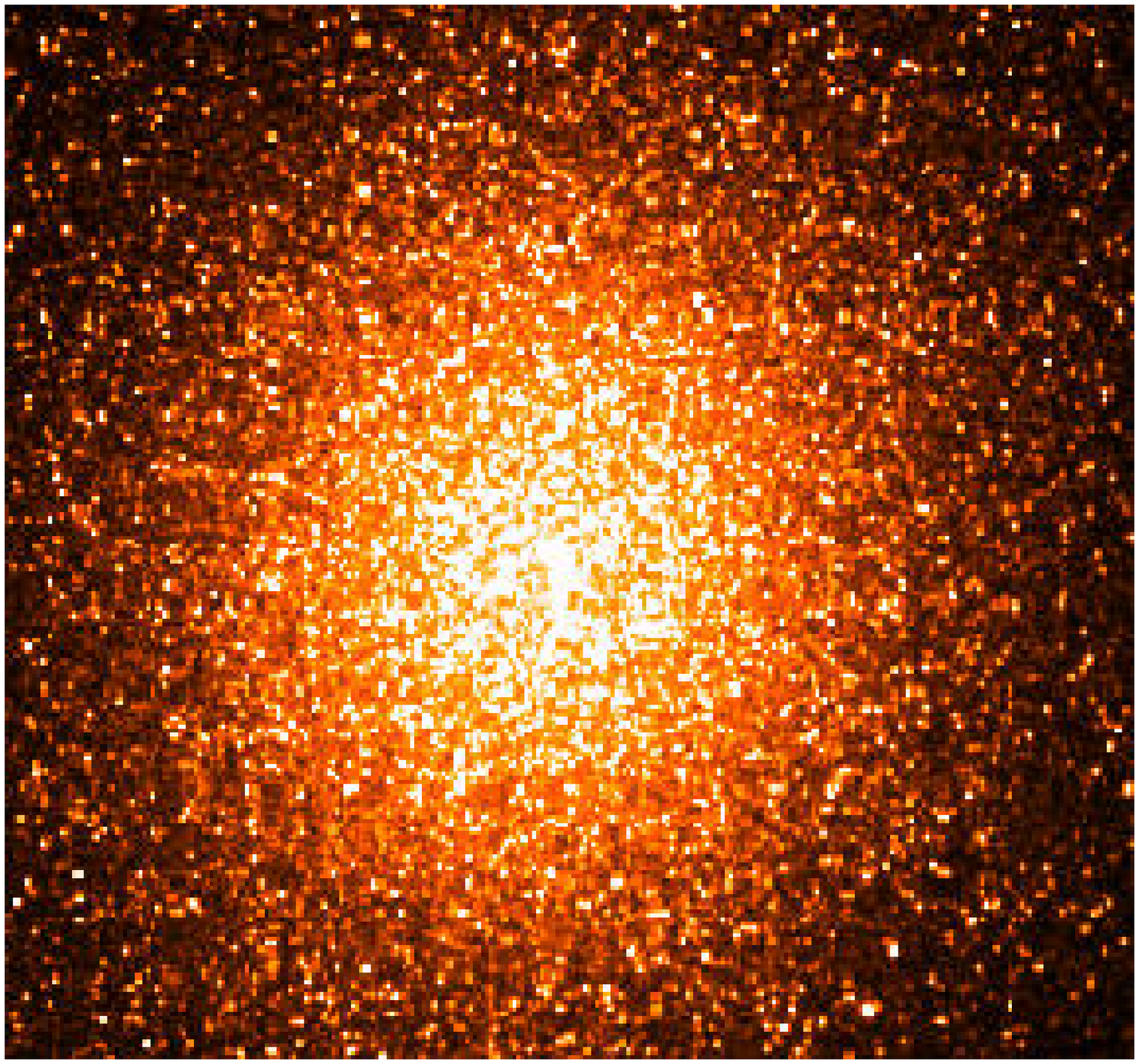}
\end{minipage}\hfill
\begin{minipage}[b]{0.33\linewidth}
\includegraphics[width=0.33\linewidth]{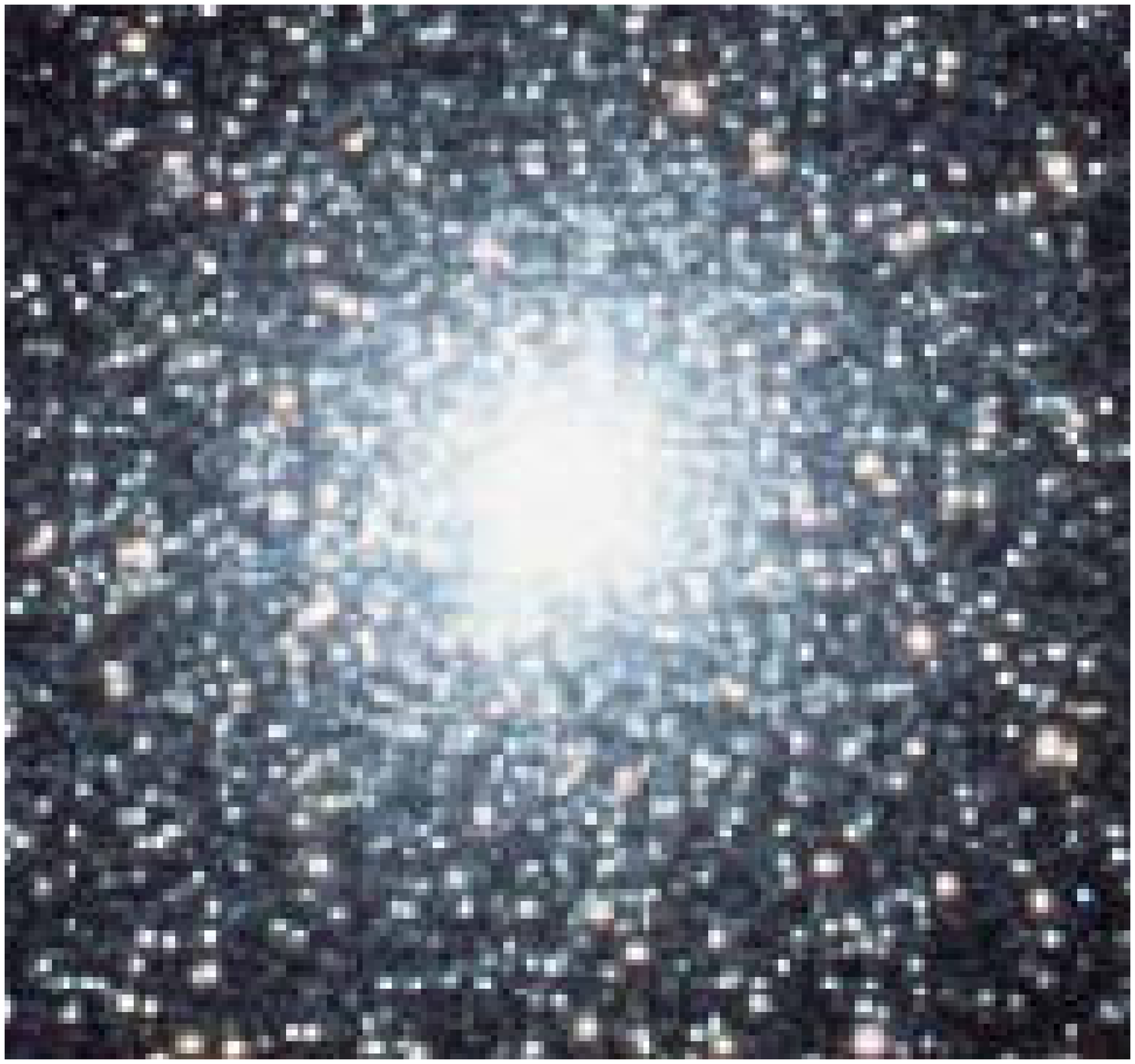}
\end{minipage}\hfill
\begin{minipage}[b]{0.33\linewidth}
\includegraphics[width=0.75\linewidth]{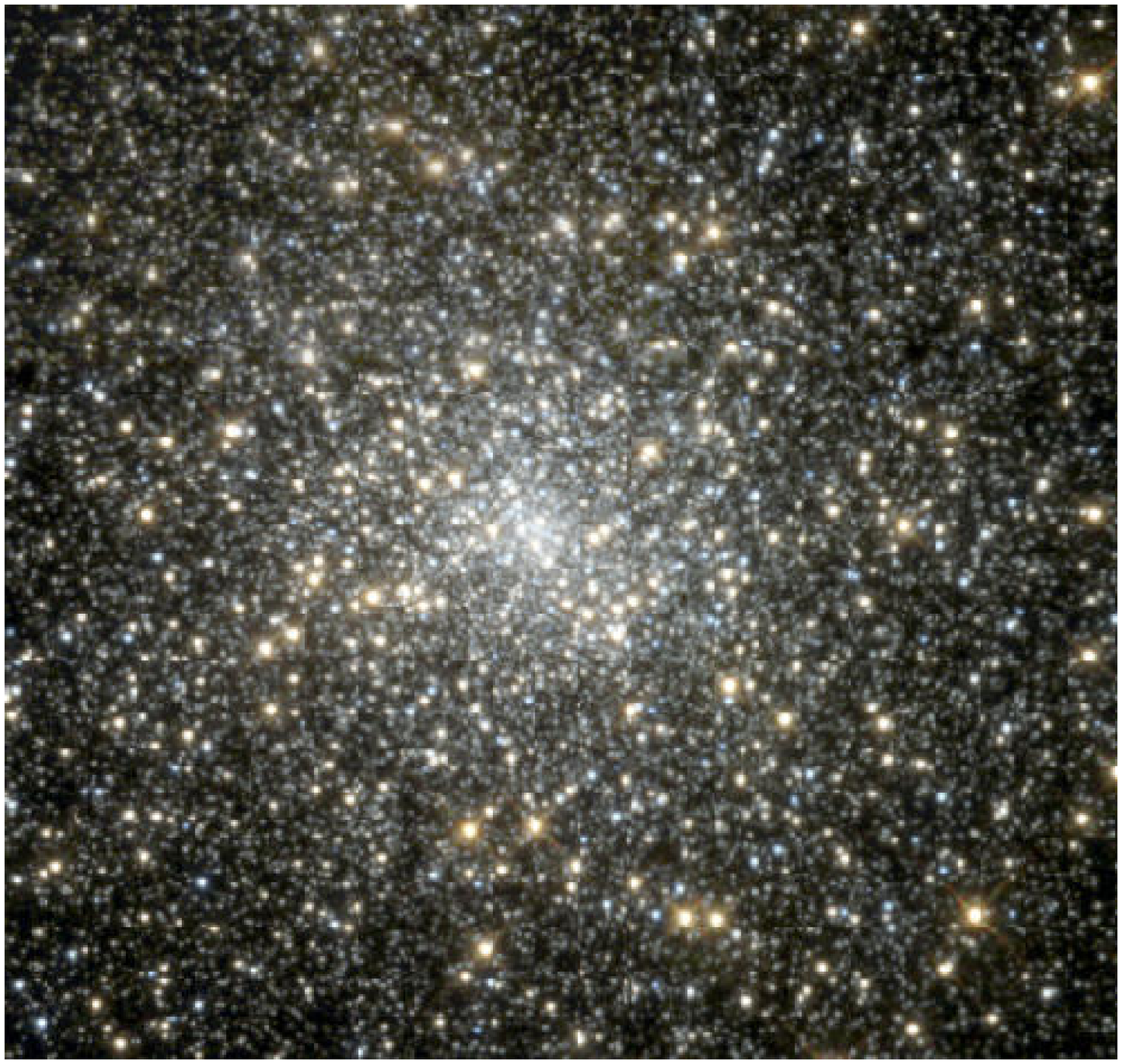}
\end{minipage}\hfill
\caption[]{Three globular clusters to illustrate
the effect of core collapse.\index{core collapse}  
{\bf Left}: the rather shallow globular clusters Omega Cen.\, (NGC
5139, image by P. Seitzer) with a concentration $\Wo \simeq
6$.\index{Omega Cen.}
{\bf Middle}: the concentrated cluster 47 Tuc.\, (NGC 104, image by
W. Keel) with concentration $\Wo \simeq 9$.\index{47\,Tucan\ae}
{\bf Right}: the core collapsed cluster M15 (NGC 7078, HST image
R. Guhathakurta) with concentration $\Wo \simeq 12$.\index{M15}  }
\label{fig:globulars}
\end{figure}

Figure\,\ref{fig:globulars} illustrates the concentration of clusters,
expressed in the structural parameter $\Wo$ (King 1966), which ranges
from about 1 for very shallow clusters to about 12 for very
concentrated clusters. The figure shows the images of three clusters
with quite different concentration ranging from $\Wo \simeq 6$ (Omega
Centauri)\index{Omega Cen.} to $\Wo\apgt 12$ for the core collapsed
globular cluster M15. The globular cluster
47\,Tuc.\,\index{47\,Tucan\ae} is highly concentrated, but not quite in
a state of core collapse.

\subsection{Fundamental time scales}

The evolution of a star cluster is dominated by two main effects; the
mutual gravitational attraction between the stars and by the evolution
of the individual stars.  Before we discuss these two ingredients in
more detail is it important do understand how they are coupled in the
ecological network.

\subsubsection{Stellar evolution}

The fundamental timescale for stellar evolution is the nuclear burning
timescale. This timescale only depends on characteristics of the stars
themselves and is unrelated to any of the dynamical cluster
parameters\footnote{At least as long as the stars do not strongly
influence each-other's evolution due to dynamical interactions.}. The
mass is the most fundamental parameter in the stellar evolution
process; more massive stars burn-up more quickly than lower mass
stars.  The main sequence lifetime\index{time scale!main sequence} of
a star with mass $m$ and luminosity $l$ can, to first order, be
approximated with
\begin{equation}
  t_{\rm ms} \simeq 1.1 \times 10^{10} {\rm year}
                    \left( {m\over \msun}\right) \left({\lsun \over l}\right).
\end{equation}

After the main sequence the star generally grown to giant dimensions,
but remains large only for approximately 10\% of its main-sequence
lifetime\footnote{As a rule of thumb, one can adopt that each
subsequent burning stage for a star takes about 10\% of the previous
burning stage; Carbon burning lasts for about 10\% of the helium
burning stage, etc.}.  Massive stars continue their evolution after
central hydrogen is exhausted by burning Carbon, Neon, Oxygen and
Silicon until an iron core forms, which ultimately collapse
catastrophically. The result is a supernova and the formation of a
compact stars.  Lower mass stars cannot process all material and shed
their envelope in a planetary nebula phase. This results in a white
dwarf.  Rather detailed stellar evolutionary
tracks\index{stellar evolution} are published in in the form of
comprehensive look-up tables (Schaller et al 1993; Meynet et al. 1994)
\nocite{1993yA&AS..96..269S} \nocite{1994A&AS..103...97M} or fitting
formulae (Eggleton, Fitchet \& Tout 1989; Hurley et al
2000)\nocite{2000astro.ph..1295H}; \nocite{1989ApJ...347..998E} using
a variety of stellar evolution codes.

\subsubsection{Dynamical timescales}

The most fundamental dynamical timescale $t_{\rm d}$ for a star
cluster is the crossing time\index{time scale!crossing} or dynamical
timescale, \index{time scale!dynamical} which for a cluster with half-mass radius $R$ and dispersion velocity \vdisp\, can be written simply as
\begin{equation}
  t_{\rm d} = R/\vdisp.
\label{Eq:tdyn}\end{equation}
We can generalize this by using the cluster half-mass radius and the
dispersion velocity to calculate the global crossing time \tcross\, of
the star cluster.

So long as stellar evolution remains relatively unimportant, the
cluster's dynamical evolution is dominated by two-body
relaxation,\index{time scale!relaxation} with proceeds via the
characteristic half-mass relaxation time scale (Spitzer,
1987)\nocite{1987degc.book.....S}
\begin{equation}
	\trlx = \left( {R^3 \over G M} \right)^{1/2} 
		{N \over 8 \lnL}.
\label{Eq:trlx}\end{equation}
Here $G$ is the gravitational constant, $M$, is the total mass of the
cluster, $N \equiv M/\mmean$ is the number of stars and $r$ is the
characteristic (half-mass) radius of the cluster.  The Coulomb
logarithm\index{Coulomb logarithm} $\lnL \simeq \ln (0.1 N) = {\rm
O}(10)$ typically. In convenient units the two-body relaxation time
becomes
\begin{equation}
	\trlx \simeq 1.9\, {\rm Myr} \left({R \over
		 1\,{\rm pc}}\right)^{3/2} \left({M \over
		1\,\msun}\right)^{1/2} \left( {1\,\msun \over
		\mm}\right) \left(\lnL \right)^{-1}\,.
\label{Eq:trlx_II}\end{equation}

\begin{table}
\caption[]{Time scales}
\begin{tabular}{lc|rrrrr} \hline
Time scale & symbol& bulge &globular& YoDeC &Open cluster& Como    \\
Star       & \tms  & 10Gyr & 10Gyr  & 10Myr &10Myr& 100 yr  \\
size       & $R$   & 100pc & 10pc&$\aplt 1$pc&10pc& 10 km  \\
mass       &$M$&$10^9$\msun&$10^6$\msun&$10^5$\msun&1000\msun&$10^5$pers \\
velocity   & \vdisp& 100\kms& 10\kms&10\kms &1\kms& 5km/h \\
\hline
relaxation & \trlx &$10^{15}$yr& 3 Gyr& 50Myr &100Myr& 10 yr   \\
crossing   & \tcrss& 100Myr& 10 Myr & 100Kyr&  1Myr& 1 day   \\
collision  & \tcoll& 10Gyr & 100 Myr& 10Kyr &100Myr& minutes \\
\hline
\multicolumn{2}{l|}{\trlx/\tms}  &$10^5$& 3  & 5       &10  & 0.1      \\
\multicolumn{2}{l|}{\tcross/\tms}&0.01& 1  & $10^{-4}$ & 0.1& 0.03     \\
\multicolumn{2}{l|}{\tcoll/\tms} & 1  & 0.1& $10^{-5}$ &10  & $10^{-7}$\\
\hline
\end{tabular}
\label{Tab:timescales} 
\end{table}

Table\,\ref{Tab:timescales} summarizes the relevant fundamental
characteristics of various types of star clusters. The parameters we
selected are the stellar evolution time scale, cluster mass, size and
velocity dispersion. With these we can compute the relaxation time and
crossing time for the various clusters. Near the bottom of the table
we present the collision time: $\tcoll \equiv 1/\Gamma_{\rm
coll}$. Here $\Gamma_{\rm coll}$ is the collision rate for a cluster
with stellar density $n$, and can be written as $\Gamma_{\rm coll} = n
\sigma v$, where $\sigma$ is the cross section\index{cross section}
for physical collisions:
\begin{equation}
   \sigma = \pi d^2 
            \left( 1 + {v^2 \over v_\infty^2} \right).
\end{equation}
Here $v_\infty$ is the relative velocity of two stars with mass $m_1$
and $m_2$ at infinity and $v$ is the relative velocity at closest
distance $d$ in a parabolic encounter, i.e: $v^2 = 2 G (m_1 + m_2)/d$.
The second term results from the gravitational attraction between the
two stars, and is referred to as gravitational
focusing.\index{gravitational focusing}

We can now estimate the timescale for a collision between two stars as
$\tcoll = 1/\Gamma_{\rm coll}$, which for a cluster with low velocity dispersions
($v_\infty \ll v$, can be written as
\begin{equation}
    \tcoll = 7 \times 10^{10} {\rm yr} \left( {10^5{\rm pc}^{-3} \over
             n } \right) \left( { v_\infty \over 10 {\rm km/s} }
             \right) \left( { \rsun \over d } \right) \left( {
             \msun \over m } \right) \ {\rm for} \ v \gg v_\infty.
\end{equation} 
For a collision between two single stars we can adopt $d \sim 3
r_\star$ (Davies, Benz \& Hills 1991).\nocite{dbh91}

The last column in table\,\ref{Tab:timescales} gives, in retrospect to
the star clusters, a comparison with the beautiful city in which this
meeting is organized.

\subsection{The effect of two-body relaxation: dynamical friction}
\label{Sect:df}
\index{dynamical friction} \index{two body relaxation}

  A star cluster in orbit around the Galactic center is subject to
  dynamical friction, in much the same way as dynamical friction
  drives massive stars toward the cluster center. This causes clusters
  to spiral into the Galactic center and stars to the cluster center.
  A star cluster is generally destroyed by the tidal field when it
  approaches the galactic center (see Gerhard
  2001).\nocite{2001ApJ...546L..39G} We derive here the dynamical
  friction time scale for a mass point in the potential of the
  Galactic center.  The derivation of the dynamical friction
  time\index{time scale!dynamical friction} scale of a star as it
  spirals to the cluster center is very similar, with the major
  exception than the cluster potential is much more complicated that
  the potential of the Galaxy. We therefore opted for showing the
  derivation for a galaxy instead.

  We assume the inpiraling object to have constant mass $M$, deferring
  the more realistic case of a time-dependent mass (for example in the
  case of a star cluster sinking to the Galactic center, see
  Eq.\ref{Eq:mass}) to McMillan \& Portegies Zwart
  (2003).\nocite{2003ApJ...596..314M}

  The drag acceleration due to dynamical friction in an infinite
  homogeneous medium with isotropic velocity distribution that is not
  self-gravitating is (equation [7-18] in Binney \& Tremaine,
  1987)\nocite{1987gady.book.....B}
  \begin{equation}
	a = -{4\pi \lnL G^2 M \rho_G(\Rgc) \over \vorb^2}
	\left[
	\erf(X) - {2X \over \sqrt{\pi}} e^{-X^2} \right]\,.
  \label{Eq:Fdf}\end{equation}
  Here $\lnL$ is the Coulomb logarithm\index{Coulomb logarithm} for
  the Galactic central region, for which we adopt $\lnL \sim \Rgc/R$,
  $\erf$ is the error function and $X \equiv \vorb/\sqrt{2}\vgc$,
  where $\vgc$ is the one-dimensional velocity dispersion of the stars
  at distance \Rgc\, from the Galactic center, and $v_c$ is the
  circular speed of the cluster around the Galactic center.

  The mass of the Galaxy lying within the cluster's orbit at distance
  {\Rgc} ($\aplt500$\,pc) from the Galactic center is (Sanders \&
  Lowinger 1972; Mezger et
  al.\,1996)\nocite{1996A&ARv...7..289M}\nocite{1972AJ.....77..292S}
  \index{Galaxy!mass}
  \begin{equation}
	\Mgal(\Rgc) = 4.25 \times 10^6 \left({\Rgc \over 1\,{\rm pc}}
		     	           \right)^{1.2}           \;\;\msun\,.
  \label{Eq:Mgal}\end{equation}
  Its derivative, the local Galactic density (see Portegies Zwart et
  al.\, 2001a) is\index{Galaxy!density}
  \begin{equation}
	\rho_G(\Rgc) \simeq 4.06\  \times 10^5 \left({\Rgc \over 1\,{\rm pc}}
		     	           \right)^{-1.8}  \;\;\msun\,{\rm pc}^{-3}\,.
  \label{Eq:rhogal}
  \end{equation}
  For inspiral through a sequence of nearly circular orbits, the
  function $\erf(X) - {2X \over \sqrt{\pi}} \exp(-X^2)$ appearing in
  Eq. \ref{Eq:Fdf} may be determined as follows.

  Following Binney \& Tremaine (p.\,226), we write the equation of
  dynamical equilibrium for stars near the Galactic center as
  \begin{equation}
	\frac{dP}{d\Rgc} = - \rho_G \frac{G\Mgal(\Rgc)}{\Rgc^2}\,,
  \label{Eq:hydro_eq}
  \end{equation}
  where $P = kT\rho/\mmean$, $\frac32kT = \frac12 \mmean\langle
  v^2\rangle$. Since $u^2 = \frac13\langle v^2\rangle$, it follows
  that $P = u^2\rho$, and Eq. \ref{Eq:hydro_eq} becomes
  \begin{equation}
	\frac{d~}{dr}(u^2\rho) = -\frac{\rho}r\,\vorb^2\,,
  \end{equation}
  where \vorb\, is the circular orbital velocity at radius $R$:
  $\vgc^2 = G\Mgal(\Rgc)/\Rgc$.  For $\Mgal \propto \Rgc^x$ (see
  Eq.\,\ref{Eq:Mgal}), and assuming that $\vgc^2\propto \vorb^2\sim
  \Rgc^{x-1}$, we find $\vgc^2\rho \sim \Rgc^{2x-4}$, so
  \begin{equation}
	r\,\frac{d~}{dr}(\vgc^2\rho) = (2x-4)\vgc^2\rho
			= -\rho \vorb^2\,,
  \end{equation}
  and hence $X = \sqrt{2-x}$.  Eq. \ref{Eq:Fdf} then becomes
  \begin{equation}
	a = -1.2 \lnL \frac{G M }{\Rgc^2} \; 
  		\left[ \erf(X) - {2X \over \sqrt{\pi}} \exp(-X^2) \right]\,.
  \label{Eq:Fdf_II}\end{equation}
  For $x = 1.2$, $X= 0.89$ and
  \begin{equation}
	a = -0.41 \lnL \frac{GM}{\Rgc^2}\,.
  \label{Eq:Fdf_IIa}\end{equation}

  Again following Binney \& Tremaine, defining $L = \Rgc \vorb$ and
  setting $dL/dt = a\Rgc$, we can integrate Eq.\,\ref{Eq:Fdf_IIa} with
  respect to time to find an inspiral time from initial radius $R_i$
  of
  \begin{eqnarray}
	\tdfg &\simeq & \frac{1.28}{\lnL}
			\frac{\Mgal(R_i)}{M}
			\left[\frac{G\Mgal(R_i)}{R_i^3}\right]^{-1/2} \\
	&\simeq& 1.4
		\left(\frac{R_i}{10 {\rm pc}}\right)^{2.1}
		\left(\frac{10^6 \msun}{M}\right) \; {\rm Myr}
		\label{Eq:tdf_GC}\label{Eq:tdfapprox}
  \end{eqnarray}
  For definiteness, we have assumed $\lnL\sim 4$ ($\Lambda \sim \Rgc/R
  \sim 100$) in Eq. \ref{Eq:tdf_GC}, corresponding to a distance of
  about 10--30\,pc from the Galactic center.


\subsection{Simulating star clusters}

Stars move around due to their mutual gravity. This principle was
first accurately described by Sir. Isaac Newton\index{Newton} in
1687 in his {\em Philosophiae naturalis principia mathematica} (an
excellent short biography can be found at \linebreak {\tt
http://www-gap.dcs.st-and.ac.uk}.

Newton's equation describe the gravitational interaction between two
stars with masses $m_1$ and $m_2$ with relative positional vector
${\bf r} = {\bf r}_2 - {\bf r}_1$, which we can written as:
\begin{equation}
  {{\rm d} {\bf r}^2 \over {\rm d}t^2} = 
    -G {m_1 + m_2 \over r^3} {\bf r}.
\label{Eq:Newton}
\end{equation}
The minus sign in the right-hand side indicates that the interaction
force (${\rm d} {\bf r}^2/{\rm d}t^2$) is attractive.

This second order differential equation can be integrated in many
ways.  At this moment Hut and Makino are in the process of writing a
series of 10 books about integrating this equation using the so called
direct N-body technique.\index{N-body!direct integration} The first three volumes
of this series are available at {\tt http://www.ArtCompSci.org}.
Other recent excellent work is published by Heggie \& Hut
(2003)\nocite{2003gmbp.book.....H} and Aarseth
(2003).\nocite{Aarseth2003} Therefore instead of worrying about the
intricacies of \Nbody\, techniques we continue directly with the core
problem.

First, however, it may be useful to explain a bit about the methods
under consideration and some of its alternatives; it is not my
intention to give a thorough overview of all the way in which you can
solve an \nbody\, system, but it is good to have some, overview. In a
direct \Nbody\, \index{N-body!direct integration}solver you integrate the
equations of motion of all stars in the system by computing the forces
from each star directly. This means that the amount of work for the
computer scales roughly with the square of the number of particles
$N$, or in units of CPU time:
\begin{equation}
  t_{\rm CPU} = \left( \!\!\!\!
                    \begin{array}{c}
		      N \\
		      2
                    \end{array}
                \!\!\! \right),
\end{equation}
which for large $N$ becomes $t_{\rm CPU} = \lim_{N\rightarrow \infty}
N^2$.  This scaling becomes even worse if one imagines that the
dynamical time unit in a star cluster is inversely proportional to $N$
(see Eq.\,\ref{Eq:tdyn}), resulting in a time complexity which
approaches $t_{\rm CPU} \propto N^3$. Due to this high computational
cost it is at this moment not possible to integrate the equations of
motion of $10^6$ stars for a Hubble time.  The largest simulations so
far have been done are $N \simeq 10^5$ stars for a Hubble time
(Baumgardt et al 2003) and of $N= 585.000$ stars for the first 12\,Myr
of the cluster (Portegies Zwart et al 2004).

One way to escape this conundrum is by integrating the equations of
motion less accurately, or to not integrate them at all but by
approximating the time evolution of the (grand)canonical
ensemble\index{canonical ensemble} of stars which forms the cluster.
The (semi)approximate methods are generally harder to code and often
impossible to assess, which makes fine-tuning to direct \Nbody\,
simulations inevitable.  The core problem here is that a star cluster
is in a delicate way not in perfect virial equilibrium.\index{virial
equilibrium} Methods which assume a vitalized state therefor have a
distinct disadvantage over methods which do not explicitly require
equilibrium.  Often the more complex numerical coding of approximate
solvers is well spend for large systems, like galaxies or cosmological
simulations, in which low precision methods are preferred because of
the shear number of 'stars' which have to be modeled.

For simulating dense star clusters the best way is probably still the
direct integration of the \Nbody\, system, though competitive
approaches have been taken (see e.g., \S\,\ref{Sect:PMcode} and
\S\,\ref{Sect:treecode}).

An extensive comparison between various types of N-body codes has been
performed in Heggie's (1998) collaborative experiment, the results of
which can be inspected at \linebreak {\tt
http://www.maths.ed.ac.uk/$\sim$douglas/experiment.html}.
\index{collaborative experiment}

Recently Spinnato et al.\, (2003) carry out a comprehensive comparison
between three very different N-body techniques. The methods they adopt
are a direct integration approach, which is, though accurate, strongly
limited in the number of particles which can be integrated. For larger
particle numbers they used a tree-code, and for the same system but
with up to several million particles they adopted a particle-mesh
technique.

\subsubsection{Particle mesh}\label{Sect:PMcode}
\index{N-body!particle mesh}

\begin{figure}
\psfig{figure=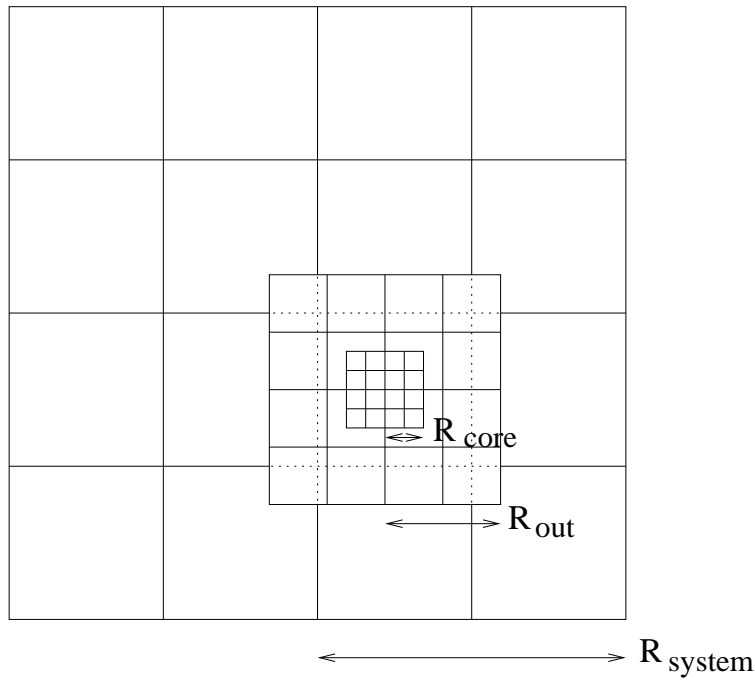,width=\linewidth,angle=0} 
\caption[]{ The different grids of {\small SUPERBOX} (Fellhauer et
al.\, 2000)\nocite{2000NewA....5..305F} for 4 cells per dimension. The
finest and intermediate grids are focused on the object of
interest. Figure from Spinnato et al (2003).}
\label{fig:PM}
\end{figure}

To perform calculations for close-to homogeneous particle
distributions a particle-mesh code is quite suitable.  Grainier
systems like dense star clusters are less suited, though some advances
have been made by increasing resolution in substructure regions.
However, caution has to be taken to make sure that the studied stellar
system does not relax, as relaxation is generally not treated
correctly in a particle-mesh technique.  A nice example is {\small
SUPERBOX}\index{N-body!{\small SUPERBOX}} (Fellhauer et al.\,
2000)\nocite{2000NewA....5..305F} in which accuracy is sacrificed for
speed.

In the particle-mesh technique densities are derived on Cartesian
grids.  Using a fast Fourier transform algorithm these densities are
converted into a grid-based potential.  Forces acting on the particles
are calculated using these grid-based potentials, making the code
nearly collision-less.  To achieve high resolution at the places of
interest several techniques to improve a better local accuracy are
used; {\small SUPERBOX} for example, incorporates two levels of
sub-grids which stay focused on the objects of interest while they are
moving through the simulated area (see Fig.~\ref{fig:PM}), providing
higher resolution where required.

Particle mesh codes, however, will always suffer from discrete effect
due to the projection of the system on a grid. The size of a grid cell
however, appears to be related directly to the softening
parameter \index{softening parameter} $\epsilon$ used in tree codes and in some
direct \Nbody\, codes.  The concept of softening as introduced by
(Aarseth 1963)\nocite{1963MNRAS.126..223A} is a technique to prevent
large angle gravitational scattering by increasing the distance
between two particles with a small parameter $\epsilon$. As long as $r
\gg \epsilon$ the effect of softening is not noticeable, but at close
distances it has a profound effect on the behavior of the system.

\begin{figure}
   \includegraphics[width=0.5
\linewidth]{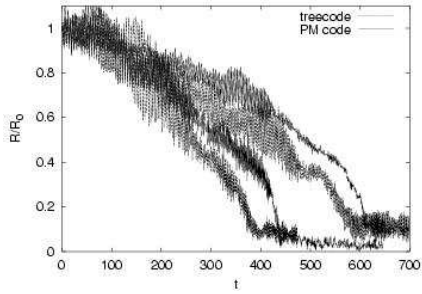}
  \caption{\label{codecomp1fig} Spiral in of a black hole (of mass
  $M_{BH} \simeq 0.0005$) to the center of the Galaxy (with mass
  $\Mgal \equiv 1$).  The simulations ware performed with 80,000
  stars, using the same initial realization for both the treecode and
  the particle-mesh code.  The particle-mesh simulations are run with
  cell size $\simeq 0.037$ (left) and $0.086$ (right); softening
  parameters in the treecode runs are resp. $0.030$ (left) and $0.060$
  (right).  }
\end{figure}

In order to understand how the cell length of the particle-mesh code
and the softening parameter $\epsilon$ of direct \Nbody\,
\index{N-body!direct integration} and treecodes relate with each other, we
compare in Fig.~\ref{codecomp1fig} the results from the particle-mesh
code {\small SUPERBOX} with the {\small GADGET} \index{N-body
code!GADGET} (Spingel et al 2001)\nocite{2001NewA....6...79S} treecode
simulations for $80\,000$ particles.  In this example Spinnato et
al.\, (2003) use a black hole of mass $0.00053$ (the mass of the inner
part of the galaxy is unity in these units) to sink to the center of
the Galaxy from a normalized distance. We can scale these numbers to
astrophysically relevant units, in which case the black hole is $\sim
65000$\,\msun\, and born at a distance of $\sim 8$\,pc from the
Galactic center.  The initial orbit of the black hole was circular,
but it still sinks slowly to the center of the Galaxy, due to
dynamical friction (see \S\,\ref{Sect:df}).
Figure\,\ref{codecomp1fig} shows the distance of the black hole to the
Galactic center as a function of time for the two computer codes.  The
results are presented for two values of the softening parameter
$\epsilon$ in the treecode and compared with two values of the
cell-size in the particle mesh code.

The in-fall of the black hole, as shown in Fig.~\ref{codecomp1fig},
depends on the values of $l$ and $\epsilon$ in a remarkably similar
way; $l$ and $\epsilon$ seem to play the same qualitative role, but
also quantitatively the results are quite similar.

\subsubsection{Tree codes}\label{Sect:treecode}
\index{N-body!treecode}

\begin{figure}
  \psfig{figure=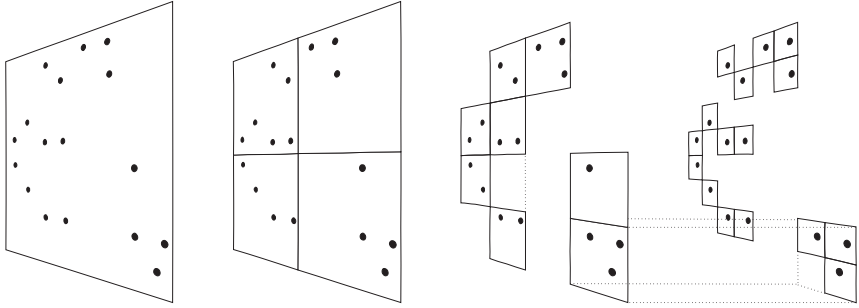,width=\textwidth,angle=90}
  \caption{\label{fig:gadget} Schematic illustration of tree-building
   in a Barnes \& Hut (1986) algorithm in two dimensions (from
   Springel, Yoshida and White, 2001).
    The
   particles are first enclosed in a single square, which is
   iteratively subdivided in squares of only half the size until a
   single particle remains per square.  }
\end{figure}

The concept of a hierarchical treecode was introduced by Apple (1985)
and Barnes \& Hut (1986),
\nocite{1985JSSC...6...85A}\nocite{1986Natur.324..446B} and is now
widely used for the simulation of (near) collision-less systems.
Figure\,\ref{fig:gadget} illustrates the concept of hierarchically
deviding the spacial coordinates in the tree code.  The force on a
given particle in a treecode is computed by considering particle
groups of ever larger size as their distance from the particle of
interest increases. Force contributions from such groups are evaluated
by truncated multipole expansions.\index{multipole expansion} The
grouping is based on a hierarchical tree data structure, which is
realized by inserting the particles one by one into initially empty
simulation cubes. Each time two particles are in the same cube, it is
split into eight 'child' cubes, whose linear size is one half of its
parent's. This procedure is repeated until each particle is in a
different cube. Hierarchically connecting such cubic cells according
to their parental relation leads to the hierarchical tree data
structure. The force on the particle of interest is then computed on
neighboring cubes, which increase in size as they are further away.
One of the interesting characteristics of tree-codes is the relatively
simple parallelization by domain decomposition of the spacial
coordinates (Olson \& Dorband 1994);\nocite{1994ApJS...94..117O} a
disadvantage is the lack of support for special purpose hardware (see
however Kawai et al 2004).  \nocite{2004ApJS..151...13K}

\nocite{2001NewA....6...79S}

\begin{figure}
\vspace*{3cm}
\centering
   \includegraphics[width=0.7\linewidth]{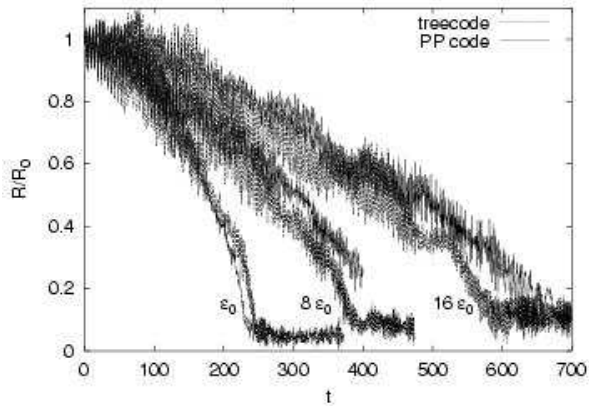}
  \caption{\label{codecompfig} Comparison of results from the PP code
  with results from the treecode, at different values of
  $\epsilon$. For all cases shown here is \mbox{$N = 80\,000$} and
  \mbox{$M_{BH} = 0.000528$}. The PP simulation with \mbox{$\epsilon =
  8\epsilon_0$} has been already shown in Fig.~\ref{obs_fig}.  the
  runs ware performed with $\epsilon \simeq 0.0037$ (left), $0.030$
  (middle) and $\epsilon  \simeq 0.060$ (right).  
  } 
\end{figure}

In the same way we compared a particle-mesh code with a treecode in
\S\,\ref{Sect:PMcode}, we can also compare the tree code with a direct
\Nbody\, code (see next section).  In Fig.\,\ref{codecompfig} such
comparison is illustrated on the same simulation of the black hole
which spirals to the Galactic center.  Both codes are run with
\mbox{$N = 80\,000$} and for a softening of $\epsilon \simeq 0.0037$,
$0.030$ and $\epsilon \simeq 0.060$. In this comparison we adopted
{\tt kira}\index{N-body!kira} from the {\tt Starlab}
\index{N-body!Starlab} package as the direct \Nbody\, code. The
latter code was also run with zero softening ($\epsilon = 0$) but for
the treecode this did not produce reliable results.

Interestingly, from the comparison between the particle-mesh code, the
tree code and the direct \Nbody\, code in the
figures\,\ref{codecomp1fig} and \ref{codecompfig} it is evident that
in this practical case all three codes produce qualitatively and
quantitatively the same results. One can wonder why then it is so
important to perform an accurate calculation, as low resolution
simulations produce results which are consistent with high precision
simulations?\footnote{It is probably worth revisiting this problem by
performing a details side-by-side analysis of two or more \Nbody\,
simulation methods (see Heggie et al 1998 and his collaborative
experiment at {\tt
http://www.maths.ed.ac.uk/$\sim$douglas/experiment.html}).
\nocite{1998HiA....11..591H}}

\subsubsection{Direct N-body}
\index{N-body!direct integration}

The most most accurate way to simulate the dynamical evolution of a
star cluster is by solving Eq.\,\ref{Eq:Newton} using direct
integration.  Direct \Nbody\, codes seem simple to write, as it is
just a matter of solving Eq.\,\ref{Eq:Newton} in small steps, but in
fact a computer program that does the job accurately and quickly is
very hard to write right. This technique was pioneered by von Hoerner
(1960, 1963) Aarseth (1963), Aarseth \& Hoyle (1964) and van Albada
1968).\nocite{1963MNRAS.126..223A} \nocite{1964ApNr....9..313A}
\nocite{1960ZA.....50..184V} \nocite{1963ZA.....57...47V}
\nocite{1968BAN....19..479V} Excellent reviews are published in the
earlier mentioned books by Hut \& Heggie (2003), Aarseth (2003) and
Hut \& Makino (2003), but see also Aarseth \& Lecar
(1975).\nocite{1975ARA&A..13....1A} A further reference to Heggie \&
Mathieu (1986)\nocite{HM1986} cannot be omitted as this valuable paper
discusses the dimension less units used in \Nbody\, techniques, in
which $M=R=G=1$.\index{N-body!units}

In a direct N-body code the forces between all stars are calculated
with numerical precision. The time complexity in this calculation is,
as discussed earlier {\large O}($N^3$).  In such a code, the particle
motion is followed using a high-order integrator often with a
predictor-corrector scheme (Makino and Aarseth
1992).\nocite{1992PASJ...44..141M} These codes can generally not work
with shared time steps\index{time step}\footnote{All particles share
the same time step.}; it saves time and gains accuracy to allow stars
in a strong encounter to be updated by individual time
steps\index{time step}\footnote{Each particle is integrated with its
own special timestep, particles in a strong interaction can then be
integrated more accurately than weakly interacting particles.}. For
simpler parallelization one generally adopts block time
steps\index{time step} so that groups strongly interacting particles
are integrated more frequently than weakly interacting stars (McMillan
1986a; 1986b; Makino
1991).\nocite{1986ApJ...307..126M}\nocite{1986ApJ...306..552M}
\nocite{1991ApJ...369..200M} Still special treatment for binaries and
higher order hierarchies are required to prevent the code to come to a
grinding halt during strong encounters.

During a time step, particle positions and velocities are first
predicted to fourth order using the acceleration and ``jerk'' (time
derivative of the acceleration), which are known from the previous
step. The new acceleration and jerk are then computed, and the motion
is corrected using the additional derivative information.

One of the great advantages of using a direct \nbody\, solver is the
simplicity at which extra effects can be incorporated. Since each star
in the cluster is represented by a particle in the code, individual
characteristics, such as stellar properties, can be accounted for
relatively easily and without loss of generality. This makes the
direct \nbody\, method preferable for simulating star clusters where
these effects are important. In our case we are interested in the
evolution of black holes, which are relatively rare objects. It would
therefore be best to utilize a technique in which we can also treat
the black holes individually. The draw back here is that black holes
are so rare that large clusters have to be simulated in order to
obtain enough statistics on the black hole population.

On the other hand the gravitational N-body problem has many
applications over a wide range of research fields, including
informatics, computational science, geology and astronomy.

\subsubsection{GRAPE family of computers}\label{Sect:GRAPE}

The enormous computational requirement for solving the \nbody\,
problem with the direct method has been effectively addressed by a
small team of researchers, who developed the GRAPE family of special
purpose computers.  GRAPE (short for {\bf GRA}vity {\bf P}ip{\bf E})
hardware was designed and built by a group of astrophysicists at the
University of Tokyo (to name only the most relevant publications:
Sugimoto et al. 1990; Fukushige et al 1991; Ito et al. 1991; Okumura
et al 1993; Taiji et al 1996; Makino et al 1997; Kawai et al. 2000;
Makino 2000; Makino et al. 2003).\nocite{1990Natur.345...33S}
\nocite{1991PASJ...43..547I} \nocite{1991PASJ...43..841F}
\nocite{1993PASJ...45..329O} \nocite{1996IAUS..174..141T}
\nocite{1997ApJ...480..432M} \nocite{2000PASJ...52..659K}
\nocite{2003PASJ...55.1163M} It may be clear that GRAPE is a very
successful endeavor.  The history and computational science of the
GRAPE project is published by Makino \& Taiji
(1998).\nocite{1998sssp.book.....M}

The GRAPE family of computer are like a graphics accelerator speeding
up graphics calculations on a workstation, without changing the
software running on that workstation, the GRAPE acts as a Newtonian
force accelerator, in the form of an attached piece of hardware. In a
large-scale gravitational N-body calculation, where $N$ is the number
of particles, almost all instructions of the corresponding computer
program are thus performed on a standard workstation, while only the
gravitational force calculations, the innermost loop, are replaced by a
function call to the special-purpose hardware.

\begin{figure}
\vspace*{3cm}
\includegraphics[width=\linewidth]{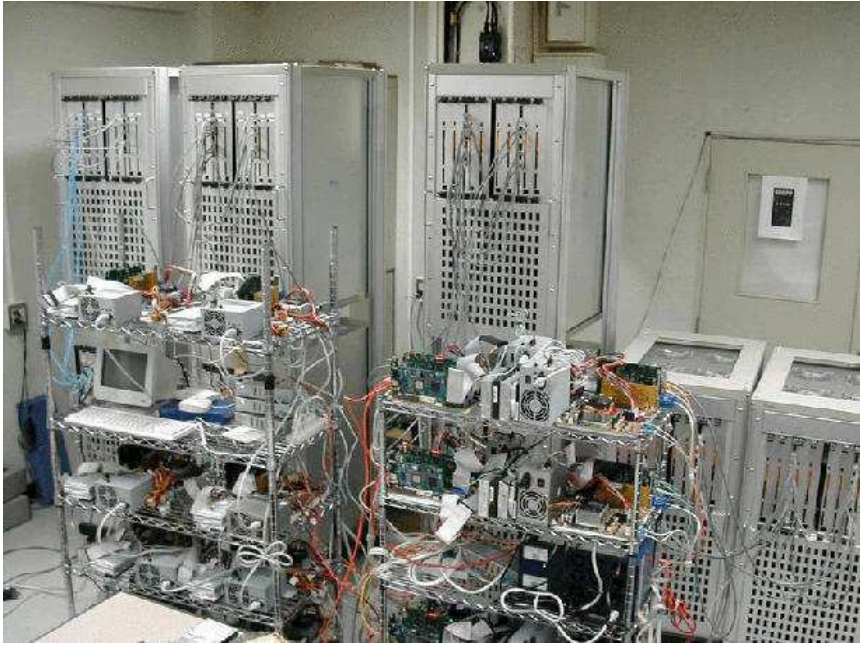}
\caption[]{ The large 64 Tflops GRAPE-6 configuration in Tokyo, in the
summer of 2003.  }
\label{fig:GRAPE}
\end{figure}

Figure\,\ref{fig:GRAPE} shows the fully configured
GRAPE-6\index{GRAPE6} at Tokyo University.

\subsection{Performing a simulations}

Before starting to simulate one may want to consider what technique is
most suitable. In our further discussion that will be the direct
\Nbody\, integrator.
\index{N-body!direct integration}
Several such computer programs are readily
available. The {\tt starlab}\index{N-body!Starlab} environment
provides a entire library of functions and routines built around the
main \Nbody\, integrator.  The package can be downloaded from
\linebreak {\tt http://www/manybody.org/starlab.html}. But also
NBODY\,1--6\index{N-body!NBODY1-6} are available by ftp via {\tt
ftp://ftp.ast.cam.ac.uk/sverre/}.  Both codes are large and very
complicated as they have been evolving to the current sophistication
over more than a decade.  But simpler alternatives are available from
a variety of sources (from example via: \linebreak {\tt
http://www.ids.ias.edu/$\sim$piet/act/comp/algorithms/starter/}.  A
fully operational parallel \nbody\, code based on the above mentioned
starter code can be obtained from \linebreak {\tt
http://carol.science.uva.nl/$\sim$spz/act/modesta/Software/index.html}.

Whatever computer code you select or even if you write one from
scratch, make sure that you test it. Test the code against other
similar codes, test it with calculations by hand, regardless how
painstaking this often is, and test it against simple problem for
which the solution is known. You should develop a feeling of the
regimes where the code can be trusted and in what cases extra care
must be taken in interpreting the results.

Let's assume that we have found an interesting problem, which we think
we can solve with the code available.  The main problem of starting a
simulation then is the selection of initial conditions. Starting with
wrong initial conditions is a complete waste of time. It is better to
spend enough time thinking about the initial conditions, until you are
convinced that they are the best choice. Possibly you want to perform
several test calculations to converge to a better understanding of the
question asked and the initial conditions required to give the most
reliable answer.

\noindent
For simulating a star cluster the primary initial conditions are:
\begin{itemize}
\item[$\bullet$] What are the basic cluster properties: mass, size, density profile?
\item[$\bullet$] How many stars did the cluster have at birth?
\item[$\bullet$] How are the stellar masses distributed?
\item[$\bullet$] What is the fraction of primordial binaries?
\item[$\bullet$] And what are the binary parameters: semi-major axis,
eccentricity, inclination, etc.
\item[$\bullet$] Do you want to include triples and higher order systems?
\item[$\bullet$] and what about the shape and strength of the external tidal
  field of the Galaxy?
\item[$\bullet$] Are there tidal shocks, spiral arms or other external
  potentials to worry about?
\item[$\bullet$] Is there anything else to add like: passing molecular clouds
  or black holes, etc.
\end{itemize}

Many of the above effects have to some extend been incorporated in
various calculations. And there is a rich scientific literature about
the relevance and effect of many of these ingredients.



\section{Theory of star cluster evolution}\label{Sect:Theory}

  Now we have set the stage and discussed the tools and the techniques
  we can continue by discussing the global evolution of star clusters,
  which is characterized by three quite distinct phases; these are
  subsequently: {\bf A} the early relaxation dominated phase, followed
  by phase {\bf B} in which the $\sim 1$\,\%\, (by number) most
  massive stars quickly evolve and lose an appreciable fraction of
  their mass. Finally, phase {\bf C} starts when stellar evolution
  slows down even more and relaxation takes over until the cluster
  dissolves. to complete the list we can define phase {\bf D} which is
  associated with the final dissolution of the cluster due to tidal
  stripping, but we will not discuss this phase in detail.
  \index{phase D}

   \subsection{Phase A: $t \aplt 10$Myr}\label{Sect:PhaseA}
\index{phase A}

   In an early stage, when stellar evolution is not yet important the
   star cluster is dominated by its own dynamical evolution; we call
   this phase A.  This stage in the evolution of the cluster is only
   relevant if $\trlx \aplt 100$\,Myr.  For most open star clusters
   and for all globular clusters this stage is
   probably\footnote{Regretfully we know very little about the early
   stages of globular clusters, and it is therefore hard to say
   whether phase A was important.} not very important, but for YoDeCs
   it is crucial, as we explain below.

   In the following discussion we assume, for clarity, that the
   location in the cluster where the stars are born is unrelated to
   the stellar mass, i.e.: there is no primordial mass
   segregation.\index{mass segregation} In that case, the early
   evolution of the star cluster is dominated by two-body relaxation,
   or to be more precise by dynamical friction.

   In young dense clusters dynamical friction implies a characteristic
   time scale $t_{\rm df}$ for a massive star in a roughly circular
   orbit to sink from the half-mass radius $R$, to the cluster center
   (Spitzer 1971; 1987):\nocite{1987degc.book.....S}
   \nocite{1971swng.conf..443S}
   \begin{equation}
      t_{\rm df} \simeq {\langle m \rangle \over 100\msun} 
                        {0.138 N \over \ln(0.11 M/100\msun) } 
                        \left( R^3 \over GM \right)^{1/2}.
   \label{Eq:tdf}
   \end{equation}
   Here $\langle m \rangle$ and $M$ are the mean stellar mass and the
   total mass of the cluster, respectively, $N$ is the number of
   stars, and $G$ is the gravitational constant.  For definiteness, we
   have evaluated $t_{\rm df}$ for a 100\,{\msun} star. Less massive
   stars undergo weaker dynamical friction, and thus must start at
   smaller radii in order to reach the cluster center on a similar
   time scale.

   Dynamical friction will have two very distinct effects on the
   cluster: 1) it tends to produce cores in hitherto core-less
   clusters, and 2) it initiates core collapse in other clusters.
   These two statements seem to contradict each other but as we will
   see below, this is not the case.

   \subsubsection{Dynamical friction induced core development}\label{Sec:DynFric}
\index{dynamical friction} Consider a gravitationally bound stellar
  system in which most of the mass is in the form of stars of mass $m
  \simeq \mmean$, but which also contains a subpopulation of more
  massive objects with masses $\mh$.  The orbits of the more massive
  objects decay due to dynamical friction.  Assume that the stellar
  density profile is initially a power-law in radius, $\rho(r) \propto
  (r/a)^{-\gamma} M/a^3$ with density scale length $a$ (Dehnen
  1993;\nocite{1993MNRAS.265..250D} Merritt et al.\,
  1994). \nocite{2004astro.ph..3331M} The orbits of the massive stars
  decay at a rate that can be computed by equating the torque from
  dynamical friction with the rate of change of the orbital angular
  momentum.  We adopt the usual approximation (Spitzer
  1987)\nocite{1987degc.book.....S} in which the frictional force is
  produced by stars with velocities less than the orbital velocity of
  the massive object.  The rate at which the orbit decays, assuming a
  fixed and isotropic stellar background, is
  \begin{equation}
     {dr\over dt} = -2{(3-\gamma)\over 4-\gamma}\sqrt{GM\over
     a}{\mh\over M} \ln\Lambda \left({r\over
     a}\right)^{\gamma/2-2}F(\gamma), 
     \label{eq:drdt}
  \end{equation}
  with
  \begin{equation}
     F(\gamma) = {2^\beta\over \sqrt{2\pi}}
     {\Gamma(\beta)\over\Gamma(\beta-3/2)}(2-\gamma)^{-\gamma/(2-\gamma)}
     \int_0^1 dy\ y^{1/2} \left(y+{2\over
     2-\gamma}\right)^{-\beta}.
  \end{equation}
  Here $\beta=(6-\gamma)/2(2-\gamma)$ and $\ln\Lambda$ is the Coulomb
  logarithm, roughly equal to 6.6 (Spinnato et
  al. 2003).\nocite{2003MNRAS.344...22S} For $\gamma=1.0$ (2.0),
  $F=0.19$ (0.43).

  If we approximate the cluster structure with an isothermal sphere,
  we find (Binney \& Tremaine 1987, Eq.\,
  7-25)\nocite{1987gady.book.....B} that a star of mass \mh\, at
  distance $R$ from the cluster center drifts inward at a rate given
  by
  \begin{equation}
	R \frac{dR}{dt} \simeq -0.43 {G \mh \over \vdisp} \lnL.
  \label{Eq:tdf_cgs}\end{equation}
  Here \vdisp\, is the clusters' velocity dispersion. 
  
  Equation\,\ref{eq:drdt} implies that the massive object comes to
  rest at the center of the stellar system in a time
  \begin{equation}
       t \approx 0.2 \sqrt{a^3\over GM} {M\over \mh}
                        \left({R_i\over a}\right)^{(6-\gamma)/2}
  \label{eq:deltat}
  \end{equation}
  with $R_i$ the initial orbital radius.

  Or, we can express the dynamical friction time in terms of the
  half-mass relaxation time by substituting Eq.\,\ref{Eq:trlx_II} in
  Eq.\,\ref{Eq:tdf_cgs} and integrate with respect to time.
  \begin{equation}
	\tdf \simeq 3.3 {\mm \over \mh} \trlx\,.
  \label{Eq:tdf_approx}\end{equation}

  To estimate the effect on the stellar density profile, consider the
  evolution of an ensemble of massive particles in a stellar system
  with initial density profile $\rho\sim R^{-2}$.  The energy released
  as one particle spirals in from radius $R_i$ to $R_f$ is
  $2\mh\sigma^2\ln(R_i/R_f)$, with $\sigma$ the 1D stellar velocity
  dispersion.  Decay will halt when the massive particles form a
  self-gravitating system of radius $\sim G\Mh/\sigma^2$ with
  $\Mh=\sum\mh$.  Equating the energy released during in-fall with the
  energy of the stellar matter initially within $r_c$, the ``core
  radius,'' gives
  \begin{equation}
         R_c\approx {2G\Mh\over\sigma^2}\ln\left({R_i\sigma^2\over
         G\Mh}\right).
  \end{equation}
  Most of the massive particles that deposit their energy within $R_c$
  will come from radii $R_i\approx {\rm a\ few}\times R_c$, implying
  $R_c\approx {\rm several}\times G\Mh/\sigma^2$ and a displaced
  stellar mass of $\sim {\rm several}\times \Mh$ (see also Watters
  2000).\nocite{2000ApJ...539..331W} If $\Mh\approx 10^{-2}M$
  (Portegies Zwart \& McMillan, 2000)\nocite{2000ApJ...528L..17P} then
  $R_c/a\approx {\rm several}\times 2 \Mh/M$ and the core radius is
  roughly $10\%$ of the effective radius.  Merritt et al (2004)
  discuss this process in more detail and apply it successfully to the
  evolution of the core radii of large Magellanic cloud star
  clusters.

  Evolution will continue as the massive particles form binaries and
  begin to engage in three-body interactions with other massive
  particles.  These superelastic encounters will eventually lead to
  the ejection of most or all of the massive particles.  Assume that
  this ejection occurs via many small 'kicks', such that almost all of
  the binding energy so released can find its way into the stellar
  system as the particle sinks back into the core after each
  ejection. The energy released by a single binary in shrinking to a
  separation such that its orbital velocity equals the escape velocity
  from the core is $\sim\mh\sigma^2\ln(4\Mh/M)$ (see also
  \S\,\ref{Sect:BHejection}).  If all of the massive particles find
  themselves in such binaries before their final ejection and if most
  of their energy is deposited near the center of the stellar system,
  the additional core mass will be
  \begin{equation} 
     M_c\approx \Mh\ln\left({M\over\Mh}\right)
  \end{equation}
  e.g. $\sim 5\Mh$ for $\Mh/M=0.01$, similar to the mass displaced by
  the initial in-fall.  The additional mass displacement takes place
  over a much longer time scale however and additional processes
  (e.g. core collapse) may compete with it.

   \begin{figure}
   \includegraphics[width=\linewidth]{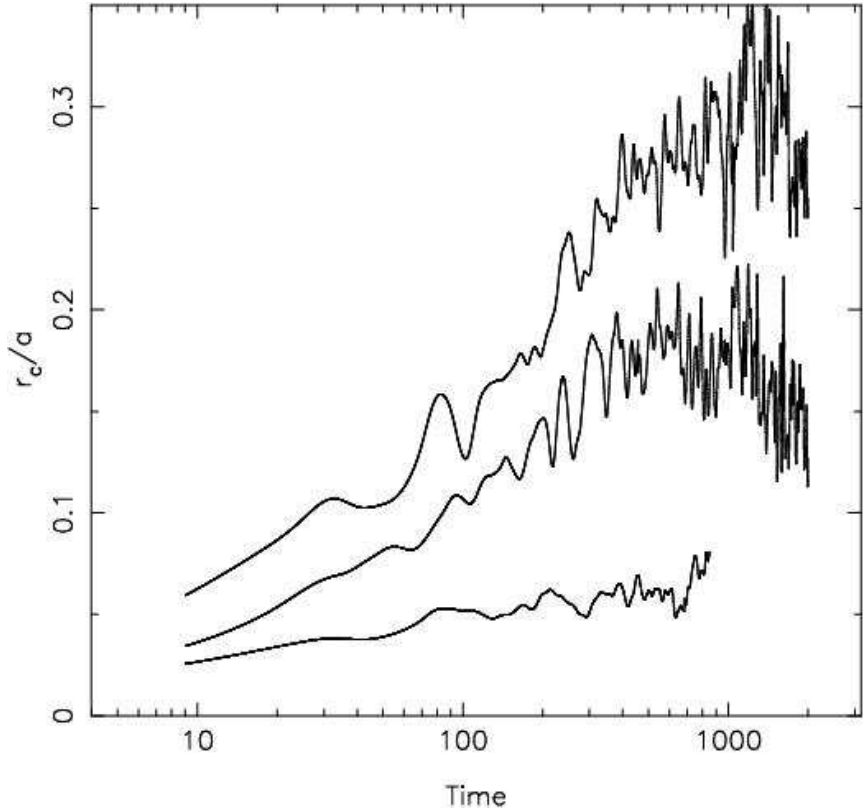}
   \caption[]{Evolution of the core radius, defined as the radius at
              which the projected density falls to one-half of its
              central value.  Each curve is the average of the various
              runs at the specified value of $\gamma$, $\gamma=1$ for
              the top curve, 1.5 for the middle curve and $\gamma = 2$
              for the bottom curve (see \S\,\ref{Sec:DynFric}).  The
              figure is taken from Merritt et al.\, 2004).  }
   \label{fig:mackey}
   \end{figure}

  \subsubsection{Dynamical friction induced core collapse}\label{Sect:DFtoCC}
\index{core collapse}\index{dynamical friction}

  The dynamical evolution of the star cluster drives it toward core
  collapse (Antonov 1962; Spitzer \& Hart, 1971a; 1971b)
  \nocite{1962spss.book.....A}\nocite{1971ApJ...164..399S}
  \nocite{1971ApJ...166..483S} in which the central density runs away
  to a formally infinite value in a finite time. In an isolated
  cluster in which all stars have the same mass, core collapse occurs
  in a time $\tcc \simeq 15\,\trlx$ (Cohn 1980; Makino 1996; Joshi et
  al 2001).\nocite{1996ApJ...471..796M} \nocite{1980ApJ...242..765C}
  \nocite{2001ApJ...550..691J}

\begin{figure}
\vspace*{3cm}
   \includegraphics[width=0.5\linewidth,angle=-90]{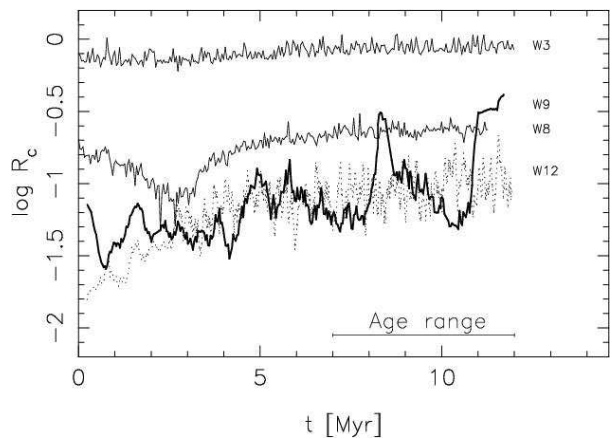}
    \caption[]{ Evolution of the core radius for four simulations of
    the star cluster MGG-11.  These calculations are performed with
    {\tt Starlab} with $c \simeq 0.67$ ($W_0=3$), $c \simeq 1.8$
    ($W_0=8$), $c \simeq 2.1$ ($W_0=9$, bold curve) and $c \simeq 2.7$
    ($W_0=12$, dotted curve) indicated along the right edge of the
    figure as $W3$, $W8$, $W9$, and $W12$ respectively.  The W9 curve
    is plotted with a heavy line to distinguish it from the curves for
    W8 and W12. The age range of the observed clusters MGG11 is
    indicated near the bottom of the figure.  In
    figure\,\ref{fig:rhocore} we present the evolution of the average
    core density for the simulations with $\Wo=3$, 8 and 9.  }
    \label{Fig:core_radius_evolution}
    \end{figure}

   If the dynamical friction time scale of a star cluster is shorter
   than the lifetime of the most massive stars, the cluster may
   experience an early phase of core collapse before the first
   supernova occurs. This can only happen if the initial half-mass
   relaxation time of the cluster is small $\trlx \aplt 100$\,Myr
   otherwise the most massive stars burn-up before they reach the
   cluster center (Portegies Zwart et al 1999; G\"{u}rkan et al
   2004).\nocite{1999A&A...348..117P} The early core collapse in a
   cluster with small relaxation time is illustrated in
   figure\,\ref{Fig:core_radius_evolution}, where we plot the core
   radius as a function of time for a number of young star clusters
   with a relaxation time of about 100\,Myr.

   The details of what exactly happens in the cluster core, and
   whether the cluster will experience gravothermal
   oscillations\index{gravothermal oscillations} probably depends
   quite critically on the initial density profile, as we will discuss
   in more detail in Sect.\,\ref{Sect:MGG11}.

   When after core collapse the most massive stars explode the core
   radius remains highly variable, but small on average.  What exactly
   happens at this stage is still ill understood. Naively one would
   expect the collapsed core to expand, as stellar mass loss drives an
   adiabatic expansion of the cluster.\index{adiabatic expansion} This
   can be been seen in the post core collapse evolution of the bottom
   solid curve in figure\,\ref{Fig:core_radius_evolution}.  Stage B
   starts when dynamical friction and relaxation cannot further drive
   the core collapse of the cluster, but expansion by stellar mass
   loss starts to dominate. Typically this happens around $\sim
   10$\,Myr.

   The cluster simulated for Fig.\,\ref{Fig:core_radius_evolution} was
   MGG11, in the starburst galaxy M82. It contained 131072 single
   stars from a Salpeter initial mass function between 1\,\msun\, and
   100\,\msun\, distributed in a King (1966)
   \nocite{1966AJ.....71...64K} model density profile with dimension
   less dept $\Wo=3$ (shallow) to $\Wo=12$ (very concentrated).  The
   half mass radius of these simulated clusters was 1.2\,pc.

   We discuss each curve in figure \ref{Fig:core_radius_evolution} in
   turn, starting at the top.  The core radius of the shallow model,
   $c \simeq 0.67$ ($W_0=3$), hardly changes with time. The
   intermediate model $c \simeq 1.8$ ($W_0=8$) almost experiences core
   collapse near $t=3$\,Myr but as stellar mass loss starts to drive
   the expansion of the core it never really experiences
   collapse. This is the moment where phase B sets in. Core
   collapse\index{core collapse} occurs in the $c \simeq 2.1$
   ($W_0=9$) model near $t=0.8$\,Myr. The $c \simeq 2.7$ ($W_0=12$)
   simulation is so concentrated that it starts virtually in core
   collapse, and the entire cluster evolution is dominated by a
   post-collapse phase. At this point it is not a-priory clear why
   core radius for models $W_0=9$ and $W_0=12$ fluctuate so much more
   violently than in the models with smaller concentration.

    \subsection{Phase B: $10{\rm Myr} \apgt t \aplt 100$Myr}
\label{phase B}\index{phase B}

    After the first few million years and until the most massive stars
    have turned into compact remnants, the cluster will be dominated
    by stellar mass loss. Since the most massive stars evolve first
    and sink most quickly to the cluster center, mass is lost from
    deep inside the potential well. A collapsed cluster may recover
    from its earlier core collapse due to stellar mass loss (see
    \S\,\ref{Sect:DFtoCC}).  The dotted curve in
    fig.\,\ref{Fig:core_radius_evolution} (for the simulations with
    $\Wo=12$) illustrates the growth of the core as a result of
    heating by dynamical friction and stellar evolution mass loss is
    quite clearly visible; it is hard to separate the two effects as
    both are taken into account in the calculation self
    consistently. For the slightly shallower initial density profile
    ($\Wo=9$) the steady growth of the core radius after collapse is
    less clear, but the rate is similar as for $\Wo=12$. This
    indicates that it is indeed mainly stellar mass loss which drives
    the expansion. At this stage we do not know why the core radius in
    the $\Wo=9$ models fluctuate more wildly than in the $\Wo=12$
    simulations. Details probably depend quite sensitively on the
    presence of an intermediate mass black hole, which could form in
    the proceeding phase A. Such massive compact object can
    effectively heat the cluster core as it forms tight multiple
    systems with other (massive) stars (Baumgardt et al
    2003).\nocite{2003IAUS..inpress}

    Few studies has been carried out for clusters with short
    relaxation time to understand this particular evolutionary stage.
    For a long relaxation time $\trlx \apgt 100$\,Myr, it is quite
    clear that the cluster expands substantially during the first Gyr
    of its lifetime (Chernoff \& Weinberg 1990; Fukushige \& Heggie
    1995; Takahashi \& Portegies Zwart 2000; Baumgardt \& Makino
    2003).\nocite{1990ApJ...351..121C} \nocite{1995MNRAS.276..206F}
    \nocite{2000ApJ...535..759T} \nocite{2003MNRAS.340..227B}

\begin{figure}
   \psfig{figure=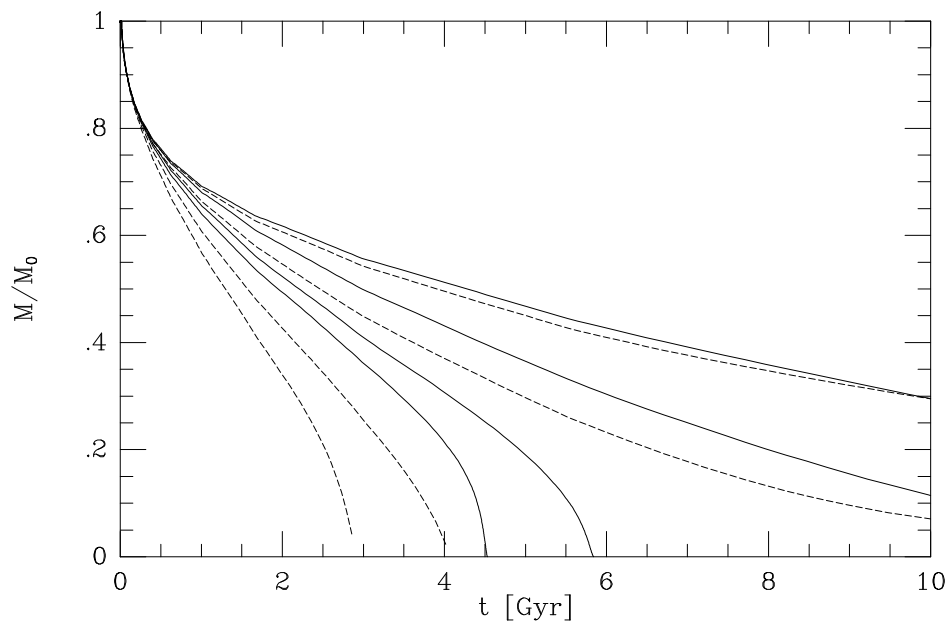,bbllx=560pt,bblly=100pt,bburx=90pt,bbury=720pt,width=\linewidth,angle=-90}
   \caption{ Mass as a function of time for a number of Fokker-Planck
             models.  The four solid lines represent the results of
             model Aa with 32k, 16k, 4k and 1k particles from left to
             right, respectively.  Dotted curves present model {\em
             Ie} for the same numbers of particles as for model {\em
             Aa} (model names refer to Takahashi \& Portegies Zwart,
             2000).  The difference between the models {\em Ie} and
             {\em Aa} hides in the way escaping stars are identified.
             }
   \label{fig:1:1998ApJ...503L..49T}
   \end{figure}

    The evolution of the cluster mass if given in
    Fig.\,\ref{fig:1:1998ApJ...503L..49T} (taken from Takahashi \&
    Portegies Zwart 2000) for a variety of particle numbers, ranging
    from 1024 to 32768. The first few 100\,Myr of all clusters is very
    similar,but at later time large differences appear.  In the first
    epoch, during the first few hundred million years, about 20\% of
    the cluster mass is lost. This mass loss is the result of stellar
    evolution and, in less extend by tidal stripping. Tidal stripping
    and relaxation become important at later time.

    \subsection{Phase C: $t \apgt 100$Myr}
    \index{phase C}

    After the most massive stars have turned into remnants, stellar
    evolution slows down, and relaxation processes can take over
    again. In fig.\,\ref{fig:1:1998ApJ...503L..49T} this phase starts
    at an age of about a Gyr. The reason that stellar mass loss slows
    down is two-fold, 1) low mass stars remain on the main sequence
    longer than high mass stars, and 2) once the star turns into a
    remnant, the mass lost in the process is relatively small; imagine
    a 12\,\msun\, star loses about 10.6\,\msun\, by stellar wind and
    in the supernova explosion, whereas a 2\,\msun\, turns into a
    0.64\,\msun\, white dwarf, losing only 1.36\,\msun\, in the
    process.

    The effects of mass lost from the evolving stars and mass lost
    from the dynamical evolution of the cluster are coupled.  If
    stellar mass loss slows down, the cluster responses to this by a
    slower expansion, which again makes it less prone to tidal
    stripping. While the importance of stellar evolution diminishes,
    relaxation gradually takes over until it becomes the
    dominant mechanism which drives the evolution of the cluster.

    The later stage of the low $N$ (1k and 2k) clusters in
    Fig.\,\ref{fig:1:1998ApJ...503L..49T} are much stronger affected
    by relaxation than the high $N$ (16k and 32k) clusters.  This
    effect was named the {\em ski-jump}\index{ski-jump problem}
    problem in Portegies Zwart et al.\,
    (1998).\nocite{1998A&A...337..363P} The transition between {\rm
    ski}-clusters (low $N$ in fig.\,\ref{fig:1:1998ApJ...503L..49T})
    and {\em jump}-clusters (high $N$) is a result of the non-linear
    interaction between the external tidal field of the parent galaxy
    and relaxation.

    At later time during phase C dynamical friction once again become
    important.  This time not driven by massive stars, as these have
    all gone supernova by now, but by the compact remnants formed in
    supernovae; black holes and neutron stars, but also heavy
    white dwarfs, blue stragglers and giants.  All these stellar
    species are generally more massive than the mean mass, and 
    therefore subject to dynamical friction. A similar process as in
    phase A starts again and the cluster may experience core collapse
    for the second time. It may therefore be possible that a cluster
    experiences two very distinct phases of core collapse, one during
    phase A, and again at a later time, during phase C (see also
    Deiters \& Spurzem 2000, 2001).\nocite{2001A&AT...20...47D}
    \nocite{2000msc..conf..204D}

    In a realistic cluster, however, there are a number of additional
    complications which are particularly important at this stage, in
    part because it may take rather long to reach a state of core
    collapse again because the stellar mass function is rather flat
    now, with a relatively small difference between the least and the
    most massive stars. The consequence is that external influences,
    like disc shocks, passing molecular clouds and the presence of an
    external tidal field, may become particularly important at this
    stage, simply because they have a lot of time to accumulate their
    effect.
    
    The slow-down of stellar evolution has a second important
    consequence, which is the termination of active binary
    evolution. Only relatively low mass stars are able to evolve off
    the main sequence, and no supernova will occur after $\sim 10^8$
    years.  It becomes therefore almost impossible to ionize hard
    binaries, which may effectively arrest the collapse of the cluster
    core (Fregeau et al.\, 2003).\nocite{2003ApJ...593..772F} The
    binaries may therefore once more\footnote{The first time binary
    interactions ware relevant was during phase A.} become
    dynamically important, and heat the cluster by interacting with
    single stars or other binaries (Heggie
    1975).\nocite{1975MNRAS.173..729H}

   \begin{figure}
   \includegraphics[width=\linewidth]{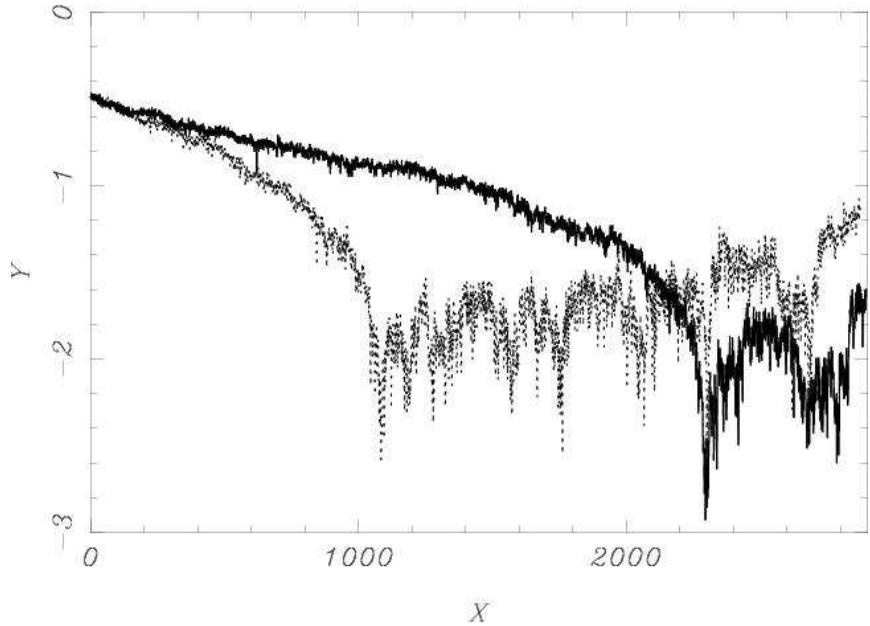}
   \caption[]{Evolution of the core radius for two star clusters well
   beyond the moment of core collapse. Time (in \Nbody\, units) is
   along the x-axis and radius ($\log r$, also in \Nbody) units along
   the y-axis. Initially each cluster contained 10,000 single stars
   distributed in a Plummer sphere with the same
   realization of positions and velocities. The solid curve was
   computed with equal masses for all stars, the dashed curve is
   computed with 20\% of the stars being twice as massive.  these
   calculations ware performed without stellar evolution and the units
   along the axis are in dimension less \Nbody\, units.
   \label{fig:CC10k}
   }
   \end{figure}

We illustrate phase C in Fig.\,\ref{fig:CC10k} with a simulation of
10,000 identical point masses initially distributed in a Plummer
sphere.  Fig.\,\ref{fig:CC10k} shows the evolution of the core radius
of this model. Core collapse occurs at $\tcc \simeq 15.2\pm
0.1\,\trlx$ \index{core collapse} (for these initial conditions $\trlx
\simeq 150$\,\Nbody\, time units).  This result is consistent with
earlier calculations of e.g., Cohn (1980) and Makino
(1996).\nocite{1996ApJ...471..796M} Doubling the mass of 20\% of the
stars reduces the core collapse time to $\tcc \simeq 7.2\,\trlx$.
Making 20\% of the stars 10 or 100 times more massive reduced the time
of core collapse further, to $\tcc \simeq 1.4\,\trlx$ and $\tcc\simeq
0.16\,\trlx$, respectively.  Clearly, the more massive stars drive the
core collapse of the cluster by dynamical friction,\index{dynamical
friction} as $\tdf \simeq {\langle m \rangle \over m} \trlx$ (see
Eq.\,\ref{Eq:tdf}, and also Watters et al.\, 2000).

    The presence of a population of relatively massive stars, such as
    black holes will therefore shorten the timescale for core collapse
    in phase C of the cluster evolution, but even in the absence of
    black holes or other heavy remnants core collapse cannot be
    prevented. In phase A this role was played by the most massive
    main-sequence stars.

    Note that if it is the population of dark remnants, black holes,
    neutron star and white dwarfs, which experience the core collapse,
    the lighter stars will not necessarily follow. So, the cluster may
    physically be in a state of core collapse, where the optical
    observer would measure a King profile (see Baumgardt et al 2003a;
    2003b) \nocite{2003ApJ...589L..25B} \nocite{2003ApJ...582L..21B}

\subsection{The consistent picture}

    In this section we have seen that a star cluster experiences three
    very distinct evolutionary phases, each of which is dominated
    either by relaxation or by stellar mass loss.

\begin{figure}
\vspace*{3cm}
   \includegraphics[width=0.6\textwidth,angle=-90]{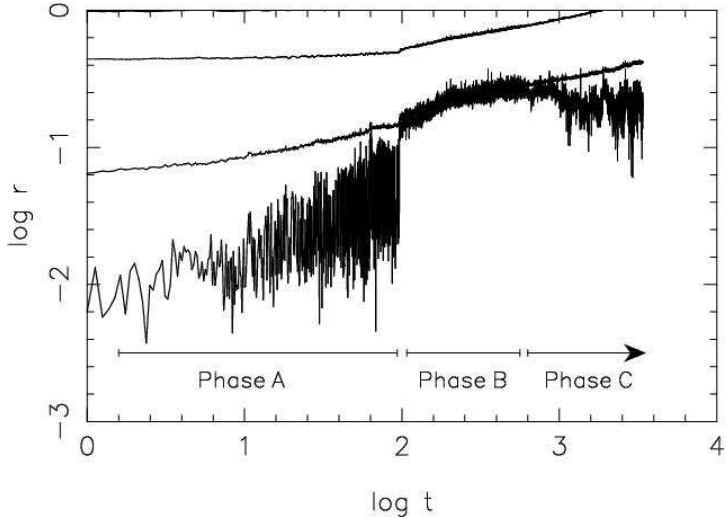}
   \caption[]{Evolution of the core radius (lower line) 10\% and the
   25\% Lagrangian radii for a 128k star cluster (axis are in \Nbody\,
   units). The various stages of cluster evolution are indicated near
   the bottom of the figure. In the early evolution of the cluster
   expansion of the core is driven by mass segregation, followed by
   phase B in which the cluster expands by stellar evolution. Near the
   end of the simulation the cluster tends to experience core collapse
   (Phase C). 
   \label{fig:Phase_ABC}
   }
   \end{figure}

    Figure.\,\ref{fig:Phase_ABC} summarized these phases in a single
    simulation which contains all relevant physics. The simulation
    presented here is carried out to illustrate the formation
    mechanism for intermediate mass black holes in dense young star
    clusters. It was performed with 131072 stars from a Salpeter
    initial mass function, 10\% (13107) of the stars form a hard
    binary system with a close companion.  The initial conditions for
    this simulation are explained in more detail in
    \S\,\ref{Sect:MGG11}. Here we only use the result of the
    simulation to illustrate the three distinct evolutionary phases in
    star cluster evolution, as with the adopted parameters each phase
    is clearly present.
        
    The three phases, A, B and C, are identified with the horizontal
    bars in fig.\,\ref{fig:Phase_ABC}. It it not a-priori clear when
    one phase stops and the next starts, and some gray area has to be
    allowed in which both, stellar evolution and relaxation, may
    temporarily have similar effect on the cluster.  These runs ware
    continued till 100\,Myr.\footnote{There was no particular reason
    to terminate this calculation at about 100\,Myr, other than that I
    got a bit tired of babysitting the run after three months straight
    on the GRAPE-6 at the University of Tokyo. In time, this cluster
    will experience core collapse again, to dissolve eventually.}
    Since this simulation was performed in isolation, the dissolution
    of the cluster will take quite a while (Baumgardt, Hut \& Heggie
    2002)\nocite{2002MNRAS.336.1069B}

\section{Black holes in star clusters}

In the previous sections we have set the stage for the evolution of
star clusters, and we have introduced the fundamental physics. We will
now continue with the evolution of black holes and their progenitors
in star clusters.

\subsection{The formation of intermediate mass black holes in phase A clusters}
\label{Sect:IMBHformation}

Young star clusters, with a half mass relaxation time $\trlx \aplt
100$\,Myr are, as we discussed in Sect.\,\ref{Sect:PhaseA}, prone to
dynamical friction, and therefore are likely to experience core
collapse before stellar mass loss drives the expansion of the cluster.
Realistic clusters have a broad range in initial stellar masses,
\index{initial mass function} generally from $m_{\rm min} \aplt
0.1$\,\msun\, to $m_{\rm max} \apgt 100$\,\msun. Adopting such a mass
function as a condition at birth, the mean mass $\mmean$ ranges then
from $\mmean \sim0.39$\,\msun\, (Salpeter
1955)\nocite{1955ApJ...121..161S} to about 0.65\,\msun\, (Scalo
1986),\nocite{scalo86} depending on the specific mass function
adopted.  Here I like to stress that there is a large variety of
initial mass function available, apart from the two above mentioned
there are the Miller \& Scalo (1979) \nocite{1979ApJS...41..513M}
\nocite{1977BAAS....9..566M}, Koupa, Tout \& Gilmore
(1990)\nocite{1990MNRAS.244...76K} with some adjustments for high mass
stars Kroupa \& Weidner (2003, see also Kroupa,
2001).\nocite{2001MNRAS.322..231K} \nocite{2003ApJ...598.1076K} These
mass functions seem to differ quite substantially, but for the
dynamical evolution of the cluster they do not make a big difference,
as for this process it is the ratio of most massive star to the mean
mass which counts (see Eq.\,\ref{Eq:tdf}).

In a multi-mass system, core collapse \index{core collapse}is driven
by the accumulation of the most massive stars in the cluster
center. This process takes place on a dynamical friction time scale
(Eq.\,\ref{Eq:tdf}).  Empirically, we find, for initial mass functions
of interest here, that core collapse (actually, the appearance of the
first persistent dynamically formed binary systems) occurs at about
(Portegies Zwart \& McMillan 2000; Fregeau et al
2002)\nocite{2002ApJ...570..171F}
\begin{equation}
	\tcc \simeq 0.2\,\trlx\,.
\end{equation}
This core collapse time is taken in the limit where stellar evolution
is unimportant, i.e.~where stellar mass loss is negligible and the
most massive stars survive until they reach the cluster center, i.e.:
what we called phase A in \S\,\ref{Sect:PhaseA}.

Experimentally we find that starting with a Scalo (1986) initial mass
function and a King (1966) $\Wo=6$ density distribution the time to
reach the first core collapse is $\tcc \simeq 0.19\pm0.08\,\trlx$.
Some of our simulations ware performed with high $W_0 \apgt 9$
concentration. Core collapse\index{core collapse} in these models
occurred within about million years, which for these simulations
corresponded to $\tcc \simeq 0.01\,\trlx$, or about 7.5\% of the core
relaxation time.  These findings are consistent with the Monte-Carlo
simulations performed by Joshi, Nave \& Rasio (2001).

The collapse of the cluster core may initiate physical collisions
between stars.  The product of the first collision is likely to be
among the most massive stars in the system, and to be in the core.
This star is therefore likely to experience subsequent collisions,
\index{collision runaway}
resulting in a collision runaway (see Portegies Zwart et al.\,
1999).\nocite{1999A&A...348..117P} The maximum mass that can be grown
in a dense star cluster if all collisions involve the same star is
\Mrun, where
\begin{equation}
	{d\Mrun \over dt} = \ncoll \dmcoll.
\label{Eq:Mrunaway}\end{equation}
Here \ncoll\, and \dmcoll\, are the average collision rate and the
average mass increase per collision (assumed independent).  We now
discuss these quantities in more detail.  Interestingly enough, G{\"
u}rkan et al.\,(2004) \nocite{2004ApJ...604..632G} performed
comparable calculations with a Monte-Carlo \Nbody\, code, which
produces qualitatively the same results. Also the calculations of
Portegies Zwart et al (2004)\nocite{2004PZetalNature} who used two
independently developed \Nbody\, codes ({\tt NBODY4} and {\tt
Starlab}), obtain similar results.

\subsubsection{The collision rate \ncoll}

A key result from the simulations of Portegies Zwart et al. (1999) was
the fact that collisions between stars generally occur in dynamically
formed (``three-body'') binaries.  The collision rate is therefore
closely related to the binary formation\index{binary formation} rate,
which we can estimate from first principles.

The flux of energy through the half-mass radius of a cluster during
one half-mass relaxation time is on the order of 10\% of the cluster
potential energy, largely independent of the total number of stars or
the details of the cluster's internal structure (Goodman
1987).\nocite{1987ApJ...313..576G} For a system without primordial
binaries this flux is produced by heating due to dynamically formed
binaries (Makino \& Hut 1990)\nocite{1990ApJ...365..208M}.  It is
released partly in the form of scattering products which remain bound
to the system, and partly in the form of potential energy removed from
the system by escapers recoiling out of the cluster (Hut \& Inagaki
1985).\nocite{1985ApJ...298..502H} Makino \& Hut argue, for an
equal-mass system, that a binary generates an amount of energy on the
order of $10^2 kT$ via binary--single-star scattering (where the total
kinetic energy of the stellar system is $\frac32N kT$).  This
quantity originates from the minimum binding energy of a binary that
can eject itself following a strong encounter.  Assuming that the
large-scale energy flux in the cluster is ultimately powered by binary
heating in the core. It follows that the required formation rate of
binaries via three-body encounters is
\begin{equation}\label{3bb_rate}
	\nbf \simeq 10^{-3} \frac{N}{\trlx}.
\end{equation}
For systems containing significant numbers of primordial binaries,
which segregate to the cluster core, equivalent energetic arguments
(Goodman \& Hut 1989) lead to a similar scaling for the net rate at
which binary encounters occur in the core.

The above arguments apply to star clusters comprising identical
point-masses.  In a cluster with a range of stellar masses, three-body
binaries generally form from stars which are more massive than
average.  After repeated exchange interactions, the binary will
consist of two of the most massive stars in the cluster.  Conservation
of linear momentum during encounters with lower mass stars means that
the binary receives a smaller recoil velocity, making it less likely
to be ejected from the cluster.  The binary must therefore be
considerably harder---$\apgt 10^3 kT$---before it is ejected following
a encounter with another star (see Portegies Zwart \& McMillan
2000).\nocite{2000ApJ...528L..17P}

However, taking the finite sizes of real stars into account, it is
quite likely that such a hard binary experiences a collision rather
than being ejected.  A strong encounter between a single star and a
hard binary generally results in a resonant interaction.  Three stars
remain in resonance until at least one of them escapes, or a collision
reduces the three-body system to a stable binary.  For harder binaries
it becomes increasingly likely that a collision occurs instead of
ejection (McMillan 1986; Gualandris et al. 2004; Fregeau et
al. 2004). \nocite{1986ApJ...306..552M} \nocite{2004astro.ph..1451G}
\nocite{2004astro.ph..1004F} In the calculations of Portegies Zwart et
al.  (1999) most binaries experience a collision at a binding energy
of order $10^2kT$, considerably smaller than the binding energy
required for ejection.  Accordingly, we retain the above estimate of
the binary formation rate (Eq. \ref{3bb_rate}) and conclude that the
collision rate per half-mass relaxation time is
\begin{equation}
	\ncoll \sim 10^{-3} f_{\rm c} {N \over \trlx}.
\label{Eq:ncoll_binary}\end{equation}
Here we introduce $f_{\rm c} \leq 1$, the effective fraction of
dynamically formed binaries that produce a collision.  Note again that
Eq.\,\ref{Eq:ncoll_binary} is valid only in the limit where stellar
evolution is unimportant.

The most massive star in the cluster is typically a member of the
interacting binary and therefore dominates the collision rate.
Subsequent collisions cause the runaway to grow in mass, making it
progressively less likely to escape from the cluster. The star which
experiences the first collision is therefore likely to participate in
subsequent collisions.  The majority of collisions thus involve one
particular object---the runaway merger---generally selected by its
high initial mass and proximity to the cluster center.

For systems containing many primordial binaries the above argument
must be modified.  Since dynamically formed binaries tend to be fairly
soft---a few $kT$--- the fraction of interactions with primordial
binaries leading to collision is comparable to the value $f_{\rm c}$
above.  However, a critical difference is that, in systems containing
many binaries, the collisions involve many different pairs of stars,
not just the binary containing the massive runaway. The total
collision rate is therefore much higher, but most collisions do not
contribute to the growth of the runaway merger. The presence of
primordial binaries have little influence on the collision
runaway.\footnote{The author realizes that the simple mention of
primordial binaries opens-up an entire discussion which would require
several pages, which I try to prevent.}

\subsubsection{Average mass increase per collision}\label{Sect:massincrease}

Once begun, the collision runaway dominates the collision cross
section.  The average mass increase per collision depends on the
characteristics of the mass function in the cluster core.  A lower
limit for stars which participate in collisions can be derived from
the degree of segregation in the cluster.  Inverting Eq.\,\ref{Eq:tdf}
results in an estimate (still assuming an isothermal sphere) of the
minimum mass of a star that can reach the cluster core in time $t$ due
to dynamical friction:
\begin{equation}
	\mdf =	1.9 \msun\,
	        \left({1\,{\rm Myr} \over t}\right)
		\left({R \over 1\,{\rm pc}}\right)^{3/2} 
		\left({m \over 1\,\msun}\right)^{1/2} 
	        \left(\lnL\right)^{-1}\,.
\label{Eq:mdf}\end{equation}
Thus, at time $t$ and for a given mass $m$, there is a maximum radius
$R$ inside of which stars of mass $m$ will have segregated to the
core.  The stars contributing to the growth of the runaway are likely
to be among those more massive than \mdf, because their number density
in the core is enhanced by mass segregation, their collision cross
sections are larger, and they contribute more to $\dmcoll$ when they
do collide.

The shape of the central mass function of a segregated cluster is not
trivial to derive\footnote{In that case, the mass function in the core
takes on a rather curious form: the mass functions for stars with
masses $m\aplt\mdf$ and $m\apgt\mdf$ have roughly the same slopes as
the initial mass function, but the more massive stars are overabundant
because they have accumulated in the cluster center.}.  In thermal
equilibrium, the central number densities of stars of different masses
would be expected to scale as
\begin{equation}
	n_0(m) \sim m^{3/2}\,\frac{dN}{dm}\,,
\end{equation}
where $dN/dm$ is the global initial mass function, which
scales roughly as $m^{-2.7}$ at the high-mass end ($m \apgt 10\msun$).
The distribution of secondary masses (i.e.~the masses of the lighter
stars participating in collisions) does not follow the above simple
relation.  Rather, we find that stars in the core do not reach thermal
equilibrium (a result generally consistent with earlier findings by
Chernoff and Weinberg 1990\nocite{1990ApJ...351..121C} and Joshi, Nave
\& Rasio 2001),\nocite{2001ApJ...550..691J} and that the dynamical
nature of the collisional processes involved means that more massive
stars tend to be consumed before lower-mass stars arrive in the core.
In addition, most collisions involve three-body binaries and
interactions with higher order multiples in a multi-mass environment.

Empirically, we find that the secondary mass distribution is quite
well fit by a power-law, $dN/dm \propto m^{-2.3}$ (coincidentally very
close to a Salpeter distribution).  Integrating this expression from a
minimum mass of {\mdf} (and ignoring the upper limit) results in a
mean mass increase per collision of
\begin{equation}
	\dmcoll \simeq 4 \mdf\,.
\label{Eq:mcoll}\end{equation}
If we neglect stars with masses less than $\mdf$ and substitute
Eq.\,\ref{Eq:trlx} into Eq.\,\ref{Eq:mdf} and Eq.\,\ref{Eq:mcoll} then
the mass increase per collision can be written as
\begin{equation}
	\dmcoll \simeq 4 {\trlx \over t} \mm \lnL\,.
\label{Eq:dmcoll}\end{equation}
This quantity remains rather constant over the entire collisional
lifetime, e.g., about 3\,Myr (see also \S\,\ref{Sect:low_density}).

\subsubsection{Lifetime of a cluster in a static tidal field}
\label{sect:ug}

With simple expressions for $\ncoll$ and $\dmcoll$ now in hand, we
return to the determination of the runaway growth rate
(Eq.~\ref{Eq:Mrunaway}).  The evaporation of a star cluster which
fills its Jacobi surface in an external potential is driven by tidal
stripping.  Portegies Zwart et al (2001a) have studied the evolution
of young compact star clusters within $\sim200$\,pc of the Galactic
center.  Their calculations employed direct {\nbody} integration,
including the effects of both stellar and binary evolution and the
(static) external influence of the Galaxy, and made extensive use of
the GRAPE-4.  They found that the mass of a typical model cluster
decreased almost linearly with time:
\begin{equation}
	M = M_{\rm 0} \left(1 - \frac{t}{\tdisr}\right)\,.
\label{Eq:mass}\end{equation}
Here $M_{\rm 0}$ is the mass of the cluster at birth and \tdisr\, is
the cluster's disruption time.  Portegies Zwart et
al.~(2001a)\nocite{2001ApJ...546L.101P} found that their model
clusters dissolved within about 30\% of the two-body relaxation time
at the tidal radius (defined by substituting the tidal radius instead
of the virial radius in Eq.\,\ref{Eq:trlx}).  In terms of the
half-mass relaxation time, this translates to $\tdisr =
1.6$--5.4\,\trlx, depending on the initial density profile (the range
corresponds to King [1966]\nocite{1966AJ.....71...64K} dimensionless
depths $\Wo=3$--7; more centrally condensed clusters live longer).

Substituting Eqs.\,\ref{Eq:ncoll_binary} and \ref{Eq:dmcoll} into
Eq.\,\ref{Eq:Mrunaway}, and defining $M_{\rm 0} = N\mmean$ to rewrite
Eq.\,\ref{Eq:mass} in terms of the number of stars in the cluster,
we find
\begin{eqnarray}
	{d\Mrun \over dt} 
		&=& 4 \times 10^{-3} f_{\rm c}
				{N \mmean \lnL \over t} 
			\nonumber \\
		&=& 4 \times 10^{-3} f_{\rm c} M_{\rm 0} \lnL
				\left(\frac{1}{t} - \frac{1}{\tdisr}\right)\,.
\end{eqnarray}
Integrating from $t=\tcc$ to $t=\tdisr$ results in
\begin{equation}
	\Mrun = \mseed + 4 \times 10^{-3} f_{\rm c}
		     M_{\rm 0} \lnL
			\left[ 
      			     \ln \left( {\tdisr \over \tcc} \right) 
				+ {\tcc \over \tdisr} - 1
			\right] \,.
\label{Eq:mrunaway}\end{equation}
Here $\mseed$ is the seed mass of the star which initiates the runaway
growth, most likely one of the most massive stars initially in the
cluster.  With $\tcc \simeq 0.2\trlx$, Eq\,\ref{Eq:mrunaway} reduces to
\begin{equation}
	\Mrun = \mseed + 4 \times 10^{-3} f_{\rm c}
			 M_{\rm 0} \kappa \lnL \,,
\label{Eq:Mrunaway_II}\end{equation}
where $\kappa \simeq \ln (\tdisr/\tcc) + \tcc/\tdisr - 1 \sim 1$.

\begin{figure}[ht]
\psfig{figure=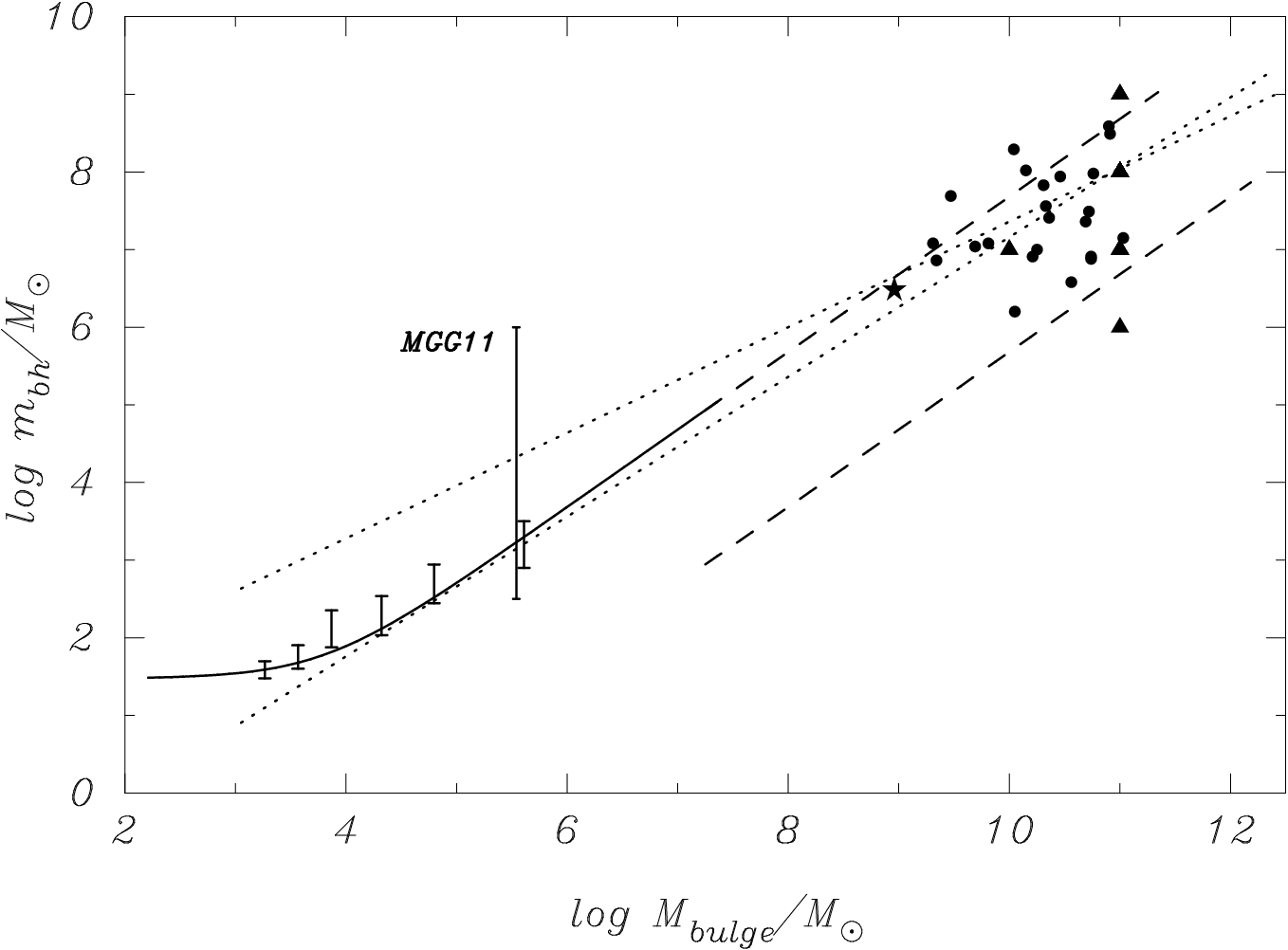,width=\columnwidth,angle=-90}
\caption[]{The mass after a period of runaway growth as a function of
the mass of the star cluster.
The solid line is $\Mrun = 30 + 8 \times 10^{-4} M_{\rm 0} \lnL$ (see
Eq.\,\ref{Eq:Mrunaway_II} with $f_{\rm c}=0.2$, $\gamma=1$ and
$\lnL=\ln M_{\rm 0}/\msun$, where $M_{\rm 0}$ is the initial mass of
the cluster or $10^6$\,\msun, whatever is smaller). This relation may
remain valid for larger systems built up from many clusters having
masses $\aplt10^6\msun$.  For clusters with $M_{\rm 0} \apgt
10^7$\,\msun\, we therefore extend the relation as a dashed line.  The
logarithmic factor, however, remains constant, as it refers to the
clusters out of which the bulge formed, not the bulge itself
(Ebisuzaki et al 2001).  The bottom dashed line shows $0.01\Mrun$.
The five error bars to the left give a summary of the results
presented by Portegies Zwart \& McMillan (2002), the two right most
error bars are taken from Portegies Zwart et al (2004).
The error bar indicating {\em MGG11} gives the result of estimates for
the mass of the intermediate mass black hole in the M82 star cluster
MGG11\index{black hole!super massive} (Matsumoto \& Tsuru 1999; McCrady
et al 2003).
The Milky Way is represented by the asterisk using the bulge mass from
Dwek (et al. 1995)\nocite{1995ApJ...445..716D} and the black hole mass
 from Eckart \& Genzel (1997)\nocite{1997MNRAS.284..576E} and Ghez
(2000).\nocite{1998ApJ...509..678G}
Bullets and triangles (upper right) represent the bulge masses and
measured black hole mass of Seyfert galaxies and Quasars, respectively
(both from Wandel 1999;
2001).\nocite{2001astro.ph..8461W}\nocite{1999ApJ...519L..39W} The
dotted lines gives the range in solutions to a least squares fits to
the bullets and triangles (Wandel 2001).
}
\label{fig:bhmass}
\end{figure}

In figure\,\ref{fig:bhmass} we present this relation in the form of a
solid curve. We comment further on the left side of this figure where
we extrapolate the relation in Eq.\,\ref{Eq:Mrunaway_II} to galactic
nuclei masses.

  The maximum mass of the runaway merger for clusters which are
  disrupted by inspiral (which of course always destroys the cluster
  before it reaches the center) may be calculated by replacing
  $\tdisr$ in Eq.\,\ref{Eq:mrunaway} by $\tdfg$.  The right-hand side
  of that equation then becomes a function of
  \begin{equation}
	\frac{\tdfg}{\tcc}
		\simeq 9.0
		\left( {R_i \over  10 \pc} \right)^{2.1} 
		\left( {0.25 \pc \over R} \right)^{3/2}
		\left( {10^5 \msun \over M} \right)^{3/2}
  \end{equation}
  We can also estimate the maximum initial distance from the Galactic
  center for which core collapse \index{core collapse} occurs (and
  hence runaway merging may begin) before the cluster disrupts by
  setting $\tdfg = \tcc$.  The result is $R_i \apgt 0.0025 \pc
  \left(RM/[\pc\,\msun] \right)^{0.71}$.  For $R = 0.25$\,pc and $M =
  10^5 \msun$, we find $R_i \apgt 3.3$\,pc.
  
\subsection{Calibration with {\nbody} simulations}\label{sect:simulations}

The development of the GRAPE (see \S\,\ref{Sect:GRAPE}) family of
special-purpose computers makes it relatively straightforward to test
and tune the above theory using direct {\nbody} calculations.  The
$N$-body calculations performed by Portegies Zwart \& McMillan (2002)
span a broad range of initial conditions in the relevant part of
parameter space.  The number of stars varied from 1k (1024) to 64k
(65536).  The initial conditions explored by Portegies Zwart et al
(2004) ranges from 131,072 to 585,000 stars. G\"{u}rkan et al (2004)
performed similar calculations using a Monte-Carlo \Nbody\, code, they
adopted a considerably larger number of particles.

Initial density profiles and velocity dispersion for the models were
taken from Heggie-Ramamani models (Heggie \& Ramamani
1995)\nocite{1995MNRAS.272..317H} with {\Wo} ranging from 1 to 7, and
from King (1966) models with $\Wo=3$ -- 15.  At birth, the
Heggie-Ramamani clusters were assumed to fill their zero-velocity
(Jacobi) surfaces in the Galactic tidal field, while the classical
King models were assumed to be isolated.  In most cases we adopted an
initial mass function between 0.1\,\msun\ and 100\,\msun\, suggested
for the Solar neighborhood by Scalo (1986) and Kroupa, Tout \& Gilmore
(1990).\nocite{scalo86} However, several calculations were performed
using power-law initial mass functions with exponents of -2 or -2.35
(Salpeter) and lower mass limits of 1\,\msun.

\subsubsection{Collision rate during phase A}
\index{collision rate}

In all calculations, the first collision occurred shortly after the
formation of the first $\apgt 10\,kT$ binary by a three-body
encounter, i.e. close to the time of core collapse.  When stars were
given unrealistically large radii (100 times larger than normal), the
first collisions occurred only slightly (about 5\%) earlier.

\begin{figure}[htbp!]
\includegraphics[width=\linewidth,angle=-90]{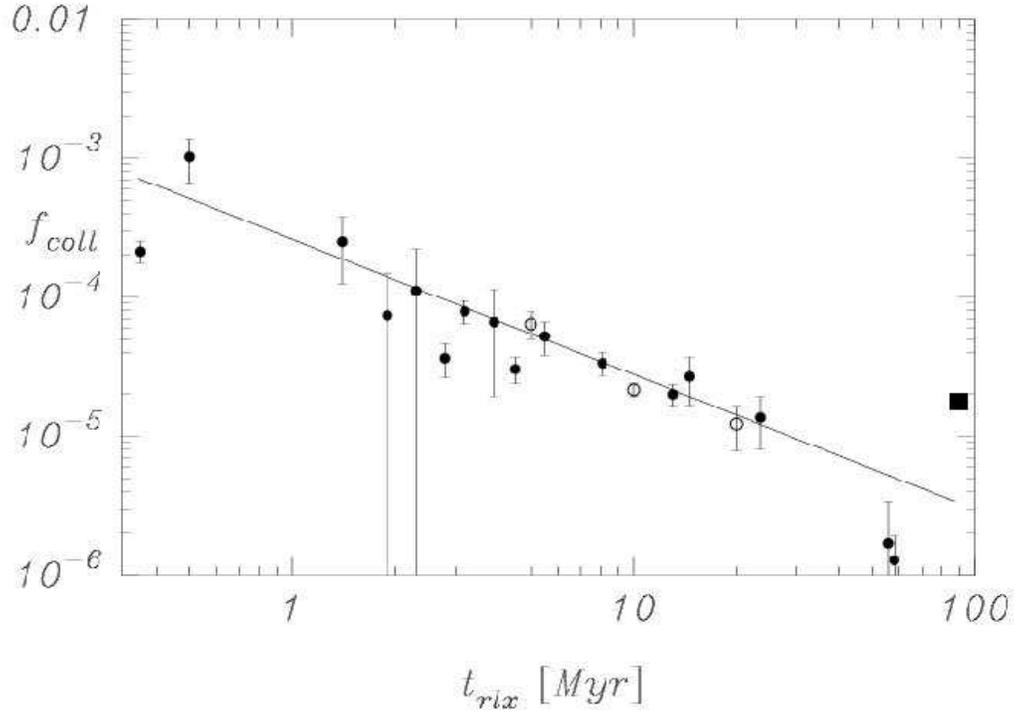}
\caption[]{Mean collision rate $\fcoll = {\Ncoll / N \tlast}$ as
function of initial relaxation time for the simulations performed by
Portegies Zwart et al. 1999; 2004 and Portegies Zwart \& McMillan
2002).  Here \tlast\, is the time of the last collision in the
cluster. The open circles give the results of systems which are
isolated from the Galactic potential (see Portegies Zwart et al 1999).
Vertical bars represent Poissonian 1-$\sigma$ errors.  The solid line
is a least squares fit to the data (see Eq.\,\ref{Eq:ncoll}).  The
strong reduction in the collision rate for cluster with an initial
relaxation time $\trlx \apgt 30$\,Myr is probably real.  Note however
the filled square to the right from the calculations of Portegies
Zwart et al (2004), which gives a higher collision rate due to the
high initial concentration of the models and the top-heavy mass
function.  }
\label{Fig:ncoll_Trlx}
\end{figure}

As discussed earlier, the first star to experience a collision was
generally one of the most massive stars in the cluster; this star then
became the target for further collisions.  In models where the core
collapse time exceeds about 3\,Myr the target star explodes in a
supernova before experiencing runaway growth.  The collision rates in
these clusters were considerably smaller than for clusters with
smaller relaxation times (see Fig.\,\ref{Fig:ncoll_Trlx}).
As discussed in more detail in \S4, the onset of stellar evolution
terminates the collision process; premature disruption of the cluster
also ends the period of runaway growth.

The first physical collisions occur at the moment of core collapse.
The cluster then enters a phase which is dominated by stellar
collisions.  In particular one single object experiences many repeated
collisions, giving rise to a collision runaway. We identify this
particular object as the {\em designated target}.  In our models
collisions tend to increase the mass of the collision product; only a
small fraction of the mass of the incoming star is lost; the
designated target therefore tends to increase in mass.

The number of collisions in the direct \Nbody\, simulations ranged
from 0 to 100.  Fig.\,\ref{Fig:ncoll_Trlx} shows the mean collision
rate {\ncoll} per star per million years as a function of the initial
half-mass relaxation time.  The solid line in
Fig.\,\ref{Fig:ncoll_Trlx} is a fit to the simulation data, and has
\begin{equation}
 	\ncoll = 2.2\times 10^{-4} {N \over \trlx}\,,
\label{Eq:ncoll}\end{equation}
for $\trlx\aplt 20-30$\,Myr, consistent with our earlier estimate
(Eq.\,\ref{Eq:ncoll_binary}) if $f_{\rm c} = 0.2$.  The quality of the
fit in Fig.\,\ref{Fig:ncoll_Trlx} is quite striking, especially when
one bears in mind the rather large spread in initial conditions for
the various models. See however the prominent square to the right,
which is about a factor $\sim 3$ above the fitted curve. This
discrepancy is mainly caused by the high average stellar mass in these
models and by the use of much more concentrated King models.

The collision runaway phase lasts until about 3.3 Myr, at which time
the designated target tends to collapse to a black hole.

\begin{figure}
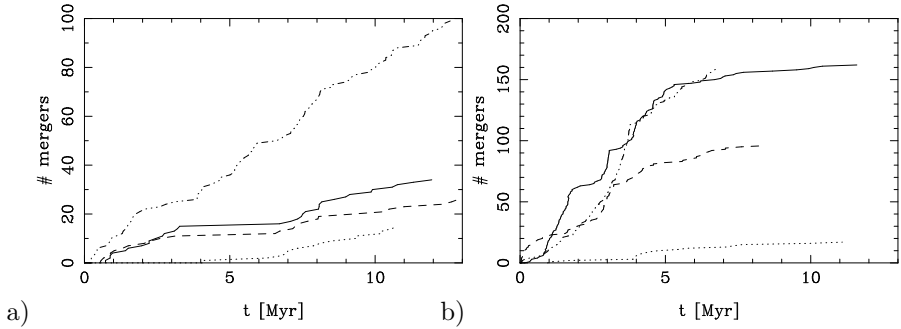

\noindent
\begin{minipage}[b]{0.50\linewidth}
a) \psfig{figure=./fig/SPZ_Tcoll.ps,width=\columnwidth,angle=-90} 
\end{minipage}
\begin{minipage}[b]{0.50\linewidth}
b) \psfig{figure=./fig/Holger_Tcoll.ps,width=\columnwidth,angle=-90} 
\end{minipage}
\caption[]{Cumulative distributions of the number of mergers which
occurred within time $t$.  Panel a) gives the results of the
calculations performed with {\tt starlab}, panel b) is from {\tt
NBODY4}.  The dotted, solid and dashed curves are for the models with
$W_0=8$ (in the right panel we show the results for $W_0=7$), $W_0=9$
and $W_0=12$. The dash-3-dotted curve in panel a) is for a model with
$W_0=12$ with 10\% primordial binaries. The dash-3-dotted curve in
panel b) is for a model with $W_0=9$ using 585 000 stars, distributed
according to a Kroupa (2001) initial mass function between 0.1\,\msun
and 100\,\msun.  }
\label{fig:Tcoll}
\end{figure}

In fig.\,\ref{fig:Tcoll} we present the cumulative distribution of the
number of mergers of some of the calculations by Portegies Zwart et al (2004). 

\begin{figure}[htbp!]
\vspace*{2cm}
\includegraphics[width=0.5\linewidth,angle=-90]{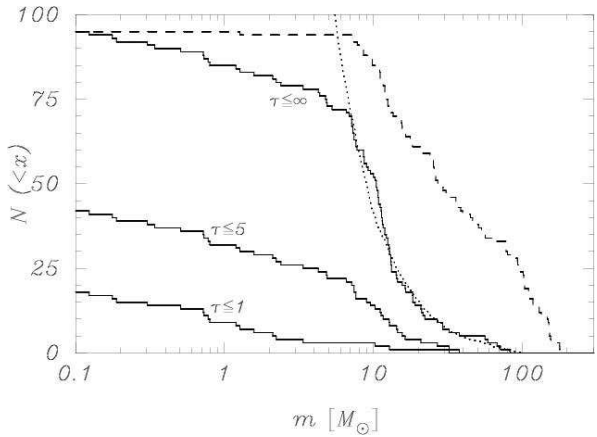}
\caption[]{Cumulative mass distributions of primary (dashed line) 
and secondary (solid lines) stars involved in collisions.  Only those
secondaries experiencing their first collision are included.  From
left to right, the solid lines represent secondary stars for which $\tau
\equiv \tcoll/\tdf \aplt 1$, 5, and $\infty$.  The numbers of
collisions included in each curve are 18 (for $\tcoll/\tdf \aplt 1$),
42 and 95 (rightmost two curves).  The dotted line gives a power-law
fit with the Salpeter exponent (between 5\,\msun\, and 100\,\msun) to
the right most solid curve ($\tau \aplt \infty$).  }
\label{Fig:dmcoll_distrn}
\end{figure}

Figure\,\ref{Fig:dmcoll_distrn} shows the cumulative mass
distributions of the primary (more massive) and secondary (less
massive) stars participating in collisions.  We include only events in
which the secondary experienced its first collision (that is, we omit
secondaries which were themselves collision products).  In addition,
we distinguish between collisions early in the evolution of the
cluster and those that happened later by subdividing our data based on
the ratio $\tau = \tcoll/\tdf$, where $\tcoll$ is the time at which a
collision occurred and $\tdf$ is the dynamical friction time scale of
the secondary star (see Eq.\,\ref{Eq:tdf}).  The solid lines in
Figure\,\ref{Fig:dmcoll_distrn} show cuts in the secondary masses at
$\tau \aplt 1$, $\tau \aplt 5$ and $\tau < \infty$ (rightmost line).
The mean secondary masses are $\mmean = 4.0\pm 4.8$\,\msun,
$8.2\pm6.5$ and $\mmean = 13.5\pm8.8$\,\msun\, for $\tau
\aplt 1$, 5 and $\infty$, respectively.

The distribution of primary masses in Figure\,\ref{Fig:dmcoll_distrn}
(dashed line) hardly changes as we vary the selection on $\tau$.  We
therefore show only the full ($\tau \aplt \infty$) data set for the
primaries.  In contrast, the distribution of secondary masses changes
considerably with increasing $\tau$.  For small $\tau$, secondaries
are drawn primarily from low-mass stars.  As $\tau$ increases, the
secondary distribution shifts to higher masses while the low-mass part
of the distribution remains largely unchanged.  The shift from
low-mass ($\aplt 8$\,\msun) to high-mass collision secondaries ($\apgt
8$\,\msun) occurs between $\tau=1$ and $\tau=5$.  This is consistent
with the theoretical arguments presented in
Sec.\,\ref{Sect:massincrease}.  During the early evolution of the
cluster ($\tau\aplt 1$), collision partners are selected more or less
randomly from the available (initial) population in the cluster core;
at later times, most secondaries are drawn from the mass-segregated
population.\index{collision runaway}

Interestingly, although hard to see in Fig.\,\ref{Fig:dmcoll_distrn},
all the curves are well fit by power laws between $\sim 8$\,\msun\,
and $\sim 80$\,\msun (0.8\,\msun\, and 30\,\msun for the leftmost
curve).  The power-law exponents are $-0.4$, $-0.5$ and $-2.3$ for
$\tau \aplt 1$, 5, and $\infty$, and $-0.3$ for the primary (dashed)
curve.  (Note that the Salpeter mass function has exponent $-2.35$.)

Figure\,\ref{fig:bhmass} shows the maximum mass reached by the runaway
collision product as function of the initial mass of the star cluster.
Only the left side ($\log M/\msun \aplt 7$) of the figure is relevant
here; we discuss the extrapolation to larger masses in
Sec.\,\ref{Sect:massive_black_holes}.  The N-body results are
consistent with the theoretical model presented in
Eq.\,\ref{Eq:mrunaway}.

\subsection{Simulating the star cluster MGG11}\label{Sect:MGG11}
\index{MGG11}

In a recent publication (Portegies Zwart et al, 2004) we simulate a
well observed star cluster in the starburst Galaxy M82.\index{M82}
Interestingly enough, this cluster has experienced a prominent phase
A\index{phase A} evolution, and is currently in a phase B. In this
\S\, I will report some of the interesting results about these
simulations. Note that we have been quoting these results in earlier
occasions in the text, but here we review the global results.

The observed mass function, as reported by McCrady et al (2003) for
the M82 star cluster MGG11 is consistent with a Salpeter power-law
(with $x = -2.35$) between 1\,\msun\, and an upper limit which
corresponds to the age of the cluster of 7 to 12\,Myr.  By that time
all stars more massive than 17--25\,\msun\, have experienced a
supernova.  For the IMF we adopt the same Salpeter slope and lower
mass limit of 1\,\msun, but extend the upper limit to 100\,\msun. This
IMF has an average mass of $\langle m \rangle \simeq 3.1\,M_\odot$.
If we assume that at an age of 12--7\,Myr all stars between
17--25\,\msun\, and 100\,\msun\, are lost from the cluster by
supernova explosions, the mean stellar mass drops to $\langle m
\rangle \simeq 2.4-2.6\,M_\odot$. With a current total mass of
$3.5\times 10^5$\,\msun\, the cluster would contain 130\,000 to
140\,000 stars.  For clarity we decided to select 128k (131072) single
stars, resulting in an initial mass of about 406,000\,\msun.

McCrady et al\, (2003) measured the projected half light
radius\index{half light radius} of the cluster, $r_{\rm hp} =
1.2$\,pc.  De-projection of the half light radius depends on the
density profile. For King (1966) models in the range $W_0 = 3-12$ it
turns out that $r_{\rm h} = 0.75-0.88 r_{\rm hp}$ (which is somewhat
larger than Spitzer's, 1987 $r_{\rm h} \sim {3\over4} r_{\rm hp}$).  A
number of initial test simulations indicate that over a time span of
7--12\,Myr the projected half light radius of the selected IMF and
number of stars, the cluster expands by about a factor 1.3. We then
adopt an initial half mass radius for our models cluster of $r_{\rm
hm} = 1.2$\,pc.

We ignore the effect of the tidal field of M82.  The star cluster is
located at a distance of about 160\,pc from the dynamical center of
the Galaxy, assuming a distance of 3.6\,Mpc (Freedman et
al.\,1994)\footnote{Freedman et al.\, measured a distance of $3.63\pm
0.34$\,Mpc to M81, the neighboring galaxy of M82, using Cepheids.}.
With the relatively small mass of M82 of $\sim 10^9$\,\msun\, the
gravitational force of the Galaxy is negligible compared to the self
gravity of the stars within the cluster.

We performed several calculations starting from King models with
different central concentrations $W_0$. We also performed one run with
10 per cent primordial binaries and one run starting with 585\,000
stars and a Salpeter IMF between 0.1\,\msun\, and 100\,\msun.

\begin{table}
\caption[]{Resulting collisions for our model clusters as a function
of the initial density profile expressed in the King parameter $W_0$
(identified on the first column) and the average core density is presented in the second column.  The third column gives the number
of collisions which occurred in 12\,Myr. The fourth and fifth columns
give the number of collisions in which one particular star participates
and the average mass of the star with which it collides. The last
three columns give the number of collision and the mean mass of the
more massive and lower mass stars participating in these collisions.
}
\begin{tabular}{ll|rrrrrr} \hline
     &	              & \multicolumn{2}{|c|}{designated target} 
                      & \multicolumn{2}{|c|}{other stars} \\
$W_0$& $\rho_c$ & $n_{\rm coll}$ & $n_{\rm coll}$ & $\langle m \rangle$ 
	              & $n_{\rm coll}$ & \multicolumn{2}{|c|}{$\langle m \rangle$} \\
     & $\msunpc$&  \multicolumn{6}{c}{Results from {\tt Starlab}} \\ 
 3   &  4.55    & 2   & 0   &  NA  &  2 & 17.3 & 5.0 \\
 7   &  5.51    &  5   & 0   &  NA  &  5 & 13.7 & 8.4 \\
 8   &  5.81    & 17   & 0   &  NA  & 17 & 19.0 & 9.7 \\
 9   &  6.58    & 36   & 21  & 48.1 & 15 & 16.4 & 6.2 \\
 12  &  8.12    & 27   & 14  & 47.9 & 12 & 33.2 & 9.0 \\   
12$^\star$&     & 101  & 25  & 41.7 & 76 & 16.6 & 14.0 \\
\hline
     &  \multicolumn{6}{c}{Results from {\tt NBODY4}} \\ 
  7  &          &  19  &  0  & NA   & 19 & 24.7 & 4.6 \\
  9  &          & 164  & 99  & 30.0 & 65 & 34.8 & 8.1 \\
 12  &          & 98   & 70  & 38.7 & 28 & 50.1 & 7.8 \\
 9$^\circ$&     & 161  & 98  & 20.9 & 63 & 31.5 & 3.1 \\
\hline
\end{tabular} \\

$^\star$: run performed with 10 per cent hard primordial binaries. \\
$^\circ$: run performed with 585,000 stars and a Kroupa (2001) IMF
between 0.1\,\msun and 100\,\msun\, (for details see Portegies Zwart
et al 2004).
\label{Tab:collisions} 
\end{table}

To qualify the results we make the distinction between clusters with a
high central concentration and clusters with low concentration. Since
the initial half mass radius is the same for all models, we varied the
density profile. For the density profile we adopted King (1966)
models. We draw the empirical distinction between high concentration
models having $W_0\apgt9$; low concentration models have $W_0\aplt 8$.
With the adopted half-mass radius and total mass the high concentration
cluster $W_0 \apgt 9$ models have core density $\log \rho_c >
6$\,\msun\,pc$^{-3}$.

\begin{figure}
\psfig{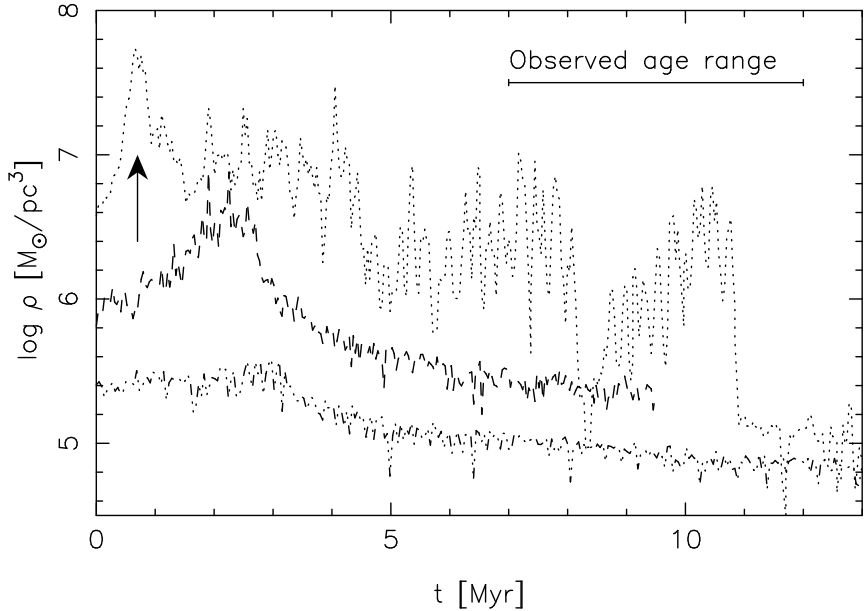} 
\caption[]{Evolution of the central density for star clusters with
$W_0=7$ (dash-3-dotted line), $W_0=8$ (dahses), $W_0=9$ (dotted
curve).  The $W_0=9$ model shows a clear core collapse at about
0.7\,Myr, indicated with the arrow.  The less concentrated models
($W_0=8$ and $W_0=7$) show a very shallow start of a core collapse
near $t\sim 3$\,Myr, but in these cases the collapse of the core is
arrested by stellar mass loss.  In
fig.\,\ref{Fig:core_radius_evolution} we present the evolution of the
core radius for the same calculations.
 }
\label{fig:rhocore}
\end{figure}

In figure\,\ref{fig:rhocore} we present the evolution of the central
density of various {\tt starlab} models with $W_0=7$, 8 and 9. As
discussed, in the low concentration models core collapse is arrested
by the copious mass loss from the evolving massive stars. In the high
concentration clusters core collapse occurs early enough that the
process is little affected by stellar mass loss.

\subsubsection{Clusters with $\log \rho_c > 6$\,\msun\,{\rm pc}$^{-3}$}

According to Portegies Zwart et al (1999), who studied similar
clusters as Portegies Zwart \& McMillan (2002), the high collision
rate is mainly the result of binaries created in 3-body encounters
during core collapse.  In our latest models, however, only about half
($0.52 \pm 0.1$) the collisions are preceded by the formation of a
binary, the other half are direct hits in which no third star was
bound to the two stars which ultimately collided. Though seemingly a
detail, it has far reaching consequences for the interpretation of the
collisional growth which plays an important role in the evolution of
these systems.

After the first epoch of runaway growth and the collapse of the
designated target, an epoch of about 3\,Myr starts in which the
collision rate drops dramatically.  This phase is visible in
fig.\,\ref{fig:Tcoll} between $\sim 3.3$\,Myr and $\sim 6$\,Myr.

This quiet phase lasts until the first M5I hyper giants appear, which,
in our stellar evolution model, happens at a turn-off mass of $\sim
25$\,\msun\, (at $\sim 6$\,Myr).  (Note that in our stellar evolution
model {\tt SeBa} \index{SeBa} stars below $\sim 25$\,\msun\, turn into
neutron stars\index{neutron star} where more massive stars become
black holes,\index{black hole} see Portegies Zwart \& Verbunt 1996) By
this time the collision rate picks up again to 2-3 per Myr. The
spectral type M0I -- M5I star dominate the collision rate; the
designated target, now an intermediate mass black hole, participates
in only about one-third of the collisions. This phase lasts until
about 9--12\,Myr, after which the rate drops below 0.3 collisions per
Myr.

\subsubsection{Clusters with $\log \rho_c < 6$\,\msun\,pc$^{-3}$}\label{Sect:low_density}

In low density cluster $W_0\aplt 8$ the initial phase of runaway
growth is absent. The subsequent collisional phase at an age $\apgt
6$\,Myr, however, occurs at a comparable rate as in the high density
simulations (see fig.\,\ref{Fig:dmcoll_distrn}).  As discussed in the
previous \S\, there is no designated target in this phase; all
collisions tend to happen between massive stars (or stellar remnants)
of which at least one component is evolved (spectral type M0I to M5I).
At a rate which becomes gradually smaller for less concentrated
clusters: 3.4 collisions per Myr for $W_0=9$, 3.2 for $W_0=8$, 1.6 for
$W_0=7$, 1.5 for $W_0=5$ and 0.5 for $W_0=3$.  Curiously enough the
$W_0=12$ models have a collision rate of only 2.3 per Myr, which is
lower than the less concentrated $W_0=9$ models.

The average star with which a collision occurs (counting the least
massive of the two stars) depends on the initial cluster density
profile, ranging from a 5.3\,\msun\, for $W_0=5$, 6.2\,\msun\, for
$W_0=7$, 9.7\,\msun\, for $W_0=8$ and $\sim 30$ for both $W_0=9$ and
$W_0=12$ (consistent with the expectations of Portegies Zwart \&
McMillan 2002)

\subsection{Black hole ejection in phase B and C cluster with $\trlx \apgt 100$\,Myr}\label{Sect:BHejection}

Here we discuss the evolution of star clusters in which the early
phase is dominated by stellar mass loss and the subsequent evolution
by the stellar interaction; $\tau_{\rm SE} < \tau_{\rm SD}$.  In this
regime or parameter space, the most massive stars evolve before the
structure of the star cluster has appreciably changed, i.e; no
intermediate mass black hole can form via the scenario discussed in
\S\,\ref{Sect:IMBHformation}.  The consequence is that the most
massive stars turn into black holes and neutron stars before they had
a chance to sink to the cluster center by dynamical friction. This
regime is valid for most globular clusters, and possibly many open
star clusters.

Upon the birth of a cluster we assume that the stars populate the
initial mass function from the hydrogen burning limit ($\sim
0.1$\,\msun) all the way to the most massive stars currently observed
in the Galaxy, which is about 100\,\msun.  Stars with Solar abundance
between 50\,\msun\, and 100\,\msun\, leave the main sequence at an age
of about 3.7\,Myr to 3.3\,Myr, to explode in a supernova a few hundred
thousand years later.  The total mass in this range is about 1\%, and
the cluster therefore loses approximately 1\% of its total mass in
less than half a million years.\footnote{For an entire star cluster a
1\% mass loss is not very dramatic, and simply causes the cluster to
expand by about the same fraction. For a cluster core, which contains
less than 5\% of the total cluster mass (for $\Wo\apgt 8$), a 1\% mass
loss may drive a substantial expansion of the core.}

The first black holes are produced at about the same time. Black hole
formation proceeds to about 7--9\,Myr, until all stars with initial
masses exceeding 20--25\,\msun (Maeder 1992; Portegies Zwart et
al. 1997).  \nocite{1992A&A...264..105M}
\nocite{1997A&A...321..207P}\nocite{1998A&A...331L..29E} have
collapsed to black holes.\index{black hole!stellar mass} Assuming a
Scalo (1986) mass distribution with a lower mass limit of 0.1\,\msun\,
and an upper limit of 100\,\msun\, about 0.071\% of the stars are more
massive than 20\,\msun, and 0.045\% are more massive than 25\,\msun.
A star cluster containing $N$ stars thus produces $\sim6\times 10^{-4}
N$ black holes.  Known Galactic black holes have masses $\mbh$ between
6\,\msun\, and 18\,\msun\ (Timmes et al.\,
1996).\nocite{1996ApJ...457..834T} For definiteness, we adopt $\mbh =
10$\,\msun.

A black hole is formed in a supernova explosion.  If the progenitor is
a single star (i.e.\ not a member of a binary), the black hole
experiences little or no recoil and remains a member of the parent
cluster (White \& van Paradijs
1996).\nocite{1996ApJ...473L..25W}\footnote{There are some arguments
which indicate that black hole may received small 'impulse' kicks,
relative to neutron stars.}  If the progenitor is a member of a
binary, mass loss during the supernova may eject the binary from the
cluster potential via the Blaauw mechanism (Blaauw
1961),\nocite{1961BAN....15..265B} where conservation of momentum
causes recoil in a binary as it loses mass impulsively from one
component.  The Blaauw velocity kick can be as large as the relative
orbital velocity of the pre-supernova binary (see e.g.\, Nelemans et
al (2001).\nocite{2001bhbg.conf..312N} The escape speed is $\sim
40\,$\kms for a young globular cluster, and somewhat smaller for
YoDeCs (see Tab\,\ref{Tab:clusters}). Such high recoil velocities are
generally achieved only if the binary loses $\aplt 50$\% of its total
mass in the supernova, and if its orbital period is initially quite
small ($\aplt 2$\,yr) (Portegies Zwart \& Verbunt
1996).\nocite{1996A&A...309..179P} The binary frequency in globular
clusters is between 5 and 40\% (Rubenstein
1997)\nocite{1997PASP..109..933R}, and less than 30\% of binaries have
orbital periods smaller than 2 years (Rubenstein \& Bailyn 1997;
1999);\nocite{1997ApJ...474..701R}\nocite{1999ApJ...513L..33R} we
assume the same distributions of orbital parameters for binaries in
YoDeCs.  We therefore estimate that no more than $\sim 10$\% of black
holes are ejected from the cluster immediately following their
formation (i.e.\ the black hole retention fraction is $\apgt
90$\%). Note the remarkable difference here with the retention
fraction of neutron stars,\index{neutron star} which is less than
about 10\% (Drukier 1996).\nocite{1996MNRAS.280..498D}

After $\sim 40$\,Myr the last supernova has occurred, the mean mass of
the cluster stars is $\mbar \sim 0.56$\,\msun\ (Scalo 1986), and black
holes are by far the most massive objects in the system.  Mass
segregation cause the black holes to sink to the cluster core in a
fraction $\sim\mbar/\mbh$ of the half-mass relaxation time scale (see
Eq.\,\ref{Eq:tdf}).  For a young populous cluster, the relaxation time
is on the order of 10\,Myr; for a globular cluster it is about 1\,Gyr
(see Tab.\,\ref{Tab:timescales}).

By the time of the last supernova, stellar mass loss has also
significantly diminished and the cluster core starts to contract,
enhancing the formation of binaries by three-body interactions.
Single black holes form binaries preferentially with other black holes
(Kulkarni et al. 1993)\nocite{1993Natur.364..421K}, while black holes
born in binaries with a lower-mass stellar companion rapidly exchange
the companion for another black hole.  The result in all cases is a
growing black-hole binary population in the cluster core.  Once
formed, the black-hole binaries become more tightly bound through
superelastic encounters with other cluster members (Heggie 1975;
Kulkarni et al. 1993; Sigurdsson \& Hernquist 1993).
\nocite{1975MNRAS.173..729H}\nocite{1993Natur.364..423S} On average,
following each close binary--single black hole encounter, the binding
energy of the binary increases by about 20\% (Hut et al.\,1992);
\nocite{1992ApJ...389..527H} roughly one third of this energy goes
into binary recoil, assuming equal mass stars.  The minimum binding
energy of an escaping black-hole binary may then be estimated as
\begin{equation}
	E_{b,{\rm min}} \sim 36\, W_0\, \frac{\mbh}{\mbar}\, kT\,,
\end{equation}
where $\frac32 kT$ is the mean stellar kinetic energy and $W_0 =
\mbar|\phi_0|/kT$ is the dimensionless central potential of the
cluster (King 1966).\nocite{1966AJ.....71...64K} By the time the black
holes are ejected, $\mbar \sim 0.4\,\msun$.  Taking $W_0 \sim 10$ we
find $E_{b,{\rm min}}\sim 10000\, kT$.

Portegies Zwart \& McMillan (2000) have tested and refined the above
estimates by performing a series of \nbody\, simulations within the
{\tt Starlab} software environment using the special-purpose computer
GRAPE-4 to speed up the calculations (see \S\,\ref{Sect:GRAPE}).  For
most (seven) of these calculations we used 2048 equal-mass stars with
1\% of them ten times more massive than the average; two calculations
were performed with 4096 stars.  One of the 4096-particle runs
contained 0.5\% black holes; the smaller black-hole fraction did not
result in significantly different behavior.  They also tested
alternative initial configurations, starting some models with the
black holes in primordial binaries with other black holes, or in
primordial binaries with lower-mass stars.

The results of these simulations may be summarized as follows.  Of a
total of 204 black holes, 62 ($\sim 30\%$) were ejected from the model
clusters in the form of black-hole binaries.  A total of 124 ($\sim
61\%$) black holes were ejected single, and one escaping black hole
had a low-mass star as a companion.  The remaining 17 ($\sim 8\%$)
black holes were retained by their parent clusters.  The binding
energies $E_b$ of the ejected black-hole binaries ranged from about
$1000\,kT$ to $10000\,kT$ in a distribution more or less flat in $\log
E_b$, consistent with the assumptions made by Hut et al.\ (1992). The
eccentricities $e$ followed a roughly thermal distribution [$p(e) \sim
2e$], with high eccentricities slightly overrepresented; the mean
eccentricity was $\langle e \rangle = 0.69 \pm 0.10$.  The 17 binaries
with the lowest binding energies ($\log_{10} E_b < 3.5$) had on
average higher eccentricities ($\langle e \rangle = 0.78 \pm 0.05$)
than the more tightly bound binaries ($\langle e \rangle = 0.62 \pm
0.11$).  About half of the black holes were ejected while the parent
cluster still retained more than 90\% of its birth mass, and $\apgt
90$\% of the black holes were ejected before the cluster had lost 30\%
of its initial mass.  These findings are in good agreement with
previous estimates that black-hole binaries are ejected within a few
Gyr, well before core collapse occurs (Kulkarni et al. 1993;
Sigurdsson \& Hernquist 1993).  Recently Merritt et al.\, (2004)
confirmed these findings with their own simulations using {\tt
NBODY6++},\index{N-body!NBODY6++} (Hemsendorf et al 2002; and can
be downloaded from {\tt
ftp://ftp.ari.uni-heidelberg.de/pub/staff/spurzem/nb6mpi/}
\nocite{2002ApJ...581.1256H} a parallel version of {\tt NBODY6}
(Aarseth 1999).\nocite{1999PASP..111.1333A}

Portegies Zwart \& McMillan (2000) have performed additional
calculations incorporating a realistic (Scalo 1986) mass function, the
effects of stellar evolution, and the gravitational influence of the
Galaxy.  These model clusters generally dissolved rather quickly
(within a few hundred Myr) in the Galactic tidal field.  We found that
clusters which dissolved within $\sim 40$\,Myr (before the last
supernova) had no time to eject their black holes; however, those that
survived for longer than this time were generally able to eject at
least one close black-hole binary before dissolution.

Based on these considerations, we conservatively estimate the number
of ejected black-hole binaries to be about $10^{-4} N$ per star
cluster, more or less independent of the cluster lifetime.

\subsubsection{Characteristics of the black-hole binary population}

The energy of an ejected binary and its orbital separation are coupled
to the dynamical characteristics of the star cluster.  For a cluster
in virial equilibrium, we have
\begin{equation}
	kT = \frac{2E_{\rm kin}}{3N}
	   = \frac{-E_{\rm pot}}{3N}
	   = {G M^2 \over 6 N \rvir}\,,
\end{equation}
where $M$ is the total cluster mass and $\rvir$ is the virial radius.
A black-hole binary with semi-major axis $a$ has
\begin{equation}
	E_b =  {G m_{\rm bh}^2 \over 2 a},
\end{equation}
and therefore 
\begin{equation}
	{E_b \over kT} = 3 N \left( {m_{\rm bh} \over M} \right)^2 
                             {\rvir \over a}.
\label{Eq:Ebhbh}\end{equation}

In computing the properties of the black-hole binaries resulting from
cluster evolution, it is convenient to distinguish three broad
categories of dense stellar systems: (1) YoDeCs, (2) globular
clusters, and (3) galactic nuclei.  Table\,\ref{Tab:clusters} lists
characteristic parameters for each category.  The masses and virial
radii of globular clusters are assumed to be distributed as
independent Gaussian with means and dispersions as presented in the
table; this assumption is supported by correlation studies (Djorgovski
\& Meylan 1994)\nocite{1994AJ....108.1292D}.
Table\,\ref{Tab:clusters} also gives estimates of the parameters of
globular clusters at birth (bottom row), based on a recent
parameter-space survey of cluster initial conditions (Takahashi \&
Portegies Zwart 2000; Baumgardt \& Makino
2003);\nocite{2000ApJ...535..759T} \nocite{2003MNRAS.340..227B}
globular clusters which have survived for a Hubble time have lost
$\apgt 60$\% of their initial mass and have expanded by about a factor
of three.  We draw no distinction between core-collapsed globular
clusters (about 20\% of the current population) and non-collapsed
globulars---the present dynamical state of a cluster has little
bearing on how black-hole binaries were formed and ejected during the
first few Gyr of the cluster's life.

The above described process causes globular clusters to be depleted of
black holes. At most one binary containing two black holes can be
present in any globular cluster.  In \S\,\ref{Sect:GWR} we will
further discuss the consequences of the ejected black holes and black
hole binaries, as the latter may become important sources for
gravitational wave detectors.\footnote{Today's globular clusters may
undergo a similar ejection phase, which can be seen in the high
proportion of recycled pulsars and the interestingly possibility of a
high merger rate of white dwarfs (Shara \& Hurley,
2002).\nocite{2002ApJ...571..830S}}

\section{Discussion and further speculations}

Now we have discussed the basic principles of star cluster dynamics
with black holes. In this section we further discuss some of the
consequences and observables to which this theory can be tested.

\subsection{Turning an intermediate mass black hole in an X-ray source}\label{Sect:IMBH_Lx}

An IMBH\index{black hole!intermediate mass} in a stellar cluster with
velocity dispersion $\sigma$ dominates the potential well within its
radius of influence $R_{\rm i}=G\mbh/\sigma^2$; inside $R_{\rm i}$ the
orbits are approximately Keplerian, and the stars are distributed
according to a power law (see Eq.\,\ref{Sec:DynFric} and also (Bahcall
\& Wolf).\nocite{1976ApJ...209..214B} N-body calculations show that
$\alpha=1.5$ (Baumgardt et al 2004),
\nocite{2003IAUS..inpress} and we assume this value in the following.

Stars can reach an orbit with peribothron of order of the tidal radius
by energy diffusion or by angular momentum diffusion. However, stellar
collisions become the dominant dynamical process within the collision
radius $r_\mathrm{coll} \sim (\Mbh/m)r$ (Frank \& Rees 1976),
\nocite{1976MNRAS.176..633F} disrupting stars within that region, and
making energy diffusion within $r_\mathrm{coll}$ implausible. For this
reason we concentrate on tidal capture of stars on very eccentric
orbits. When the star arrives at peribothron a certain amount of
energy $\Delta E_t$ is invested in raising tides, causing the orbit to
circularize. It is hard to know how the star dissipates the tides
after the repetitive encounters with the IMBH. Two extreme models of
``squeezars'' (stars that are ``squeezed'' by the tidal field) are
studied by Alexander \& Morris (2003),\nocite{2003ApJ...590L..25A}
namely ``cold squeezars'', which puff up, or ``hot squeezars'', which
are heated only in their outer layers and radiate their excess energy
effectively.

Hopman et al (2003)\nocite{2004ApJ...604L.101H} calculated
that the probability that an intermediate mass black hole captures a
stellar companion via this process (assuming the ``hot
squeezar'' model) has a reasonable probability, making it likely
that we could in fact observe one or two of such objects.  Once
captured the companion star is likely to fill its Roche-lobe while
still on the main-sequence.

From the moment of first Roche-lobe contact the evolution of the
binary is further determined by mass transfer from a lower mass
(secondary) star to the much more massive (primary) IMBH. This process
is driven by the thermal expansion of the donor and the loss of
angular momentum from the binary system.  Mass transfer then implies
that the donor fills its Roche lobe ($R_{\rm don} = R_{\rm Rl}$) and
continues to do so, i.e., $\Rdot_{\rm don} = \Rdot_{\rm Rl}$. We also
assume that, as long as $L_{\rm disc}<L_{\rm Edd}$ all the mass lost
from the donor via the Roche lobe is accreted by the black hole,
$\Mdot = -\mdot$. If the mass transfer rate exceeds the Eddington
limit, the remaining mass is lost from the binary system with the
angular momentum of the accreting star. 

We estimate the X-ray Luminosity during mass transfer using the simple
model of K\"ording, Falcke \& Markoff
(2002).\nocite{2002A&A...382L..13K} They argue that in the hard state
($\mdot > \mcrit$) $L_x \simeq 0.1\mdot c^2$, where at lower accretion
rates the source becomes transient (see also Kalogera et al,
2004)\nocite{2004ApJ...603L..41K} with an average luminosity of $L_x
\simeq 0.1 c^2\mdot^2/\mcrit$.  For \mcrit\, we use the relation
derived by Dubus et al (1999)\nocite{1999MNRAS.303..139D}

\begin{figure}
\psfig{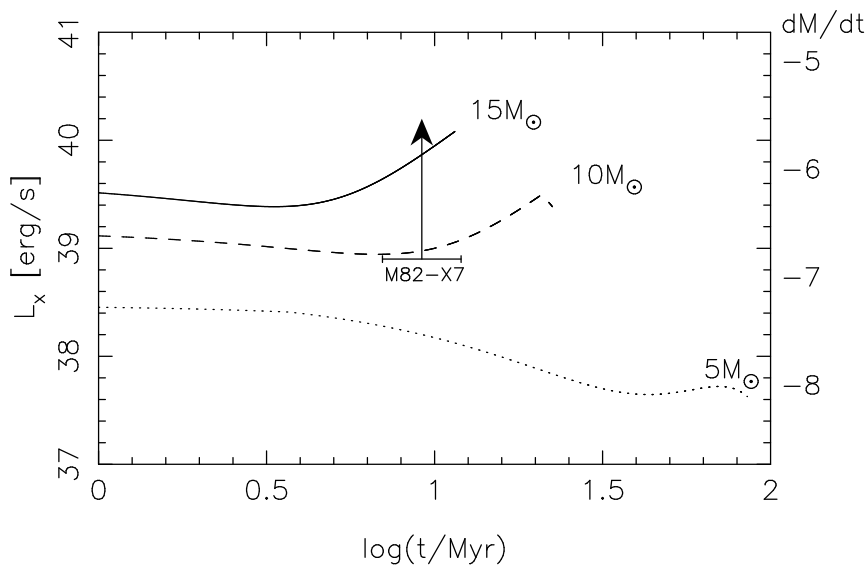} 
\caption[]{ Estimated X-ray luminosity (log erg/s) as a function of time
            for a $1000\,\msun\,$ accreting black hole. The solid,
            dashed and dotted curves are for a $15\,\msun,
            10\,\msun\,$ and a $5\,\msun\,$ donor which started
            Roche-lobe overflow at the zero-age main-sequence. (Figure
            from Hopman, Portegies Zwart \& Alexander, 2004)
            \label{fig:Lx}
}
\end{figure}

\subsection{Speculation on the formation of supermassive black holes}
\label{Sect:massive_black_holes}\index{black hole!super massive}

A million-solar-mass star cluster formed at a distance of $\aplt
30$\,pc from the Galactic center can spiral into the Galactic center
by dynamical friction before being disrupted by the tidal field of the
Galaxy (see Gerhard 2001).\nocite{2001ApJ...546L..39G} Only the
densest star clusters survive to reach the center.  These clusters are
prone to runaway growth and produce massive compact objects at their
centers.  Upon arrival at the Galactic center, the star cluster
dissolves, depositing its central black hole there.  Black holes from
in-spiraling star clusters may subsequently merge to form a
supermassive black hole.  Ebisuzaki et al.~(2001) have proposed that
such a scenario might explain the presence of the central black hole
in the Milky way galaxy.

If we simply assume that bulges and central supermassive black holes
are formed from disrupted star clusters, this model predicts a
relation between black hole and bulge masses in galaxies similar to
the expression (Eq.\,\ref{Eq:Mrunaway_II}) connecting the mass of an
IMBH to that of its parent cluster.  However, the ratio of stellar
mass to black-hole mass might be expected to be smaller for galactic
bulges than for star clusters, because not all star clusters produce a
black hole and not all star clusters survive until the maximum black
hole mass is reached.  We would expect, however, that the general
relation between the black hole mass and that of the bulge remains
valid.

Figure\,\ref{fig:bhmass} shows the relation between the black hole
mass and the bulge mass for several Seyfert
galaxies\index{Seyfert galaxies} and quasars.\index{quasars} The
expression derived in Sec.\,\ref{sect:ug} and the results of these
\nbody\, calculations (Sec.\,\ref{sect:simulations}) are also
indicated.  The solid and dashed lines (Eq.\,\ref{Eq:Mrunaway_II}) fit
the \nbody\, calculations and enclose the area of the measured black
hole mass--bulge masses.  On the way, the solid curve passes though
the two black-hole mass estimates, for M82 star cluster MGG-11.  We
note that the observed relation between bulge and black hole masses
has a spread of two orders of magnitude.  If this bold extrapolation
really does reflect the formation process of bulges and central black
holes, this spread could be interpreted as a variation in the
efficiency of the runaway merger process.  In that case, only about
one in a hundred star clusters reaches the galactic center, where its
black hole is deposited.

The dashed curve is an extrapolation beyond $10^7$\,\msun\, the range
where we think that Eq.\,\ref{Eq:Mrunaway_II} is applicable.  There
appears to be a very interesting relation for galactic nuclei between
the velocity dispersion and the mass of the central black hole
(Ferrarese \& Merritt 2000;\nocite{2000ApJ...539L...9F} Gebhardt et al
2000, 2001);\nocite{2000ApJ...539L..13G} \nocite{2001ApJ...555L..75G}
and Merritt \& Ferrarese (2001).\nocite{2001ApJ...547..140M} I do not
wish to claim that the process described in Eq.\,\ref{Eq:Mrunaway_II}
and the formation of supermassive black holes in galactic nuclei (see
however Ebisuzaki et al 2001),\nocite{2001ApJ...562L..19E} but just
point out how well the same theory explains the relation between
nuclear black hole mass and the velocity dispersion in the nucleus. At
the moment of writing there more than a dozen alternative explanations
for this interesting relation (see for example Haehnelt et al.\,
2000;\nocite{2000MNRAS.318L..35H} Dokushaev et al
2002;\nocite{2001AstL...27..759D} Boroson et al
2002);\nocite{2002AAS...201.9801B} Green et al
(2004);\nocite{2004cbhg.sympE..22G}), and this number is growing
almost daily\footnote{We saw a similar interesting wild-grown of
theories in the gamma-ray burst community a few years back.}.

\subsection{Is the globular cluster M15 a special case?}\index{M15}

The presence of an intermediate mass black hole in the core collapsed
globular cluster M15 has been vividly debated over the last several
decades (Illingworth \& King, 1977a,
1977b)\nocite{1977BAAS....9..343I} \nocite{1977ApJ...218L.109I} and is
still discussed to the present day (van der Marel et al
2002).\nocite{2002AJ....124.3255V} If M15 contains an intermediate
mass black hole, as has recently been suggested by some observers
(Gerrisen et al, 2002\nocite{2002AJ....124.3270G}; later retracted and
subsequently corrected in Gerrisen 2003)\nocite{2003AJ....125..376G}
and argued against by theorists (Baumgardt et al
2003),\nocite{2003ApJ...582L..21B} it is unlikely to have formed via
the scenario discussed in \S\,\ref{Sect:IMBHformation}.  The cluster's
initial relaxation time probably exceeded our upper limit of $\sim
100$\,Myr. The current half-mass relaxation time of M15 is about
2.5\,Gyr (Harris 1996),\nocite{1996yCat.7195....0H} which is far more
than our 100\,Myr limit for forming a massive central object from a
collision runaway (see \S\,\ref{Sect:IMBHformation} for details).

An intermediate mass black hole in the globular cluster M15 may have
been formed by a different scenario.  An alternative is provided by
Miller \& Hamilton (2001),\nocite{2001astro.ph..6188M} who describe
the formation of massive ($\sim 10^3$\,\msun) black holes in star
clusters with relatively long relaxation times.  In their model the
black hole grows very slowly over a Hubble time via occasional
collisions with other black holes, in contrast to the model described here,
in which the runaway grows much more rapidly, reaching a
characteristic mass of about 0.1\% of the total birth mass of the
cluster within a few megayears.

One possible way around M15's long relaxation time may involve the
cluster's rotation.  Gebhardt (2000; 2001; private communication) has
measured radial velocities of individual stars in the crowded central
field, down to two arcsec of the cluster center.  He finds that, both
in the central part of the cluster ($R < 0.1R_{\rm hm}$) and outside
the half mass radius, the average rotation velocity is substantial
($\vrot \apgt 0.5\vdisp$).  Rotation is quickly lost in a cluster
(Baumgardt et al 2003),\nocite{2003ApJ...589L..25B} so to explain a
current rotation, M15's initial rotation rate must probably have been
even larger than observed today (see Einsel \& Spurzem
1999).\nocite{1999MNRAS.302...81E} Hachisu (1978; 1982)
\nocite{1979PASJ...31..523H}\nocite{1982PASJ...34..313H} found, using
gaseous cluster models, that an initially rotating cluster tends to
evolve into a 'gravo-gyro catastrophe'\index{gravo-gyro catastrophe}
which drives the cluster into core collapse far more rapidly than
would occur in a non-rotating system.  If the gravo-gyro-driven core
collapse occurred within 25\,Myr, a collision runaway might have
initiated the growth of an intermediate mass black hole in the core of
M15.

\subsection{The gravitational wave signature of dense star clusters}\label{Sect:GWR}
\index{gravitational waves}

The peak amplitude of the gravitational wave form produced by
black-hole inspiral is (Peters 1964)\nocite{peters64}
\begin{equation}  
	h = 8.0 \times 10^{-20} 
	          \left({M_{\rm chirp} \over \msun}\right)^{5/6}	
	          \left({20 \,{\rm ms} \over P_{\rm orb}} \right)^{1/6}
	          \left({1 {\rm Mpc} \over d}\right)\,.
\end{equation}
Here the ``chirp'' mass is $M_{\rm chirp} = (m_1
m_2)^{3/5}/(m_1+m_2)^{1/5}$ for a binary with component masses $m_1$
and $m_2$.  The frequency of the gravitational wave is $2/P_{\rm
orb}$, where $P_{\rm orb}$ is the orbital period.  The first
LIGO\index{LIGO} interferometer is expected to achieve a sensitivity
$h$ of $10^{-20}$ to $10^{-21}$ at its most sensitive frequency around
100\,Hz (Abramovici et al.\, 1992)\nocite{1992Sci...256..325A},
corresponding to an orbital period of about 20\,ms.  The details of
the wave form and the recovery of the signal from the noisy data
complicates matters somewhat.  With a sensitivity of $h = 5 \times
10^{-21}$, black-hole binaries will be detectable out to a distance of
about 100\,Mpc, and intermediate mass black holes to several kpc
distances.  Within a volume of $\frac{4}{3}\pi d^3$ we then expect a
detection rate for the first generation of interferometers of about
0.3\,$h_0^3$ per year.  For $h_0 \sim 0.72$ (Freedman et
al.\,2001),\nocite{2001ApJ...553...47F} this results in about one
detection event per decade.  LIGO-II is tentatively expected to see
out to an effective distance about ten times farther than LIGO-I and
be operational between 2005 to 2007 (K.~Thorne, private
communication).  This would result in a 1000 times higher detection
rate, or several events per month.

The current best estimate of the maximum distance within which LIGO-I
can detect an inspiral event is
\begin{equation}  
	R_{\rm eff} = 18\,{\rm Mpc}\ 
			\left(\frac{M_{\rm chirp}}{\msun}\right)^{5/6}
\end{equation}
(K.~Thorne, private communication). For neutron star inspiral, $m_1 =
m_2 = 1.4\,\msun$, so $M_{\rm chirp} = 1.22\,\msun$, $R_{\rm eff} =
21$ Mpc.  For black-hole binaries with $m_1 = m_2 = \mbh = 10\,\msun$,
we find $M_{\rm chirp} = 8.71\,\msun$, $R_{\rm eff} = 109$ Mpc.

\subsubsection{The gravitational wave signature for intermediate 
               mass black holes} 
\index{gravitational waves}
\index{black hole!intermediate mass}

In \S\,\ref{Sect:IMBH_Lx} we discussed a model for producing X-rays
from an intermediate mass black hole, in this \S\, we continue that
discussion but then in relation to gravitational wave detectors.

\begin{figure}
\psfig{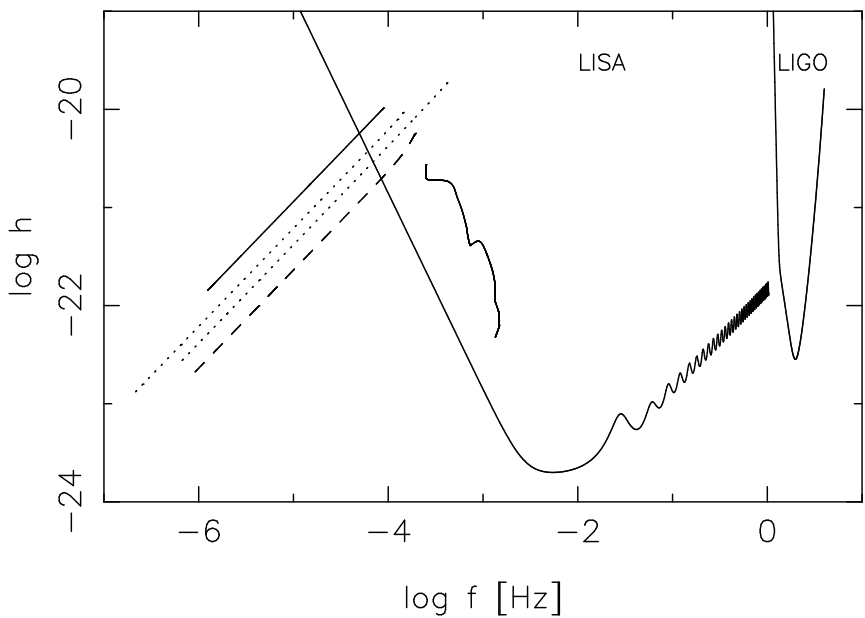} 
\caption{A selected sample of binaries in the gravitational wave
         strain and frequency domain with the LISA and LIGO noise
         curves over-plotted.  We assumed a distance to the source of
         10kpc.  The lower solid curve to the right is for a
         2\,\msun\, donor, the dashes for a 5\,\msun\, donor, the
         dotted curves are for a 10\,\msun\, donor with a 100\,\msun\,
         black hole (lower dotted curve) and 1000\,\msun\, black hole
         (upper dotted curve). The upper solid curve is for a
         15\,\msun\, donor. Except for the lower dotted curve are all
         calculations for a 1000\,\msun\, donor. Notice the enormous
         difference for the 2\,\msun\, donor, which is caused by the
         lack of growth in size of the donor.
         \label{fig:tMdotEdd_GWR}
}
\end{figure}

Figure\,\ref{fig:tMdotEdd_GWR} presents the evolution of the
gravitational wave signal for several systems which start mass
transfer at zero age.  Assuming a standard distance of 10\,Kpc, the
binaries which start mass transfer at birth emit gravitational waves
at a strain of about $\log h \simeq -20.2$ (almost independent of the
donor mass). The frequency of the gravitational waves is barely
detectable for the {\em LISA}\index{LISA} space based antennae (see
Figure\,\ref{fig:tMdotEdd_GWR}). During mass transfer the binary moves
out of the detectable frequency band, as the orbital period increases.
Once the donors has turned into a white dwarf, as happens for the
5\,\msun\, and 10\,\msun\, donors, the emission of gravitational waves
brings the two stars back into the relatively high frequency regime
and it becomes detectable again for the {\em LISA} antennae. This
process, however takes far longer than a Hubble time.

\begin{figure}[htbp!]
\includegraphics[width=0.5\linewidth,angle=-90]{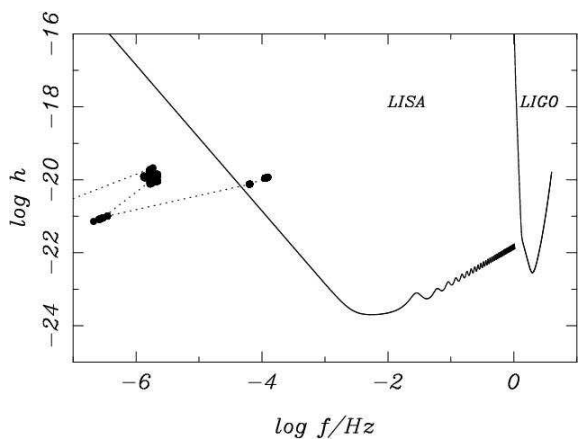}
\caption[]{Evolutionary track in gravitational wave space (frequency
and strain) for an intermediate mass black hole formed in a young and
dense star cluster at a distance of 1kpc. the initial conditions were
to mimic the M82 star cluster MGG-11 (Portegies Zwart et
al.\,2004). The bullets indicate the moments of output one million
years apart. They are connected with a dotted line.  The intermediate
mass black hole acquires a stellar mass black hole in a close binary
near the end of the simulation at about 90\,Myr which puts it in the
LISA band.}
\label{Fig:GWRtrackIMBH}
\end{figure}

The binary in which a 2\,\msun\, main-sequence star starts to transfer
to a 1000\,\msun\, black hole, however, remains visible as a bright
source of gravitational waves for its entire lifetime (see
Figure\,\ref{fig:tMdotEdd_GWR}). The reason for this striking result
is the curious evolution of the donor as it transfers mass at a slow
rate.  The 2\,\msun\, main-sequence donor becomes fully mixed after
the hydrogen fraction exceed about 66 per cent. As a result, the star
remains rather small in size and therefore the orbital period during
mass transfer remains roughly constant. This binary remains visible in
the {\em LIGO} frequency regime for its entire main-sequence lifetime,
only the gravitational wave strain drops as the donor is slowly
consumed by the intermediate mass black hole.  Mass transfer in this
evolution stage is rather slow causing the X-ray source to be
transient. Such a transient could result in an interesting synchronous
detection of X-rays and gravitational waves.

\begin{figure}[htbp!]
\psfig{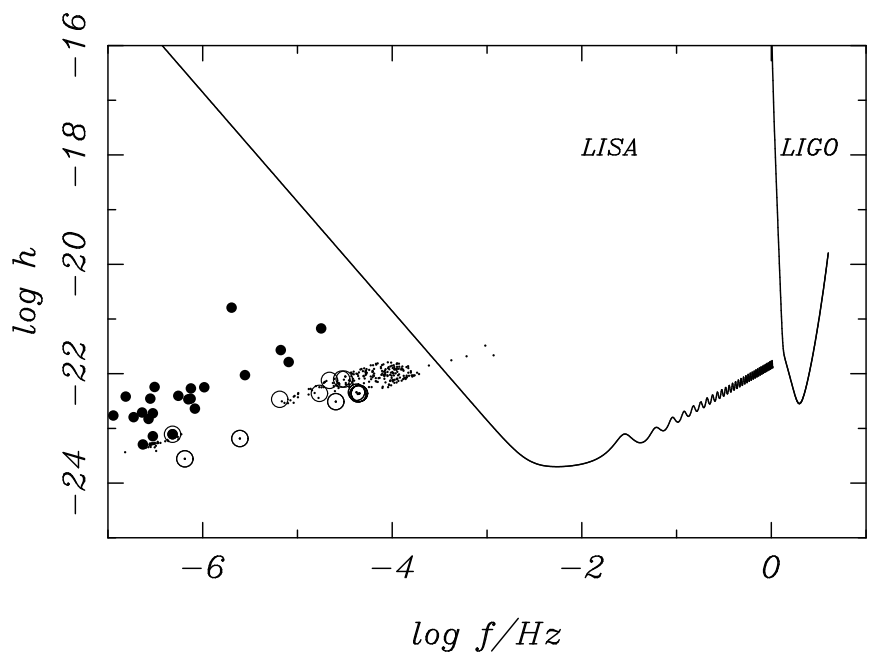}
\caption[]{Population of compact stars in binaries in gravitational
wave space (frequency and strain) in a simulated star cluster at a
distance of 1kpc. The initial conditions were to mimic the M82 star
cluster MGG-11 (Portegies Zwart et al.\,2004). The cluster was
100\,Myr when this image was taken.  The bullets indicate binaries
with one or two black holes, open circles indicate neutron stars and
small dots are for white dwarfs.  Where symbols overplot exactly, as is
quite common for a circle with a small dot in the middle indicate
close binaries with two different compact objects, in these cases they
are (ns, wd) binaries.  Only a few white dwarf binaries may be visible
in the LISA band, black hole and neutron star binaries are generally
too wide to be detectable.  The intermediate mass black hole is not
shown here, but in figure.\,\ref{Fig:GWRtrackIMBH}.  }
\label{Fig:GWRbinaries}
\end{figure}

The star cluster contains many of binaries, some of which may have
orbital periods small enough, and component masses large enough to be
visible by advanced gravitational wave detectors.  In
figure\,\ref{Fig:GWRbinaries} we show the population of binaries in
gravitational-wave space (strain $h$ versus frequency $f$) for the
compact binaries for one of our MGG-11 simulations (see
\S\,\ref{Sect:MGG11}) at an age of 100\,Myr. This simulation model had
initially 20\% of its 128k stars in a binary system. The majority of
systems is not even on this graph, as they comprise of main-sequence
stars, giants or other 'large' stellar object. Only binaries with at
least one compact objects are presented, and of these only the white
dwarf binaries make it in the {\em LISA} band. Many of these systems,
however, will be unobservable in realistic star clusters, as the star
cluster is much further than the here adopted 1\,kpc or the system
ends-up in the confusion limited noise of the white dwarf population
from the Milky-way Galaxy (see e.g.\, Hogan \& Bender 2001;
\nocite{2001PhRvD..64f2002H} Seto 2002; \nocite{2002MNRAS.333..469S}
Nelemans et al, 2004, but see Benaquista et al
2001).\nocite{2004astro.ph.12193N} \nocite{2001CQG...18..4025B}

\subsubsection{The merger rate of stellar mass black holes}\label{Sect:BHmergerrate}

In \S\,\ref{Sect:BHejection} we discussed the evolution of relatively
low density clusters through phase B and C. These clusters tend to
eject their lack holes, in part in the form of binaries consisting of
two black holes. The orbital parameters of these binaries are such
that merger may occur in a relatively short time scale; well withing a
Hubble time. We now will discuss the merger rate per unit volume
${\cal R}$ which we predict from various types of star clusters.

A conservative estimate of the merger rate of black-hole binaries
formed in globular clusters is obtained by assuming that globular
clusters in other galaxies have characteristics similar to those found
in our own.  Using the galaxy density in the local universe (see
Tab.\,\ref{Tab:galaxies}) the result is
\begin{equation}
	{\cal R}_{GC} = 5.4\times 10^{-8} h^3\; 
	                {\rm yr}^{-1}\,{\rm Mpc}^{-3}.
\end{equation}

Irregular galaxies, starburst galaxies, early type spirals and blue
elliptical galaxies all contribute to the formation of young dense
clusters.  In the absence of firm measurements of the numbers of
YoDeCs in other galaxies, we simply use the same values of $S_N$ as
for globular clusters (see from Tab.\,\ref{Tab:galaxies}).  The space
density of such clusters is then
\begin{equation}
	\phi_{\rm YoDeC}  = 3.5\,h^3 \; {\rm Mpc^{-3}},
\end{equation}
and the black hole merger rate is
\begin{equation}
      {\cal R}_{\rm YoDeC} = 2.1 \times 10^{-8} h^3\;
				{\rm yr}^{-1}\,{\rm Mpc}^{-3}.
\end{equation}

For purposes of comparison, we also estimate the contribution to the
total merger rate from galactic nuclei, neglecting obvious
complicating factors such as the presence of central supermassive
black holes (Sage 1994),\nocite{1994Natur.369..345S} which may inhibit
the formation, hardening, and ejection of black-hole binaries.  The
space density of galactic nuclei is only\footnote{Each galaxy may
contain hundreds of globulars but has only one nucleus.}  about
\begin{equation}
	\phi_{GN}  = 0.012\,h^3 \; {\rm Mpc^{-3}},
\end{equation}
and the corresponding contribution to the black hole merger
rate\index{black hole! merger rate} is
\begin{equation}
       {\cal R}_{GN} = 2.5 \times 10^{-9} h^3\; 
	               {\rm yr}^{-1}\,{\rm Mpc}^{-3}, 
\end{equation}
which is negligible compared to the other rates. 

Based on the assumptions outlined above, our estimated total merger
rate per unit volume of black-hole binaries is
\begin{equation}
	{\cal R} = 7.5 \times 10^{-8} h^3\; {\rm yr}^{-1}\,{\rm
	           Mpc}^{-3}.
\label{total-merger-rate}
\end{equation}
However, this may be a considerable underestimate of the true rate.
First, as already mentioned, our assumed number ($\sim10^{-4}N$) of
ejected black-hole binaries is quite conservative.  Second, the
correlation between orbital eccentricity and binding energy and the
excess of high-eccentricity binaries mentioned earlier both favor more
rapid inspiral, causing a larger fraction of the black-hole binaries
to merge.  Third, the observed population of globular clusters
naturally represents only those clusters that have survived until the
present day.  The study by Takahashi \& Portegies Zwart (2000) and
Baumgardt \& Makino (2003) indicates that $\sim$ 50\% of globular
clusters dissolve in the tidal field of the parent galaxy within a few
billion years of formation.  We have therefore underestimated the
total number of globular clusters, and hence the black-hole merger
rate, by about a factor of two.  Fourth, a very substantial
underestimate stems from the assumption that the masses and radii of
present-day globular clusters are representative of the initial
population.  When estimated initial parameters
(Table\,\ref{Tab:clusters}, bottom row) are used, the total merger
rate increases by a further factor of six.  Taking all these effects
into account, we obtain a net black-hole merger rate of
\begin{equation}
	{\cal R} \sim 6 \times 10^{-7} h^3\; 
	           {\rm yr}^{-1}\,{\rm Mpc}^{-3}.
\label{best_guess}
\end{equation}
We note that this rate is significantly larger than the current best
estimates of the neutron-star merger rate (e.g.\, Phinney 1991;
Tutukov \& Yungelson 1994; Burgay et al. 2003; and many
others). \nocite{1991ApJ...380L..17P} \nocite{1994MNRAS.268..871T}
\nocite{2003Natur.426..531B} Since black hole mergers are also
``visible'' to much greater distances, we expect that black hole
events will dominate the LIGO detection rate.

\subsubsection{Gravitational inspiral}

An approximate formula for the merger time of two stars due to the
emission of gravitational waves is given by Peters (1964):
\nocite{peters64}
\begin{equation}
 t_{\rm mrg} \approx 150\, {\rm Myr}\; 
	\left( {\msun \over \mbh} \right)^{3}
	\left( {a \over \rsun} \right)^4 (1-e^2)^{7/2} \;.
\label{Eq:tmrg}\end{equation}
The sixth column of Table\,\ref{Tab:clusters} lists the fraction of
black-hole binaries which merge within a Hubble time due to
gravitational radiation, assuming that the binary binding energies are
distributed flat in $\log E_b$ between $1000\,kT$ and $10000\,kT$,
that the eccentricities are thermal, independent of $E_b$, and that
the universe is $\sim 13$\,Gyr old ($H_0=72\pm8$
km\,s$^{-1}$\,Mpc$^{-1}$; Freedman et al.\,
2001).\nocite{2001ApJ...553...47F} The final column of the table lists
the contribution to the total black-hole merger rate from each cluster
category.

For black-hole binaries with $m_1 = m_2 \simeq 10\,\msun$ we expect a
LIGO-I detection rate of about 3\,$h^3$ per year.  For $h \sim 0.72$
(Freedman et al.\,2001),\nocite{2001ApJ...553...47F} this results in
about one detection event annually.  LIGO-II should become operational
by 2007, and is expected to have $R_{\rm eff}$ about ten times greater
than LIGO-I, resulting in a detection rate 1000 times higher, or 2--3
events per day.

Black-hole binaries ejected from galactic nuclei, the most massive
globular clusters (masses $\apgt 5 \times 10^6\,\msun$), and globular
clusters which experience core collapse soon after formation tend to
be very tightly bound, and merge within a few million years of
ejection.  These mergers therefore trace the formation of dense
stellar systems with a delay of a few Gyr (the typical time required
to form and eject binaries), making these systems unlikely candidates
for LIGO detections, as the majority merged long ago.  This effect
reduces the current merger rate somewhat, but more sensitive future
gravitational wave detectors may be able to see some of these early
universe events.  In fact, we estimate that the most massive globular
clusters contribute about 50\% of the total black hole merger rate.
However, while their black-hole binaries merge promptly upon ejection,
the longer relaxation times of these clusters mean that binaries tend
to be ejected much later than in lower mass systems.  Consequently, we
have retained these binaries in our final merger rate estimate
(Eq. \ref{best_guess}).

Finally, we have assumed that the mass of a stellar black hole is
10\,\msun.  Increasing this mass to 18\,\msun\ decreases the
calculated merger rate by about 50\%---higher mass black holes tend to
have wider orbits.  However, the larger chirp mass increases the
signal to noise, and the distance to which such a merger can be
observed increases by about 60\%.  The detection rate on Earth
therefore increases by about a factor of three.  For 6\,\msun\ black
holes, the detection rate decreases by a similar factor.  For
black-hole binaries with component masses $\apgt 12$\,\msun\, the
first generation of detectors will be more sensitive to the merger
itself than to the spiral-in phase that precedes it (Flanagan \&
Hughes 1998a, 1998b).\nocite{1998PhRvD..57.4566F}
\nocite{1998PhRvD..57.4535F} Since the strongest signal is expected
from black-hole binaries with high-mass components, it is critically
important to improve our understanding of the merger waveform.  Even
for lower-mass black holes (with $m_{bh}\apgt 10\,\msun$), the
inspiral signal comes from an epoch when the black holes are so close
together that the post-Newtonian expansions used to calculate the wave
forms are unreliable.  The wave forms of this ``intermediate binary
black hole regime'' (Brady et al. 1998)\nocite{1998PhRvD..57.2101B}
are only now beginning to be explored.

\section{Concluding remarks}\label{sect:discussion}

We approach the end of the chapter on the ecology of black holes in
dense star clusters, as part of the Como advances school in
astrophysics. Here we will summarize a few of the key-concepts
discussed in this chapter.

Star clusters go through three (or four) evolutionary phases called A,
B and C, each of which is dominated by either stellar mass loss or
relaxation (see \S\,\ref{Sect:Theory}). In this discussion we assumed
that external influences are not particularly competitive in effect,
but if they are, we call it phase D. Phase D generally results in an
early termination of the cluster.

Black holes tend to behave differently in each of these phases.  In
phase A, it is possible to built-up a {\large O}(1000)\,\msun\, star
which ultimately collapses to a black hole of intermediate mass (see
\S\,\ref{Sect:IMBHformation}).

This black hole may subsequently capture a stellar companion by tidal
effects and turn into a bright X-ray source (see
\S\,\ref{Sect:IMBH_Lx}).

In clusters where a runaway collision is prevented by stellar
evolution, the most massive stars collapse to stellar mass black holes
in usual supernovae (see \S\,\ref{phase B}).

The stellar mass black holes sink to the cluster center by dynamical
friction, pair off in binaries and eject each other from the stellar
system. This scenario elegantly explains the absence of black hole
X-ray transients in globular clusters (see \S\,\ref{Sect:BHejection}).

The ejected black hole binaries spiral-in due to gravitational wave
radiation and ultimately coalesce. Upon coalescence they produce a
strong burst of gravitational waves which can be detected to a
distance of $\sim 100$\,Mpc (see \S\,\ref{Sect:BHmergerrate}).

\bigskip\bigskip
\noindent{\bf Acknowledgments}

First of all I am grateful to the organizers (Monica Colpi, Francesco
Haardt, Vittorio Gorini, Ugo Moschella and Aldo Treves) for the
wonderful week in Como.  It is a great pleasure to name and thank my
collaborators who helped developing the work on which this chapter is
made, they are:
Tal Alexander,
Ortwin Gerhard,
Alessia Gualandris,
Mark Hemsendorf,
Clovis Hopman,
Nate McCrady,
David Merritt,
Slawomir Piatek,
Piero Spinnato,
Rainer Spurzem.
Special thanks goes to Jun Makino and the University of Tokyo for
allowing me to use his great GRAPE-6 hardware, to Piet Hut and
Steve McMillan for their valuable help and support.
In addition I would like to acknowledge the hospitality of American
Museum for Natural History, das Astronomisches Rechen Institute,
Drexel University, the Institute for Advanced Study, the University of
Basel and of course to Tokyo University for the use of their GRAPE-6
facilities
This work was supported by NASA ATP grants NAG5-6964 and NAG5-9264, by
the Royal Netherlands Academy of Sciences (KNAW), the Dutch
organization of Science (NWO), the Netherlands Research School for
Astronomy (NOVA).

\input ./book.ind

\begin{thebibliography}{}

\bibitem{Aarseth2003}
{Aarseth}, S.~A. 2003,
\newblock {Gravitational N-body simulations},
\newblock Cambdridge University press, 2003

\bibitem{1963MNRAS.126..223A}
{Aarseth}, S.~J. 1963, \mnras, 126, 223

\bibitem{1999PASP..111.1333A}
{Aarseth}, S.~J. 1999, \pasp, 111, 1333

\bibitem{1964ApNr....9..313A}
{Aarseth}, S.~J., {Hoyle}, F. 1964, Astrophysica Norvegica, 9, 313

\bibitem{1975ARA&A..13....1A}
{Aarseth}, S.~J., {Lecar}, M. 1975, \araa, 13, 1

\bibitem{1992Sci...256..325A}
{Abramovici}, A., {Althouse}, W.~E., {Drever}, R.~W.~P., {Gursel}, Y.,
  {Kawamura}, S., {Raab}, F.~J., {Shoemaker}, D., {Sievers}, L., {Spero},
  R.~E., {Thorne}, K.~S. 1992, Science, 256, 325

\bibitem{2003ApJ...590L..25A}
{Alexander}, T., {Morris}, M. 2003, \apjl, 590, L25

\bibitem{1962spss.book.....A}
{Antonov}, V.~A. 1962,
\newblock {Solution of the problem of stability of stellar system Emden's
  density law and the spherical distribution of velocities},
\newblock Vestnik Leningradskogo Universiteta, Leningrad: University, 1962

\bibitem{1985JSSC...6...85A}
{Apple}, A.~W. 1985, SIAM, J. Sci. Stat. Comp, 6, 85

\bibitem{1976ApJ...209..214B}
{Bahcall}, J.~N., {Wolf}, R.~A. 1976, \apj, 209, 214

\bibitem{1986Natur.324..446B}
{Barnes}, J., {Hut}, P. 1986, \nat, 324, 446

\bibitem{2002MNRAS.336.1069B}
{Baumgardt}, H., {Hut}, P., {Heggie}, D.~C. 2002, \mnras, 336, 1069

\bibitem{2003ApJ...582L..21B}
{Baumgardt}, H., {Hut}, P., {Makino}, J., {McMillan}, S., {Portegies Zwart}, S.
  2003a, \apjl, 582, L21

\bibitem{2003MNRAS.340..227B}
{Baumgardt}, H., {Makino}, J. 2003, \mnras, 340, 227

\bibitem{2003ApJ...589L..25B}
{Baumgardt}, H., {Makino}, J., {Hut}, P., {McMillan}, S., {Portegies Zwart}, S.
  2003b, \apjl, 589, L25

\bibitem{2003IAUS..inpress}
Baumgardt, H., Makino, J., {Portegies Zwart} 2003,
\newblock in Scientific Highlights of the IAU XXVth General Assembly (Eds. D.
  Richstone, P. Hut)

\bibitem{2001CQG...18..4025B}
{Benacquista}, M.~J., {Portegies Zwart}, S.~F., Rasio, F. 2001, {Class. Quantum
  Grav.}, 18, 4025

\bibitem{1987gady.book.....B}
{Binney}, J., {Tremaine}, S. 1987,
\newblock Galactic dynamics,
\newblock Princeton, NJ, Princeton University Press, 1987, 747 p.

\bibitem{1961BAN....15..265B}
{Blaauw}, A. 1961, BAN, 15, 265

\bibitem{2002AAS...201.9801B}
{Boroson}, T.~A. 2002, Bulletin of the American Astronomical Society, 34, 1265

\bibitem{1997A&A...323..139B}
{Bouvier}, J., {Rigaut}, F., {Nadeau}, D. 1997, \aap, 323, 139

\bibitem{1998PhRvD..57.2101B} {Brady}, P.~R., {Creighton}, T.,
{Cutler}, C., {Schutz}, B.~F. 1998, Phys. Rev. D 57, 2101

\bibitem{2003Natur.426..531B}
{Burgay}, M., {D'Amico}, N., {Possenti}, A., {Manchester}, R.~N., {Lyne},
  A.~G., {Joshi}, B.~C., {McLaughlin}, M.~A., {Kramer}, M., {Sarkissian},
  J.~M., {Camilo}, F., {Kalogera}, V., {Kim}, C., {Lorimer}, D.~R. 2003, \nat,
  426, 531

\bibitem{1990ApJ...351..121C}
{Chernoff}, D.~F., {Weinberg}, M.~D. 1990, \apj, 351, 121

\bibitem{1980ApJ...242..765C}
{Cohn}, H. 1980, \apj, 242, 765

\bibitem{2001astro.ph..6188M}
{Coleman Miller}, M., {Hamilton}, D.~P. 2001,
\newblock in 9 pages, submitted to MNRAS.,  6188

\bibitem{dbh91}
Davies, M., Benz, W., Hills, J. 1991, \apj, 381, 449

\bibitem{1993MNRAS.265..250D}
{Dehnen}, W. 1993, \mnras, 265, 250

\bibitem{2000msc..conf..204D}
{Deiters}, S., {Spurzem}, R. 2000,
\newblock in ASP Conf. Ser. 211: Massive Stellar Clusters, p.~204

\bibitem{2001A&AT...20...47D}
{Deiters}, S., {Spurzem}, R. 2001, Astronomical and Astrophysical Transactions,
  20, 47

\bibitem{1994AJ....108.1292D}
{Djorgovski}, S., {Meylan}, G. 1994, \aj, 108, 1292

\bibitem{2001AstL...27..759D}
{Dokuchaev}, V.~I., {Eroshenko}, Y.~N. 2001, Astronomy Letters, 27, 759

\bibitem{1996MNRAS.280..498D}
{Drukier}, G.~A. 1996, \mnras, 280, 498

\bibitem{1999MNRAS.303..139D}
{Dubus}, G., {Lasota}, J., {Hameury}, J., {Charles}, P. 1999, \mnras, 303, 139

\bibitem{1995ApJ...445..716D}
{Dwek}, E., {Arendt}, R.~G., {Hauser}, M.~G., {Kelsall}, T., {Lisse}, C.~M.,
  {Moseley}, S.~H., {Silverberg}, R.~F., {Sodroski}, T.~J., {Weiland}, J.~L.
  1995, \apj, 445, 716

\bibitem{2001ApJ...562L..19E}
{Ebisuzaki}, T., {Makino}, J., {Tsuru}, T.~G., {Funato}, Y., {Portegies Zwart},
  S., {Hut}, P., {McMillan}, S., {Matsushita}, S., {Matsumoto}, H., {Kawabe},
  R. 2001, \apjl, 562, L19

\bibitem{1997MNRAS.284..576E}
{Eckart}, A., {Genzel}, R. 1997, \mnras, 284, 576

\bibitem{1989ApJ...347..998E}
{Eggleton}, P.~P., {Fitchett}, M.~J., {Tout}, C.~A. 1989, \apj, 347, 998

\bibitem{1999MNRAS.302...81E}
{Einsel}, C., {Spurzem}, R. 1999, \mnras, 302, 81

\bibitem{1998A&A...331L..29E}
{Ergma}, E., {van den Heuvel}, E.~P.~J. 1998, \aap, 331, L29

\bibitem{2000NewA....5..305F}
{Fellhauer}, M., {Kroupa}, P., {Baumgardt}, H., {Bien}, R., {Boily}, C.~M.,
  {Spurzem}, R., {Wassmer}, N. 2000, New Astronomy, 5, 305

\bibitem{2000ApJ...539L...9F}
{Ferrarese}, L., {Merritt}, D. 2000, \apjl, 539, L9

\bibitem{1999ApJ...514..202F}
{Figer}, D.~F., {McLean}, I.~S., {Morris}, M. 1999, \apj, 514, 202

\bibitem{1998PhRvD..57.4535F}
{Flanagan}, {\' E}.~{\' E}., {Hughes}, S.~A. 1998a, Phys. Rev. D, 57, 4535

\bibitem{1998PhRvD..57.4566F}
{Flanagan}, {\' E}.~{\' E}., {Hughes}, S.~A. 1998b, Phys. Rev. D, 57, 4566

\bibitem{1976MNRAS.176..633F}
{Frank}, J., {Rees}, M.~J. 1976, \mnras, 176, 633

\bibitem{2001ApJ...553...47F}
{Freedman}, W.~L., {Madore}, B.~F., {Gibson}, B.~K., {Ferrarese}, L., {Kelson},
  D.~D., {Sakai}, S., {Mould}, J.~R., {Kennicutt}, R.~C., {Ford}, H.~C.,
  {Graham}, J.~A., {Huchra}, J.~P., {Hughes}, S.~M.~G., {Illingworth}, G.~D.,
  {Macri}, L.~M., {Stetson}, P.~B. 2001, \apj, 553, 47

\bibitem{2004astro.ph..1004F}
{Fregeau}, J.~M., {Cheung}, P., {Zwart}, S.~F.~P., {Rasio}, F.~A. 2004, ArXiv
  Astrophysics e-prints

\bibitem{2003ApJ...593..772F}
{Fregeau}, J.~M., {G{\" u}rkan}, M.~A., {Joshi}, K.~J., {Rasio}, F.~A. 2003,
  \apj, 593, 772

\bibitem{2002ApJ...570..171F}
{Fregeau}, J.~M., {Joshi}, K.~J., {Portegies Zwart}, S.~F., {Rasio}, F.~A.
  2002, \apj, 570, 171

\bibitem{2001ApJ...554..548F}
{Fryer}, C.~L., {Kalogera}, V. 2001, \apj, 554, 548

\bibitem{1995MNRAS.276..206F}
{Fukushige}, T., {Heggie}, D.~C. 1995, \mnras, 276, 206

\bibitem{1991PASJ...43..841F}
{Fukushige}, T., {Ito}, T., {Makino}, J., {Ebisuzaki}, T., {Sugimoto}, D.,
  {Umemura}, M. 1991, \pasj, 43, 841

\bibitem{2004ApJ...604..632G}
{G{\" u}rkan}, M.~A., {Freitag}, M., {Rasio}, F.~A. 2004, \apj, 604, 632

\bibitem{2004A&A...416..917G}
{Galleti}, S., {Federici}, L., {Bellazzini}, M., {Fusi Pecci}, F., {Macrina},
  S. 2004, \aap, 416, 917

\bibitem{2000ApJ...539L..13G}
{Gebhardt}, K., {Bender}, R., {Bower}, G., {Dressler}, A., {Faber}, S.~M.,
  {Filippenko}, A.~V., {Green}, R., {Grillmair}, C., {Ho}, L.~C., {Kormendy},
  J., {Lauer}, T.~R., {Magorrian}, J., {Pinkney}, J., {Richstone}, D.,
  {Tremaine}, S. 2000, \apjl, 539, L13

\bibitem{2001ApJ...555L..75G}
{Gebhardt}, K., {Bender}, R., {Bower}, G., {Dressler}, A., {Faber}, S.~M.,
  {Filippenko}, A.~V., {Green}, R., {Grillmair}, C., {Ho}, L.~C., {Kormendy},
  J., {Lauer}, T.~R., {Magorrian}, J., {Pinkney}, J., {Richstone}, D.,
  {Tremaine}, S. 2001, \apjl, 555, L75

\bibitem{2001ApJ...546L..39G}
{Gerhard}, O. 2001, \apjl, 546, L39

\bibitem{2002AJ....124.3270G}
{Gerssen}, J., {van der Marel}, R.~P., {Gebhardt}, K., {Guhathakurta}, P.,
  {Peterson}, R.~C., {Pryor}, C. 2002, \aj, 124, 3270

\bibitem{2003AJ....125..376G}
{Gerssen}, J., {van der Marel}, R.~P., {Gebhardt}, K., {Guhathakurta}, P.,
  {Peterson}, R.~C., {Pryor}, C. 2003, \aj, 125, 376

\bibitem{1998ApJ...509..678G}
{Ghez}, A.~M., {Klein}, B.~L., {Morris}, M., {Becklin}, E.~E. 1998, \apj, 509,
  678

\bibitem{1987ApJ...313..576G}
{Goodman}, J. 1987, \apj, 313, 576

\bibitem{2004cbhg.sympE..22G}
{Green}, R.~F., {Nelson}, C.~H., {Boroson}, T. 2004,
\newblock in Coevolution of Black Holes and Galaxies, from the Carnegie
  Observatories Centennial Symposia. Carnegie Observatories Astrophysics
  Series. Edited by L. C. Ho, 2004. Pasadena: Carnegie Observatories,
  http://www.ociw.edu/ociw/symposia/series/symposium1/proceedings.html

\bibitem{2004astro.ph..1451G}
{Gualandris}, A., {Zwart}, S.~P., {Eggleton}, P.~P. 2004, ArXiv Astrophysics
  e-prints

\bibitem{1996AJ....111..267G}
{Guhathakurta}, P., {Yanny}, B., {Schneider}, D.~P., {Bahcall}, J.~N. 1996,
  \aj, 111, 267

\bibitem{1979PASJ...31..523H}
{Hachisu}, I. 1979, \pasj, 31, 523

\bibitem{1982PASJ...34..313H}
{Hachisu}, I. 1982, \pasj, 34, 313

\bibitem{2000MNRAS.318L..35H}
{Haehnelt}, M.~G., {Kauffmann}, G. 2000, \mnras, 318, L35

\bibitem{1996AJ....112.1487H}
{Harris}, W.~E. 1996a, \aj, 112, 1487

\bibitem{1996yCat.7195....0H}
{Harris}, W.~E. 1996b, VizieR Online Data Catalog, 7195, 0

\bibitem{2003ApJ...591..288H}
{Heger}, A., {Fryer}, C.~L., {Woosley}, S.~E., {Langer}, N., {Hartmann}, D.~H.
  2003, \apj, 591, 288

\bibitem{2003gmbp.book.....H}
{Heggie}, D., {Hut}, P. 2003,
\newblock {The Gravitational Million-Body Problem: A Multidisciplinary Approach
  to Star Cluster Dynamics},
\newblock The Gravitational Million-Body Problem: A Multidisciplinary Approach
  to Star Cluster Dynamics, by Douglas Heggie and Piet Hut.~ Cambridge
  University Press, 2003, 372 pp.

\bibitem{1975MNRAS.173..729H}
{Heggie}, D.~C. 1975, \mnras, 173, 729

\bibitem{1992Natur.359..772H}
{Heggie}, D.~C. 1992, \nat, 359, 772

\bibitem{1998HiA....11..591H}
{Heggie}, D.~C., {Giersz}, M., {Spurzem}, R., {Takahashi}, K. 1998, Highlights
  in Astronomy, 11, 591

\bibitem{HM1986}
{Heggie}, D.~C., {Mathieu}, R. 1986, \mnras, in P. Hut, S. McMillan (eds.),
  Lecture Not. Phys 267, Springer-Verlag, Berlin

\bibitem{1995MNRAS.272..317H}
{Heggie}, D.~C., {Ramamani}, N. 1995, \mnras, 272, 317

\bibitem{2002ApJ...581.1256H}
{Hemsendorf}, M., {Sigurdsson}, S., {Spurzem}, R. 2002, \apj, 581, 1256

\bibitem{1997MNRAS.285..613H}
{Heyl}, J., {Colless}, M., {Ellis}, R.~S., {Broadhurst}, T. 1997, \mnras, 285,
  613

\bibitem{2001PhRvD..64f2002H}
{Hogan}, C.~J., {Bender}, P.~L. 2001, Phys. Rev. D, 64, 062002

\bibitem{2004ApJ...604L.101H}
{Hopman}, C., {Portegies Zwart}, S.~F., {Alexander}, T. 2004, \apjl, 604, L101

\bibitem{2000astro.ph..1295H}
{Hurley}, J.~R., {Pols}, O.~R., {Tout}, C.~A. 2000,
\newblock in 29 pages, 20 figures, submitted for publication in MNRAS.,  1295

\bibitem{1985ApJ...298..502H}
{Hut}, P., {Inagaki}, S. 1985, \apj, 298, 502

\bibitem{1992ApJ...389..527H}
{Hut}, P., {McMillan}, S., {Romani}, R.~W. 1992, \apj, 389, 527

\bibitem{1977ApJ...218L.109I}
{Illingworth}, G., {King}, I.~R. 1977a, \apjl, 218, L109

\bibitem{1977BAAS....9..343I}
{Illingworth}, G.~D., {King}, I.~R. 1977b, \baas, 9, 343

\bibitem{1991PASJ...43..547I}
{Ito}, T., {Ebisuzaki}, T., {Makino}, J., {Sugimoto}, D. 1991, \pasj, 43, 547

\bibitem{2001ApJ...550..691J}
{Joshi}, K.~J., {Nave}, C.~P., {Rasio}, F.~A. 2001, \apj, 550, 691

\bibitem{2002A&A...382L..13K}
{K{\" o}rding}, E., {Falcke}, H., {Markoff}, S. 2002, \aap, 382, L13

\bibitem{2004ApJ...603L..41K}
{Kalogera}, V., {Henninger}, M., {Ivanova}, N., {King}, A.~R. 2004, \apjl, 603,
  L41

\bibitem{2000PASJ...52..659K}
{Kawai}, A., {Fukushige}, T., {Makino}, J., {Taiji}, M. 2000, \pasj, 52, 659

\bibitem{2004ApJS..151...13K}
{Kawai}, A., {Makino}, J., {Ebisuzaki}, T. 2004, \apjs, 151, 13

\bibitem{1966AJ.....71...64K}
{King}, I.~R. 1966, \aj, 71, 64

\bibitem{2001MNRAS.322..231K}
{Kroupa}, P. 2001, \mnras, 322, 231

\bibitem{1990MNRAS.244...76K}
{Kroupa}, P., {Tout}, C.~A., {Gilmore}, G. 1990, \mnras, 244, 76

\bibitem{2003ApJ...598.1076K}
{Kroupa}, P., {Weidner}, C. 2003, \apj, 598, 1076

\bibitem{1993Natur.364..421K}
{Kulkarni}, S.~R., {Hut}, P., {McMillan}, S. 1993, \nat, 364, 421

\bibitem{1992A&A...264..105M}
{Maeder}, A. 1992, \aap, 264, 105

\bibitem{1991ApJ...369..200M}
{Makino}, J. 1991, \apj, 369, 200

\bibitem{1996ApJ...471..796M}
{Makino}, J. 1996, \apj, 471, 796

\bibitem{1992PASJ...44..141M}
{Makino}, J., {Aarseth}, S.~J. 1992, \pasj, 44, 141

\bibitem{2003PASJ...55.1163M}
{Makino}, J., {Fukushige}, T., {Koga}, M., {Namura}, K. 2003, \pasj, 55, 1163

\bibitem{1990ApJ...365..208M}
{Makino}, J., {Hut}, P. 1990, \apj, 365, 208

\bibitem{1998sssp.book.....M}
{Makino}, J., {Taiji}, M. 1998,
\newblock {Scientific simulations with special-purpose computers : The GRAPE
  systems},
\newblock Scientific simulations with special-purpose computers : The GRAPE
  systems /by Junichiro Makino \& Makoto Taiji.~Chichester ; Toronto : John
  Wiley \& Sons, c1998.

\bibitem{1997ApJ...480..432M}
{Makino}, J., {Taiji}, M., {Ebisuzaki}, T., {Sugimoto}, D. 1997, \apj, 480, 432

\bibitem{1999ApJ...512L...9M}
{McLaughlin}, D.~E. 1999, \apjl, 512, L9

\bibitem{1986ApJ...306..552M}
{McMillan}, S. L.~W. 1986a, \apj, 306, 552

\bibitem{1986ApJ...307..126M}
{McMillan}, S. L.~W. 1986b, \apj, 307, 126

\bibitem{2003ApJ...596..314M}
{McMillan}, S.~L.~W., {Portegies Zwart}, S.~F. 2003, \apj, 596, 314

\bibitem{2001ApJ...547..140M}
{Merritt}, D., {Ferrarese}, L. 2001, \apj, 547, 140

\bibitem{2004astro.ph..3331M}
{Merritt}, D., {Piatek}, S., {Zwart}, S.~P., {Hemsendorf}, M. 2004, ArXiv
  Astrophysics e-prints

\bibitem{1994A&AS..103...97M}
{Meynet}, G., {Maeder}, A., {Schaller}, G., {Schaerer}, D., {Charbonnel}, C.
  1994, \aaps, 103, 97

\bibitem{1996A&ARv...7..289M}
{Mezger}, P.~G., {Duschl}, W.~J., {Zylka}, R. 1996, A\&A Rev, 7, 289

\bibitem{1977BAAS....9..566M}
{Miller}, G.~E., {Scalo}, J.~M. 1977, \baas, 9, 566

\bibitem{1979ApJS...41..513M}
{Miller}, G.~E., {Scalo}, J.~M. 1979, \apjs, 41, 513

\bibitem{2001bhbg.conf..312N}
{Nelemans}, G., {Tauris}, T.~M., {van den Heuvel}, E.~P.~J. 2001,
\newblock in Black Holes in Binaries and Galactic Nuclei, p.~312

\bibitem{2004astro.ph.12193N}
{Nelemans}, G., {Yungelson}, L.~R., {Portegies Zwart}, S.~F. 2004, ArXiv
  Astrophysics e-prints (astro-ph/0312193)

\bibitem{1993PASJ...45..329O}
{Okumura}, S.~K., {Makino}, J., {Ebisuzaki}, T., {Fukushige}, T., {Ito}, T.,
  {Sugimoto}, D., {Hashimoto}, E., {Tomida}, K., {Miyakawa}, N. 1993, \pasj,
  45, 329

\bibitem{1994ApJS...94..117O}
{Olson}, K.~M., {Dorband}, J.~E. 1994, \apjs, 94, 117

\bibitem{peters64}
Peters, P.~C. 1964, Phys. Rev., 136, 1224

\bibitem{1991ApJ...380L..17P}
{Phinney}, E.~S. 1991, \apjl, 380, L17

\bibitem{1998MNRAS.299..955P}
{Pinfield}, D.~J., {Jameson}, R.~F., {Hodgkin}, S.~T. 1998, \mnras, 299, 955

\bibitem{2004PZetalNature}
{Portegies Zwart}, S.~F., {Baumgardt}, H., {Hut}, P., {Makino}, J., {McMillan},
  S. 2004, {Nature}, 428, 724

\bibitem{1998A&A...337..363P}
{Portegies Zwart}, S.~F., {Hut}, P., {Makino}, J., {McMillan}, S.~L.~W. 1998,
  \aap, 337, 363

\bibitem{1999A&A...348..117P}
{Portegies Zwart}, S.~F., {Makino}, J., {McMillan}, S.~L.~W., {Hut}, P. 1999,
  \aap, 348, 117

\bibitem{2001ApJ...546L.101P}
{Portegies Zwart}, S.~F., {Makino}, J., {McMillan}, S.~L.~W., {Hut}, P. 2001,
  \apjl, 546, L101

\bibitem{2000ApJ...528L..17P}
{Portegies Zwart}, S.~F., {McMillan}, S.~L.~W. 2000, \apjl, 528, L17

\bibitem{1996A&A...309..179P}
{Portegies Zwart}, S.~F., {Verbunt}, F. 1996, \aap, 309, 179

\bibitem{1997A&A...321..207P}
{Portegies Zwart}, S.~F., {Verbunt}, F., {Ergma}, E. 1997, \aap, 321, 207

\bibitem{1998A&A...329..101R}
{Raboud}, D., {Mermilliod}, J.~C. 1998, \aap, 329, 101

\bibitem{1997PASP..109..933R}
{Rubenstein}, E.~P. 1997, \pasp, 109, 933

\bibitem{1997ApJ...474..701R}
{Rubenstein}, E.~P., {Bailyn}, C.~D. 1997, \apj, 474, 701

\bibitem{1999ApJ...513L..33R}
{Rubenstein}, E.~P., {Bailyn}, C.~D. 1999, \apjl, 513, L33

\bibitem{1994Natur.369..345S}
{Sage}, L. 1994, \nat, 369, 345

\bibitem{1955ApJ...121..161S}
{Salpeter}, E.~E. 1955, \apj, 121, 161

\bibitem{1972AJ.....77..292S}
{Sanders}, R.~H., {Lowinger}, T. 1972, \aj, 77, 292

\bibitem{scalo86}
Scalo, J.~M. 1986, Fund. of Cosm. Phys., 11, 1

\bibitem{2002Natur.419..694S}
{Sch{\" o}del}, R., {Ott}, T., {Genzel}, R., {Hofmann}, R., {Lehnert}, M.,
  {Eckart}, A., {Mouawad}, N., {Alexander}, T., {Reid}, M.~J., {Lenzen}, R.,
  {Hartung}, M., {Lacombe}, F., {Rouan}, D., {Gendron}, E., {Rousset}, G.,
  {Lagrange}, A.-M., {Brandner}, W., {Ageorges}, N., {Lidman}, C., {Moorwood},
  A.~F.~M., {Spyromilio}, J., {Hubin}, N., {Menten}, K.~M. 2002, \nat, 419, 694

\bibitem{1993yA&AS..96..269S}
{Schaller}, G., {Schaerer}, D., {Meynet}, G., {Maeder}, A. 1993, VizieR On-line
  Data Catalog: J/A+AS/96/269.~ Originally published in: 1992A\&AS...96..269S,
  96, 269

\bibitem{2002MNRAS.333..469S}
{Seto}, N. 2002, \mnras, 333, 469

\bibitem{2002ApJ...571..830S}
{Shara}, M.~M., {Hurley}, J.~R. 2002, \apj, 571, 830

\bibitem{1993Natur.364..423S}
{Sigurdsson}, S., {Hernquist}, L. 1993, \nat, 364, 423

\bibitem{2003MNRAS.344...22S}
{Spinnato}, P.~F., {Fellhauer}, M., {Portegies Zwart}, S.~F. 2003, \mnras, 344,
  22

\bibitem{1971swng.conf..443S}
{Spitzer}, L. 1971,
\newblock in Pontificiae Academiae Scientiarum Scripta Varia, Proceedings of a
  Study Week on Nuclei of Galaxies, held in Rome, April 13-18, 1970, Amsterdam:
  North Holland, and New York: American Elsevier, 1971, edited by D.J.K.
  O'Connell., p.443, p.~443

\bibitem{1987degc.book.....S}
{Spitzer}, L. 1987,
\newblock Dynamical evolution of globular clusters,
\newblock Princeton, NJ, Princeton University Press, 1987, 191 p.

\bibitem{1971ApJ...164..399S}
{Spitzer}, L.~J., {Hart}, M.~H. 1971a, \apj, 164, 399

\bibitem{1971ApJ...166..483S}
{Spitzer}, L.~J., {Hart}, M.~H. 1971b, \apj, 166, 483

\bibitem{2001NewA....6...79S}
{Springel}, V., {Yoshida}, N., {White}, S.~D.~M. 2001, New Astronomy, 6, 79

\bibitem{1990Natur.345...33S}
{Sugimoto}, D., {Chikada}, Y., {Makino}, J., {Ito}, T., {Ebisuzaki}, T.,
  {Umemura}, M. 1990, \nat, 345, 33

\bibitem{1996IAUS..174..141T}
{Taiji}, M., {Makino}, J., {Fukushige}, T., {Ebisuzaki}, T., {Sugimoto}, D.
  1996,
\newblock in IAU Symp. 174: Dynamical Evolution of Star Clusters: Confrontation
  of Theory and Observations, p.~141

\bibitem{2000ApJ...535..759T}
{Takahashi}, K., {Portegies Zwart}, S.~F. 2000, \apj, 535, 759

\bibitem{1996ApJ...457..834T}
{Timmes}, F.~X., {Woosley}, S.~E., {Weaver}, T.~A. 1996, \apj, 457, 834

\bibitem{1994MNRAS.268..871T}
{Tutukov}, A.~V., {Yungelson}, L.~R. 1994, \mnras, 268, 871

\bibitem{1968BAN....19..479V}
{van Albada}, T.~S. 1968, BAN, 19, 479

\bibitem{1984PASP...96..329V}
{van den Bergh}, S. 1984, \pasp, 96, 329

\bibitem{1995AJ....110.2700V}
{van den Bergh}, S. 1995, \aj, 110, 2700

\bibitem{2002AJ....124.3255V}
{van der Marel}, R.~P., {Gerssen}, J., {Guhathakurta}, P., {Peterson}, R.~C.,
  {Gebhardt}, K. 2002, \aj, 124, 3255

\bibitem{1960ZA.....50..184V}
{von Hoerner}, S. 1960, Zeitschrift Astrophysics, 50, 184

\bibitem{1963ZA.....57...47V}
{von Hoerner}, S. 1963, Zeitschrift Astrophysics, 57, 47

\bibitem{1999ApJ...519L..39W}
{Wandel}, A. 1999, \apjl, 519, L39

\bibitem{2001astro.ph..8461W}
{Wandel}, A. 2001,
\newblock in submitted to the ApJ,  8461

\bibitem{2000ApJ...539..331W}
{Watters}, W.~A., {Joshi}, K.~J., {Rasio}, F.~A. 2000, \apj, 539, 331

\bibitem{1996ApJ...473L..25W}
{White}, N.~E., {van Paradijs}, J. 1996, \apjl, 473, L25+

\end{thebibliography}
\end{document}